\documentclass[10pt,journal,compsoc]{IEEEtran}
\usepackage{listings}
\usepackage{ifthen}
\usepackage[table]{xcolor}
\usepackage{xspace}
\usepackage{comment}
\usepackage{graphicx}
\usepackage{amssymb}
\usepackage{amsmath}
\usepackage{balance}
\usepackage{amsfonts}
\usepackage{mathtools}
\usepackage{inconsolata}
\usepackage[T1]{fontenc}
\mathchardef\mhyphen="2D
\usepackage[linesnumbered,vlined,boxed,ruled,nofillcomment,noend]{algorithm2e}
\usepackage{setspace}
\usepackage{etoolbox}
\usepackage[hidelinks]{hyperref}
\usepackage[most]{tcolorbox}
\tcbset{%
    bugreportstyle/.style={
        enhanced,
        colback=white,
        colframe=black,
        fonttitle=\itshape,
        colbacktitle=white,
        coltitle=black,
        boxrule=1pt,
        boxsep=0pt,
        left=4pt,
        right=4pt,
        before skip=4pt,
        after skip=0pt,
        separator sign none,
        attach boxed title to top left={%
            yshift*=-\tcboxedtitleheight/2, 
            xshift=5mm},
        boxed title style={colframe=black}
    }
}
\newtcbtheorem{bugreport}{}{bugreportstyle}{number}
\usepackage{fancyvrb}
\usepackage{fvextra}
\usepackage{multirow}
\usepackage{makecell}
\usepackage{bm}
\usepackage{hhline}
\usepackage{pifont}
\newcommand{\cmark}{\ding{51}\xspace}%
\newcommand{\xmark}{\ding{55}\xspace}%
\usepackage{enumitem}
\usepackage{tikz}

\DeclareRobustCommand{\Arrow}[1]{%
\parbox{#1}{\tikz{\draw[->](0,0)--(#1,0);}}
}

\makeatletter
\patchcmd{\@algocf@start}% <cmd>
  {-1.5em}% <search>
  {0pt}% <replace>
  {}{}% <success><failure>
\makeatother

\SetKwBlock{Event}{upon event}{}
\SetKw{Continue}{continue}
\SetKw{Break}{break}
\SetAlgoSkip{}
\SetInd{3pt}{3pt}%set indentation size for algorithms

\include{macro}

\newcommand{\tech}{\textsc{EBug}\xspace}
\newcommand{\fusion}{\textsc{Fusion}\xspace}

\makeatletter
\newcommand{\shorteq}{%
  \settowidth{\@tempdima}{=}% Width of hyphen
  \resizebox{\@tempdima}{\height}{=}%
}
\makeatother

\makeatletter
\newcommand{\shortminus}{%
  \settowidth{\@tempdima}{-}% Width of hyphen
  \resizebox{\@tempdima}{\height}{-}%
}
\makeatother

\newcolumntype{C}[1]{>{\centering\arraybackslash}p{#1}}

\usepackage{etoolbox}

\makeatletter
\patchcmd\algocf@Vline{\vrule}{\vrule \kern-0.4pt}{}{}
\patchcmd\algocf@Vsline{\vrule}{\vrule \kern-0.4pt}{}{}
\makeatother
\ifCLASSOPTIONcompsoc
  \usepackage[nocompress]{cite}
\else
  \usepackage{cite}
\fi
\ifCLASSINFOpdf
\else
\fi
\begin{document}

\title{Enhancing Mobile App Bug Reporting via Real-time Understanding of Reproduction Steps}
%\author{Anonymous Author(s)}
%\begin{comment}
\author{Mattia Fazzini,~\IEEEmembership{Member,~IEEE,}
	   Kevin Moran,~\IEEEmembership{Member,~IEEE,}
	   Carlos Bernal-Cardenas,~\IEEEmembership{Student Member,~IEEE,}
	   Tyler Wendland,~\IEEEmembership{Student Member,~IEEE,}
	   Alessandro Orso,~\IEEEmembership{Fellow,~IEEE,} and
	   Denys Poshyvanyk,~\IEEEmembership{Member,~IEEE}
        % <-this % stops a space
\IEEEcompsocitemizethanks{
\IEEEcompsocthanksitem{M. Fazzini and T.Wendland are with the Department
of Computer Science \& Engineering, University of Minnesota, Minneapolis,
MN, 55455.\protect\\
E-mail: [mfazzini,wendl155]@umn.edu}
\IEEEcompsocthanksitem{K. Moran is with the Department
of Computer Science, George Mason University, Fairfax,
VA, 22030.\protect\\
E-mail: kpmoran@gmu.edu}
\IEEEcompsocthanksitem{C. Bernal Cardenas, and D. Poshyvanyk are with the Department of Computer Science, William \& Mary, Williamsburg, VA, 23185.\protect\\
E-mail: [cebernal,denys]@cs.wm.edu}
\IEEEcompsocthanksitem{A. Orso is with the School of Computer Science, College of Computing, Georgia Institute of Technology, Atlanta,
GA, 30332.\protect\\
E-mail: orso@cc.gatech.edu}}
\thanks{Manuscript received March, 2021}}%; revised August 26, 2015.}}
%\end{comment}


\markboth{IEEE Transactions on Software Engineering,~Vol.~\#, No.~\#,~2021}%
{Fazzini \MakeLowercase{\textit{et al.}}: Enhancing Bug Reporting via Real-time Natural Language Understanding and Predictive Modeling}



\IEEEtitleabstractindextext{%
\begin{abstract}
One of the primary mechanisms by which developers receive feedback about in-field failures of software from users is through bug reports. Unfortunately, the quality of manually written bug reports can vary widely due to the effort required to include essential pieces of information, such as detailed reproduction steps (S2Rs). Despite the difficulty faced by reporters, few existing bug reporting systems attempt to offer automated assistance to users in crafting easily readable, and conveniently reproducible bug reports. To address the need for proactive bug reporting systems that actively aid the user in capturing crucial information, we introduce a novel bug reporting approach called \tech. \tech assists reporters in writing S2Rs for mobile applications by analyzing natural language information entered by reporters in real-time, and linking this data to information extracted via a combination of static and dynamic program analyses. As reporters write S2Rs, \tech is capable of automatically suggesting potential future steps using predictive models trained on realistic app usages. To evaluate \tech, we performed two user studies based on $20$ failures from $11$ real-world apps.  The empirical studies involved ten participants that submitted ten bug reports each and ten developers that reproduced the submitted bug reports. In the studies, we found that reporters were able to construct bug reports $31$\% \textit{faster} with \tech as compared to the state-of-the-art bug reporting system used as a baseline.  \tech's reports were also \textit{more reproducible} with respect to the ones generated with the baseline. Furthermore, we compared \tech's prediction models to other predictive modeling approaches and found that, overall, the predictive models of our approach outperformed the baseline approaches. Our results are promising and demonstrate the feasibility and potential benefits provided by proactively assistive bug reporting systems.
\end{abstract}

}

\maketitle

\IEEEdisplaynontitleabstractindextext

% \fi
\IEEEpeerreviewmaketitle

%Paper Sections

\section{Introduction}
\label{sec:intro}

Developers rely on bug reports to identify and fix problems in software. These bug resolution activities have been shown to constitute a major part of the software maintenance process, which in turn, typically accounts for a majority of the development effort~\cite{25Tassey:NIST}. There are generally two major types of reports that developers must manage: those produced by automated systems and those manually constructed by users.
\revision{For in-field software failures that can be automatically recognized and captured, \ie those that are revealed by a \textit{known oracle} (\eg a crash or a program assertion failure), researchers have developed automated techniques that capture fine-grained failure information and devised approaches to reproduce the failures based on the captured information. However, prior research that empirically analyzed issue trackers in open source projects found that software problems identified by known oracles accounted for less than 50\% of the studied issues~\cite{Tan:EMSE'14}.
This finding illustrates that a majority of software problems in the studied open source systems stemmed from functional issues that must be reported \textit{manually}.}

Despite the prevalence and importance of manually reported bugs, most existing issue tracking systems largely rely upon loosely structured, free-from text entry to capture information from reporters. The lack of structure generally results in reports whose quality is largely dependent upon the experience and thoroughness of the reporter. Past studies have illustrated that the most important information for developers in bug reports is reproduction steps, which is also the most difficult information for reporters to provide, resulting in reports that developers are unable to reproduce~\cite{3Bettenburg:FSE08}. The difficulties often faced in manual bug reporting largely stem from the \textit{cognitive and lexical} gap that exists between reporters and developers~\cite{Moran:FSE15}. That is, there is typically a rift between the knowledge of the reporter of a bug and that of a developer, who has a wealth of domain specific knowledge about a given software system. In the current landscape of issue tracking systems, the task of bridging this gap falls almost entirely upon the stakeholders: either a reporter must spend extensive effort in crafting a report, or a developer must strive to ``translate'' the low-quality information provided in a poorly written report to map it to the code and ultimately reproduce and fix the issue.

To mitigate this problem, we investigate the potential of offloading part of the cognitive burden of bug reporting on the reporting system itself. Specifically, we aim to \textit{proactively assist} reporters in defining high-quality bug reproduction steps (S2Rs), so as to facilitate the reporting process and provide developers with more useful information, thus reducing the cognitive load on all the stakeholders. As past work showed, designing such a bug reporting system is a tenuous balancing act. On the one hand, in order to elicit more detailed information from reporters, a bug reporting system must impose some structure on the information being collected. On the other hand, doing so can make a system more difficult to use for reporters, which would ultimately limit its practical applicability and usefulness.

Prior work on the \fusion bug reporting interface has clearly illustrated this tension~\cite{Moran:FSE15}. \fusion leveraged information gleaned from static and dynamic analysis to populate a series of dropdown menus having the objective to help reporters in providing S2Rs. That work demonstrated that an enhanced reporting system with populated, structured fields can increase the quality of reports as measured by their reproducibility. However, the higher quality of \fusion's reports came at a cost, as it generally took reporters \textit{longer} to write reports than with traditional free-form, text-based reporting systems. Longer report creation times may indicate a higher cognitive load for reporters, which is likely to hurt adoption in practice. In this paper, we further investigate these trade-offs in intelligent bug reporting systems, with the aim of designing a system that finds a sweet spot between ease of use and effectiveness in capturing high quality bug reports.

To this end, we introduce the \tech reporting technique. At the core of \tech there are (1) a real-time engine for understanding S2Rs written in natural language and (2) a predictive model capable of suggesting likely S2Rs, trained on real application usages. Supporting this core functionality are automated static and dynamic analyses capable of extracting a detailed GUI model from a given mobile app, including screenshots and GUI-related metadata. The primary means of interaction with \tech is via a ``smart'' unstructured text field, wherein reporters are tasked with writing S2Rs in natural language. When a user begins writing, \tech automatically recognizes different components of a given S2R (\ie action, target GUI-component) and attempts to ``auto-complete'' missing information in a manner akin to Google's \textit{smart compose} feature in Gmail~\cite{smart-compose}. \tech's natural language understanding and predictive model are capable of recognizing and suggesting both entire S2Rs as well as individual components thereof. We developed \tech's predictive modeling using a novel application of \textit{n}-gram language modeling to sequences of actions from natural app usages collected by end-users. We provide a video demonstration of \tech in its online appendix~\cite{appendix}.

\revision{The main assumption for \tech to work effectively is that reported S2Rs describe GUI actions in the app. The description of an S2R needs to be provided through text but does not need to follow a pre-imposed structure. Forthermore, \tech does not make assumptions on the type of failure reported by the user as it focuses on the S2Rs associated with the failure.}

We evaluated \tech in two user studies that involved both bug reporting and bug reproduction tasks, based on $20$ failures from $11$ real-world apps. Specifically, the studies included ten participants submitting ten bug reports each and ten developers reproducing the submitted bug reports. \revision{The ten participants that submitted the bug reports were not developers but four of the participants have industry exposure through internships as software engineers. The participants that reproduced bug reports were all developers. These developers have experience in app development, but were not the developers of the apps considered in the study.} Comparing bug creation and bug reproduction against the state-of-the-art \fusion approach, we found that participants were able to create high quality reports quicker, as compared to \fusion, and that the resulting reports were more reproducible that those created with \fusion.

\noindent
This paper makes the following contributions:

\begin{itemize}
	\item{\tech, a bug reporting system for mobile apps that leverages automated natural language understanding, static and dynamic analyses, and predictive modeling to facilitate the reporting process.}

%%mf: I removed the following two items as R3 feels strong about the fact that and evaluation study is part of  every scientific contribution
%	\item{A comprehensive user study to evaluate the effectiveness and efficiency of \tech
%	compared to the \fusion bug reporting system. In the study, we also evaluate and compare the two approaches in terms of user experience. The results of this study demonstrate that \tech allows for faster creation of bug reports that are more reproducible compared to \fusion.}
%	\item{An empirical investigation of predictive modeling approaches, whose results demonstrate that  \tech's predictive models outperform the baseline approaches when predicting S2Rs or portion thereof.}

	\item{An implementation of \tech for Android apps that is publicly available, together with the artifacts and the infrastructure we used in our evaluation~\cite{appendix}.}
	
	\item{\revisiontwo{An evaluation that provides initial evidence of the efficiency, effectiveness, and usefulness of \tech.}}
	
\end{itemize}

\section{Terminology \& Motivating Example}
\label{sec:term-motivating}

This section introduces some relevant terminology and presents an example that we use to motivate our approach.

\subsection{Terminology}
\label{sec:terminology}

Given a bug report that describes a failure for an app, we informally use the terms \textit{relevant failure} and \textit{relevant app} to indicate the failure and the app. We use the term \textit{steps to reproduce (S2R)} to indicate the textual description, in the bug report, of an operation that should be performed on the relevant app to reproduce the relevant failure (\eg Enter ``Transaction'' in the ``Description'' text box). We use the terms \textit{GUI action} (or simply \textit{action}) and \textit{GUI interaction} (or simply \textit{interaction}) interchangeably to indicate the operation performed on relevant app's GUI. A GUI action is composed by its \textit{type} (\eg typing something in the GUI), the \textit{GUI element} (or simply \textit{element}) affected by the action (\eg the text box in the GUI having the label ``Description''), and, if present, the action's parameters (\eg the ``Transaction'' text). We use the term \textit{target GUI element} (or simply \textit{target}) to indicate the GUI element affected by the action.

{
\setlength{\fboxrule}{0pt}
\begin{figure}[t]
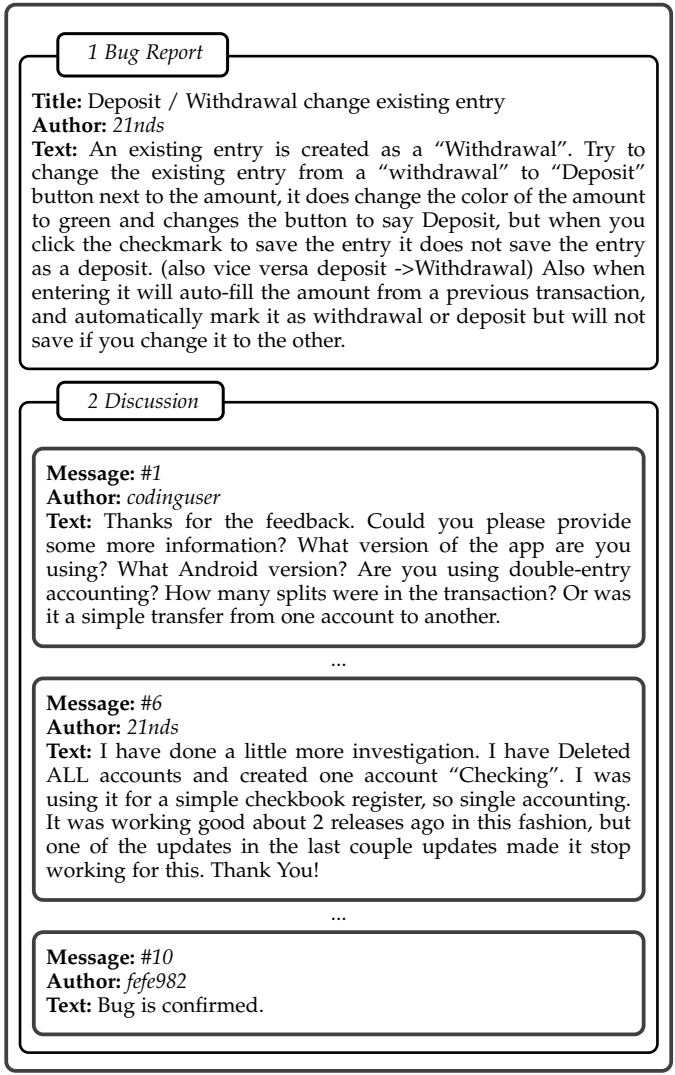

\centering
\footnotesize
\makebox[\columnwidth]{
%first example
\framebox{
\begin{minipage}[t]{\columnwidth}
\begin{tcolorbox}[colback=white, boxsep=0pt, left=4pt, right=4pt, before skip=4pt,after skip=0pt]
%\textbf{Title:} Deposit / Withdrawal change existing entry
\begin{bugreport}{Bug Report}{}

%\begin{tcolorbox}[colback=white, boxsep=0pt, left=4pt, right=4pt, before skip=4pt,after skip=0pt]
\textbf{Title:} Deposit / Withdrawal change existing entry

\textbf{Author:} \textit{21nds}

\textbf{Text:} An existing entry is created as a ``Withdrawal''. Try to change the existing entry from a ``withdrawal'' to ``Deposit'' button next to the amount, it does change the color of the amount to green and changes the button to say Deposit, but when you click the checkmark to save the entry it does not save the entry as a deposit. (also vice versa deposit ->Withdrawal) Also when entering it will auto-fill the amount from a previous transaction, and automatically mark it as withdrawal or deposit but will not save if you change it to the other.
%\end{tcolorbox}
\end{bugreport}
\begin{bugreport}{Discussion}{}
\begin{tcolorbox}[colback=white, boxsep=0pt, left=4pt, right=4pt, before skip=4pt,after skip=0pt]
\textbf{Message:} \textit{\#1}

\textbf{Author:} \textit{codinguser}

\textbf{Text:} Thanks for the feedback. Could you please provide some more information?
What version of the app are you using? What Android version?
Are you using double-entry accounting?
How many splits were in the transaction? Or was it a simple transfer from one account to another.
\end{tcolorbox}
\center{...}
\begin{tcolorbox}[colback=white, boxsep=0pt, left=4pt, right=4pt, before skip=4pt,after skip=0pt]
\textbf{Message:} \textit{\#6}

\textbf{Author:} \textit{21nds}

\textbf{Text:} I have done a little more investigation.
I have Deleted ALL accounts and created one account ``Checking''. I was using it for a simple checkbook register, so single accounting. It was working good about 2 releases ago in this fashion, but one of the updates in the last couple updates made it stop working for this. Thank You!
\end{tcolorbox}
\center{...}
\begin{tcolorbox}[colback=white, boxsep=0pt, left=4pt, right=4pt, before skip=4pt,after skip=0pt]
\textbf{Message:} \textit{\#10}

\textbf{Author:} \textit{fefe982}

\textbf{Text:} Bug is confirmed.
\end{tcolorbox}
\end{bugreport}
\end{tcolorbox}
\caption{Bug report and reproduction discussion from \textsc{GnuCash} app.}
\label{fig:bugreport}
\end{minipage}
}
}
\vspace{-15pt}
\end{figure}
}

\subsection{Motivating Example}
\label{sec:motivating}

Our motivating example, shown in Fig.~\ref{fig:bugreport}, is a real bug report~\cite{2020_github_deposit} for \textsc{GnuCash}~\cite{2020_github_gnucash}, a widely used real-world app to track expenses that has been installed over 100K times~\cite{2020_googleplay_gnucash}. In the figure, the section labeled \textit{Bug Report} contains the bug report submitted by user \textit{21nds}, which contains the same information as the one in \textsc{GnuCash}'s issue tracking system~\cite{2020_github_deposit}. The report describes a bug that manifested itself when the user changed the type of an account's transaction from \texttt{Withdrawal} to \texttt{Deposit}, and the app failed to store this change. The bug report is followed by a discussion (section labeled \textit{Discussion} in Fig.~\ref{fig:bugreport}), which was necessary for developers to understand and reproduce the bug. In the first message (message \#1), an app developer (\textit{codinguser}) asked the user to provide additional information on the bug. After a few messages among developers (message \#6), the user reported that she deleted all accounts, created a \textit{single accounting} account, and the transaction that she created in this account was the one leading to the bug described in the original report. After the user provided these additional details, an app developer (\textit{fefe982}) could confirm the bug (message \#10). The bug report took six days and ten discussion messages to be reproduced and confirmed.

In the bug report, the user provided a description containing some of the S2Rs but forgot to report some essential S2Rs that are necessary to reproduce the bug. These circumstances caused additional and unnecessary work for developers. Our bug reporting approach, \tech, aim to decrease this extra work by guiding users in thoroughly and efficiently reporting S2Rs so that developers can more easily reproduce reported bugs.
\section{The EBug Approach}
\label{sec:approach}

In this section, we present \tech, an approach for enhancing mobile app bug reporting. The basic idea behind \tech is to help users write high-quality bug reports by processing natural language S2Rs in real-time (\ie as S2Rs are typed in the report) and guiding its users in adding S2Rs (or portions thereof) that make the bug reports complete and accurate.

\begin{figure*}[t]
	\centering
	\centerline{\includegraphics[width=\textwidth]{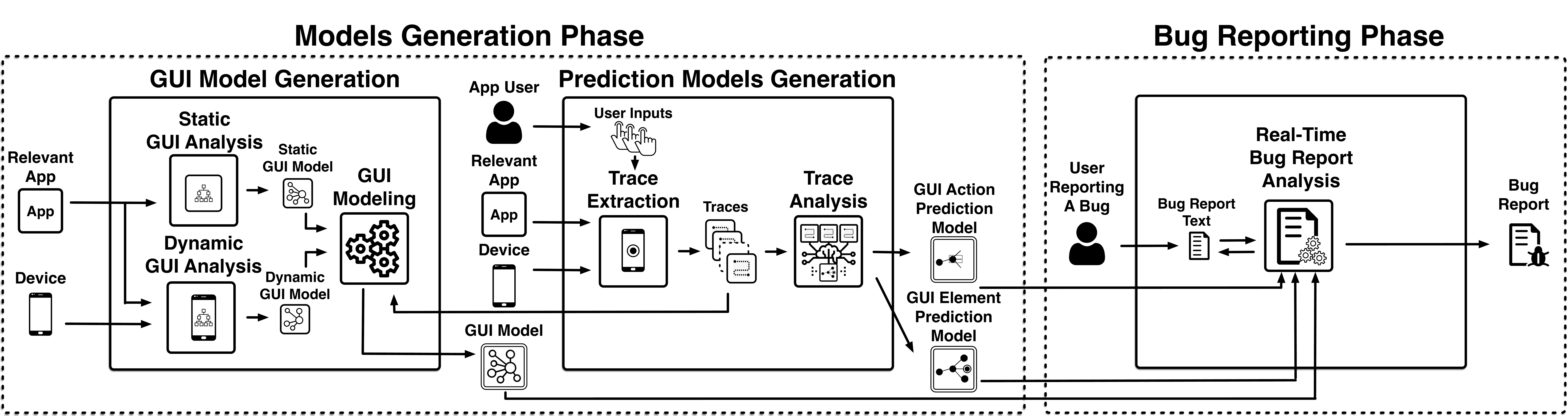}}
	\vspace{-5pt}
	\caption{High-level overview of the \tech approach.} \label{fig:approach}
	\vspace{-15pt}
\end{figure*}

Fig.~\ref{fig:approach} provides an overview of \tech's workflow and its two main phases: \textit{models generation} and \textit{bug reporting}.

The models generation phase operates offline and generates three app-related models (depicted with solid, double borders in Fig.~\ref{fig:approach}). First, in its \textit{GUI model generation} step, this phase takes as input the \textit{relevant app} and a \textit{device} and combines static and dynamic analysis to compute the \textit{GUI model} of the app.
Then, in its \textit{prediction models generation} step, this phase processes execution traces generated by the \textit{app users} and creates two statistical models: the \textit{GUI action prediction model} and the \textit{GUI element prediction model}.

%The bug reporting phase uses the generated models to help users, as they type their bug reports, generate accurate and complete S2Rs.
%More precisely,
When a user is reporting a failure, the bug reporting phase processes the S2Rs in real-time using natural language processing, maps the content of the S2Rs to the GUI model of the relevant app, and uses the prediction models to suggest S2Rs (or portions thereof). The final output of this phase is a \textit{bug report} containing S2Rs aimed at reproducing the failure experienced by the user. The rest of this section describes the two phases of \tech in detail.

\subsection{Models Generation Phase}
\label{sec:mode-gen-phase}

\subsubsection{GUI Model Generation}
\label{sec:gui-mod-gen}

The GUI model ($\mathit{GM}$) captures which GUI actions are possible in the different screens of the relevant app and encodes GUI properties associated with actions. \tech generates $\mathit{GM}$ offline, as it might not be always practical to analyze S2Rs while the user is typing them in the bug report. For example, a solution that dynamically analyzes the GUI of the app while the user is typing the S2Rs might need to perform time-consuming operations for backtracking the exploration (\eg restarting the app~\cite{Choi:OOPSLA13}) when the user edits/deletes previously typed S2Rs, which might significantly affect the performance of the approach.

\tech builds the $\mathit{GM}$ of the relevant app by (1) computing a \textit{static GUI model} ($\mathit{GM_{s}}$) through static analysis, (2) extracting a \textit{dynamic GUI model} ($\mathit{GM_{d}}$) through a dynamic analysis, and (3) combining $\mathit{GM_{s}}$ and $\mathit{GM_{d}}$ into $\mathit{GM}$. The use of both static and dynamic analysis allows \tech to achieve high-coverage in modeling the GUI of the app and its corresponding GUI actions.
%%mf: removed to shortend the paper
%\tech also uses $\mathit{GM}$ to provide visual feedback to app users as they write S2Rs in the bug reporting phase.

The GUI models computed by \tech ($\mathit{GM_{s}}$, $\mathit{GM_{d}}$, and $\mathit{GM}$) are directed graphs $\mathit{G}$~=~$(N, E)$, where $N$ is the set of nodes in $\mathit{G}$ and $E \subseteq N \times N$. \tech creates two types of nodes. A node $s \in S$ corresponds to a screen in the app, whereas a node $v \in V$ corresponds to a GUI element. The set of nodes $N$ contains the sets $S$ and $V$, therefore having $N$~=~$S \cup V$. Both types of nodes have properties that are stored in the nodes as tuples. The properties of a screen $s$ consist of its name and screenshot ($\langle \mathit{name}, \mathit{screenshot} \rangle$). 
The properties of an element $v$ consist of its text, identifier, type, and screenshot ($\langle \mathit{text}, \mathit{id}, \mathit{type}, \mathit{screenshot} \rangle$). (If one or more of these properties is missing or cannot be computed for a given element, \tech stores an empty value for it.)
\tech uses the text and the identifier of an element to map S2Rs to GUI actions in the app during the bug reporting phase. We selected these properties as they are particularly suitable for characterizing and identifying GUI elements from textual information in bug reports~\cite{2018_issta_fazzini_automatically}. 

\tech computes two types of edges $E$~=~$C \cup T$. A containment edge $c \in C \subseteq S \times V$ captures the containment relationship between a screen and a GUI element ($s \xrightarrow{c} v$), whereas a transition edge $t \in T \subseteq V \times S$ connects a GUI element to a screen ($v \xrightarrow{t} s$) and models the transition to $s$ after exercising $v$. Transition edges have two properties $t$ ($\langle \mathit{a\mhyphen type}, \mathit{t\mhyphen type} \rangle$) that capture the ($\mathit{a\mhyphen type}$) of action performed on $v$ to trigger the transition to $s$, and the ($\mathit{t\mhyphen type}$) of analysis---static or dynamic---used to extract the information from the app. Finally, \tech models clicks, long clicks, scrolls, text typing as actions in the model. It also represents screen rotations in the model by associating to each screen $s$ a screen rotation operation. Specifically, screen rotations are represented with a dummy GUI element $v$ contained in $s$ whose transition $t$ reaches $s$. The rest of this section describes how \tech computes $\mathit{GM}$.

To build the static GUI model $\mathit{GM_{s}}$, \tech leverages a technique presented in related work~\cite{2015_ase_yang_static} that captures (1) the creation and propagation of GUI elements in the code, (2) the relationship between screens and elements, (3) what actions can be performed on the elements, and (4) the effects of source code operations on the elements.
\revision{To capture this information, the static analysis used by \tech~\cite{2015_ase_yang_static} builds and processes a constraint graph. The nodes in the constraint graph model statements creating and affecting the behavior of screens and GUI elements. Edges represent constraints on the flow of values between statements. The analysis identifies properties and relationships associated with screens and GUI elements by propagating values in the graph representing screens and GUI element objects. To propagate the values, the analysis uses a fixed-point algorithm. After running the fixed-point algorithm, the analysis extracts properties and relationships of screens and GUI elements by analyzing the values flowing to specific statements in the graph.}
Based on the results of the static analysis, \tech first adds all the screens $S_{s}$ to $\mathit{GM_{s}}$; it then adds all the element nodes $V_{s}$, while creating a containment edge $c_{s}$ between an element's screen and the element; it finally adds transition edges $T_{s}$ between nodes in $V_{s}$ and $S_{s}$ (setting $\mathit{t\mhyphen type}$~=~$\texttt{STATIC}$) based on the identified transitions.

\tech builds the dynamic GUI model $\mathit{GM_{d}}$ by dynamically exploring the relevant app through a depth-first traversal (DFT) that starts from the \textit{initial/main} screen of the app and navigates through the app by interacting with the GUI elements on each screen. 
\revision{Specifically, the traversal clicks on all the clickable components and types pre-defined text (\ie ``Test'') on all editable elements that are reachable through the traversal. (The pre-defined value of text inputs could be extended in future work to use semantically relevant inputs as done in related work~\cite{2020_ast_wnwarang_testing}.) Before each step of the traversal is executed, the technique extracts relevant information about the current screen and its GUI elements. The technique then executes the action associated with each GUI component in a depth-first manner. During the exploration, the technique also captures the transitions between different screens by tracking the visited screens. In the traversal, if a GUI element is clicked and that operation would bring the traversal to a screen not belonging to the app (\eg clicking a web link that would launch a browser app), the technique would execute a back command in order to continue the exploration in the relevant app. If the traversal exits the app and reaches the home screen of the device/emulator for any reason, the technique re-launches the app and continues the traversal using the depth-first strategy.}
During GUI exploration, there may be cases in which there are no GUI actions are available on the screen and \tech must backtrack to make progress. 
When this occurs, our approach restarts the app (while resetting its data) and navigates to the screen containing the GUI element that should be exercised next. This step is possible because \tech keeps track of visited screens and exercised GUI elements.

 \begin{figure*}[t]
	\centering
	\begin{minipage}[t]{.5\textwidth}
	\centerline{\includegraphics[width=.99\textwidth]{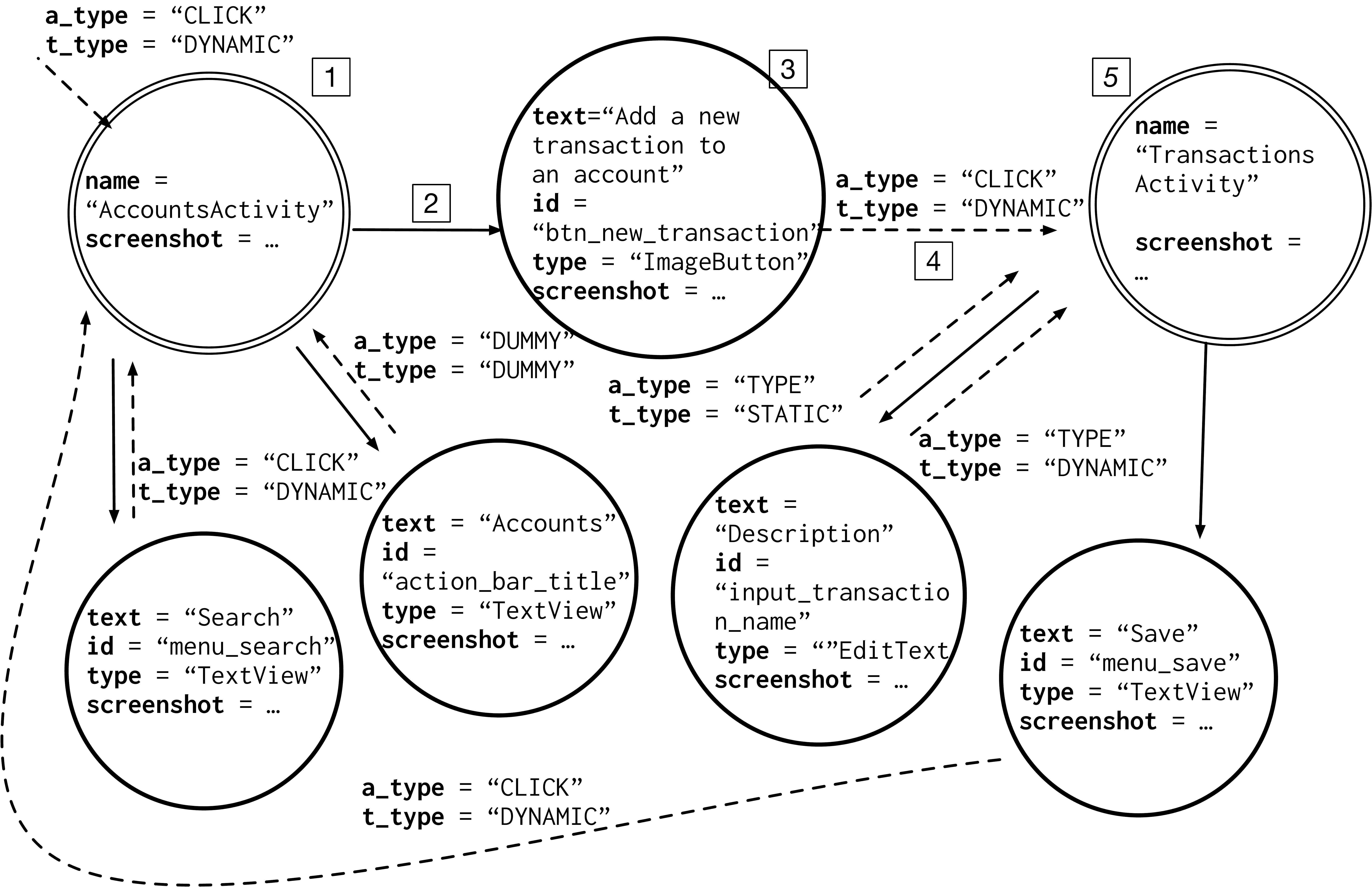}}
	\begin{minipage}{.9\textwidth}
	\caption{Portion of the GUI model graph (GM) that includes parts of the \texttt{AccountsActivity} and \texttt{TransactionsActivity} screens.} 			\label{fig:guimodel}
	\end{minipage}
	\end{minipage}
	\begin{minipage}[t]{.245\textwidth}
	\centerline{\includegraphics[trim=0 900 0 0,clip,width=.98\textwidth]{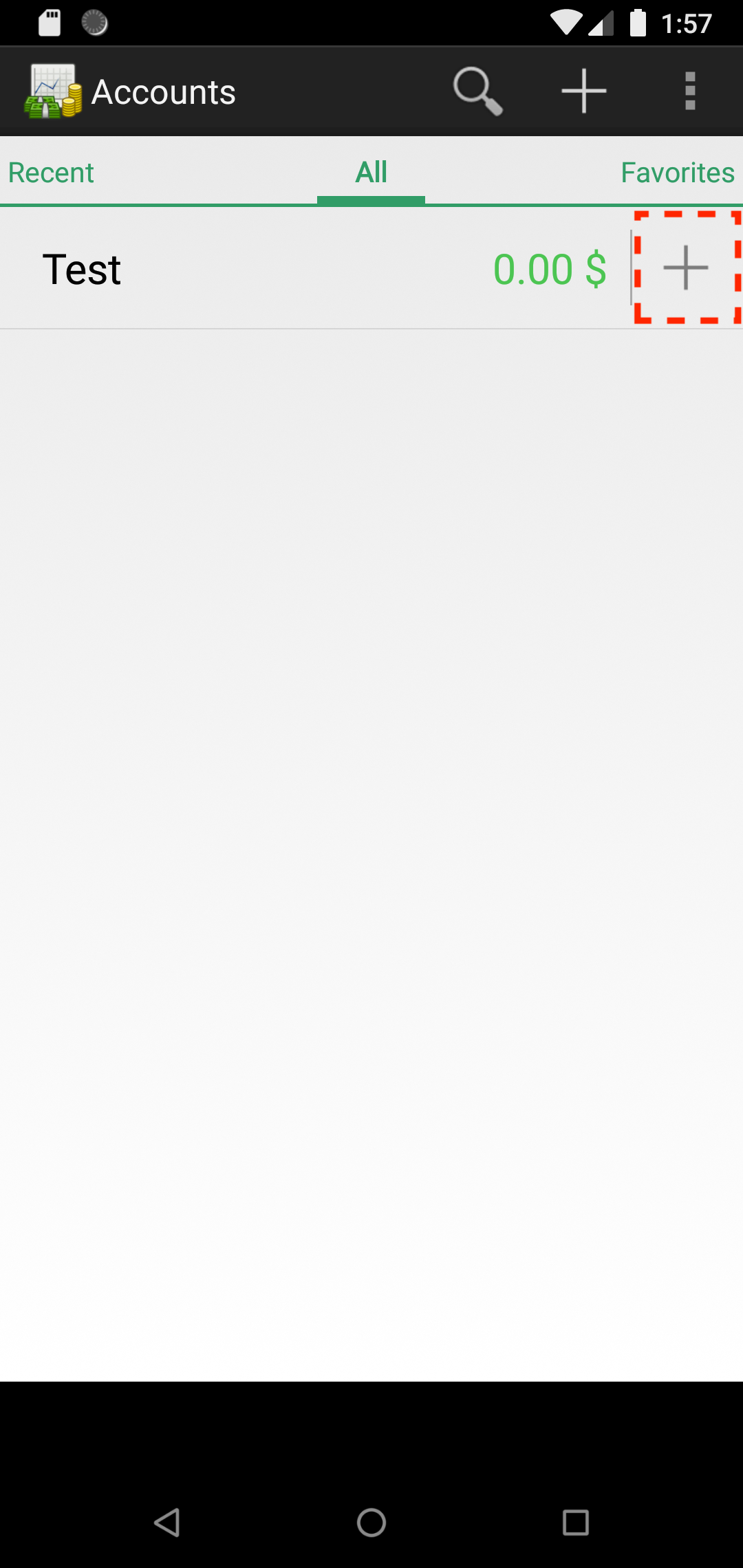}}
	\begin{minipage}{.9\textwidth}
	\caption{\texttt{AccountsActivity} screen in the \textsc{GnuCash} app.}
	\label{fig:accountsscreen}
	\end{minipage}
	\end{minipage}
	\begin{minipage}[t]{.245\textwidth}
	\centerline{\includegraphics[trim=0 900 0 0,clip,width=.98\textwidth]{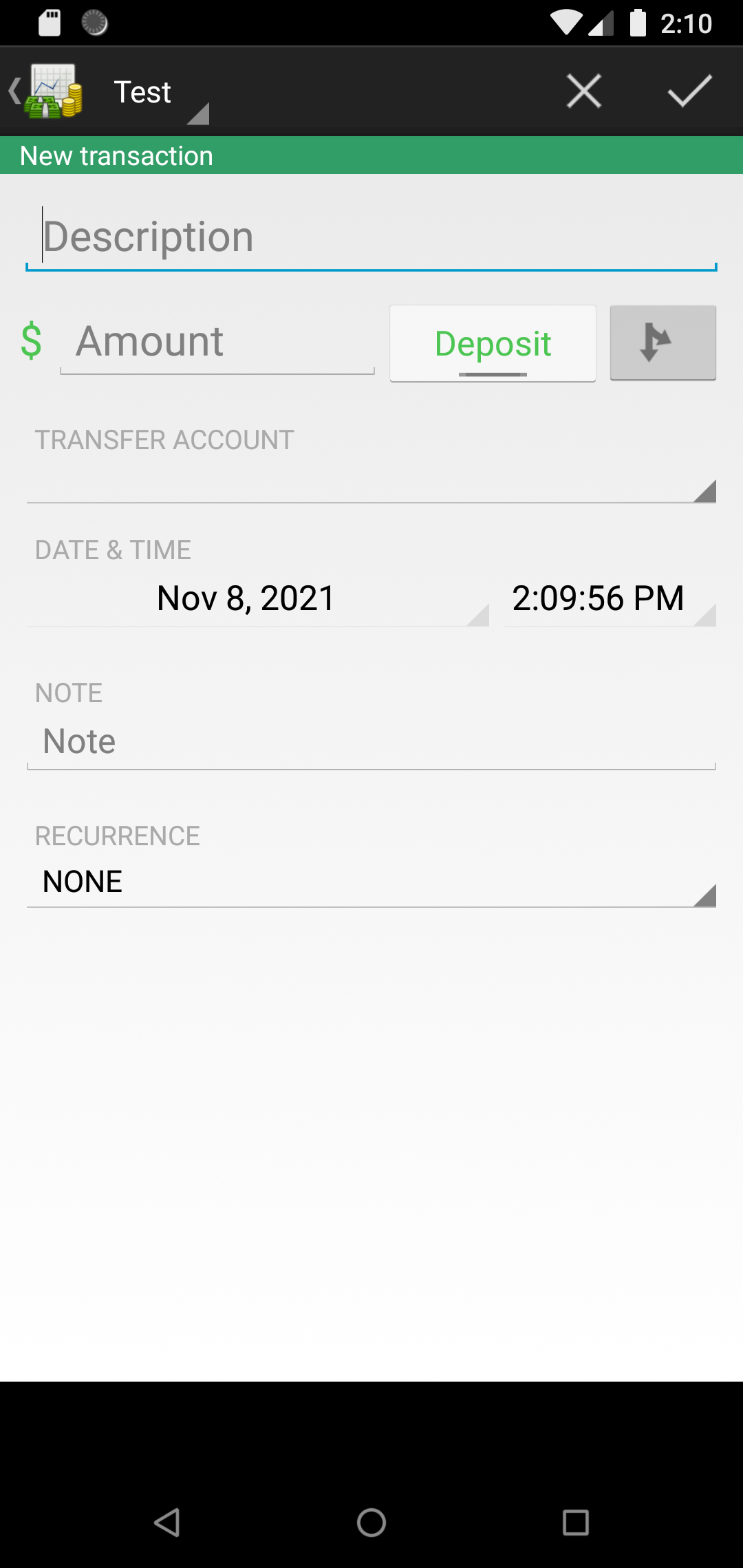}}
	\begin{minipage}{.9\textwidth}
	\caption{\texttt{TransactionsActivity} screen in the \textsc{GnuCash} app.}
	\label{fig:transactionsscreen}
	\end{minipage}
	\end{minipage}
	\vspace{-10pt}
\end{figure*}

When \tech visits a new screen $s_{d}$ (\ie $s_{d} \notin S_{d}$), it adds $s_{d}$ to $\mathit{GM_{d}}$. It then adds all the elements $\{v^{1}_{d},...,v^{n}_{d}\}$ in $s_{d}$ to $\mathit{GM_{d}}$. Finally, it creates a containment edge $c^{i}_{d}$ from $s_{d}$ to $v^{i}_{d}$ for each element $v^{i}_{d}$ (with $i \in \{1,...,n\}$) just added to $\mathit{GM_{d}}$. \tech determines that two screens are the same if their GUI trees (composed of their GUI elements) are the same. In computing this information, \tech does not consider the elements contained in ``container elements'' (\eg list containers), as the number of contained elements can change dynamically. However, if a new element is present in such containers, this element is added to $\mathit{GM_{d}}$. When \tech identifies that a screen $s_{d}$ is different from the previously visited screen $s^{p}_{d}$, it adds a transition edge from $v^{p}_{d}$ to $s_{d}$ where $v^{p}_{d}$ is is the element from $s^{p}_{d}$ exercised by the approach (setting $\mathit{t\mhyphen type}$~=~$\texttt{DYNAMIC}$). While adding a node or an edge to $\mathit{GM_{d}}$ \tech also collects the node's or edge's properties. For example, when \tech adds a GUI element to the graph, the approach also saves the element's screenshot. Finally, for elements that do not have a $\mathit{text}$ property, \tech tries to extract this information from the screenshot of the element using optical character recognition (based on the algorithm implemented in~\cite{2020_github_tesseract}).

\tech creates $\mathit{GM}$ by computing the union~\cite{1969_harary_graph} of $\mathit{GM_{s}}$ and $\mathit{GM_{d}}$ that is, $\mathit{GM}$~=~$\mathit{GM_{s}} \cup \mathit{GM_{d}}$~=~$(N_{s} \cup N_{d}, E_{s} \cup E_{d})$. While computing the union, \tech identifies matching nodes in the two graphs. In this process, the approach does not consider the $\mathit{screenshot}$ property. However, \tech uses the $\mathit{screenshot}$ values from the nodes in $\mathit{GM_{d}}$ when the approach adds the nodes in $\mathit{GM}$. After taking the union of the two graphs, if a node $v \in V$ does not have any transition to any node in $S$, \tech creates a ``dummy'' %transition $v \xrightarrow{t} s$ where
 $s \in S \land s \xrightarrow{c} v \land c \in C$ and $t$~=~$\langle \texttt{DUMMY}, \texttt{DUMMY} \rangle$ to denote that after performing an action on $v$ the screen does not change. During the bug reporting phase, \tech uses $\mathit{GM}$ to identify the GUI actions described by the S2Rs.

Fig.~\ref{fig:guimodel} provides a graphical representation for a portion of the $\mathit{GM}$ computed from \textsc{GnuCash}, which we presented in the motivating example (Sec.~\ref{sec:term-motivating}). This portion of the graph models the screens depicted in Fig.~\ref{fig:accountsscreen} and~\ref{fig:transactionsscreen}. Nodes with double borders are screens, while the remaining nodes are the elements of the screens. Edges with solid lines are containment edges, while the other edges are transitions.
%%@Alex: removed to shorten the paper but I also feel it would be good to have it
%The node with the label $\fbox{1}$ represents the \texttt{AccountsActivity} screen (portrayed in Fig.~\ref{fig:accountsscreen}). Users can press the ``plus'' image button (highlighted with a dashed red line in the figure) to access the \texttt{TransactionsActivity} screen (depicted in Fig.~\ref{fig:transactionsscreen}), where users can create and save a new transaction. The node in the graph labeled with $\fbox{3}$ models the image button. As it is possible to see from Fig.~\ref{fig:guimodel}, the $\mathit{GM}$ suitably represent that this node is contained (edge with label $\fbox{2}$) in the \texttt{AccountsActivity} screen and a click (edge labeled with $\fbox{4}$) on the image icon brings the user to the \texttt{TransactionsActivity} screen (node labeled with $\fbox{5}$).

\subsubsection{Prediction Models Generation}
\label{sec:pred-mod-gen}

In this step, \tech builds two statistical models to capture how users interact with the relevant app and predict likely interactions given a certain context (\ie a sequence of interactions). The two models are the GUI action and the GUI element prediction models (GAPM and GEPM in short, respectively). The bug reporting phase uses the GAPM to provide recommendations for not-yet-typed S2Rs. \tech leverages the GEPM to, instead, help users complete partially written S2Rs.

\tech computes the two models from user-generated execution traces. \tech can collect these traces when developers distribute the relevant app for alpha or beta testing, the app runs in the field, or during in-house user testing. \tech produces a trace by intercepting the actions performed on the app's GUI and logging them in the trace. The approach identifies clicks, long clicks, scrolling actions, text typing actions, and screen rotations. For each user execution, \tech stores actions as a sequence of tuples in the trace. For each action, \tech stores the tuple:
$\langle\mathit{s\mhyphen name},~\mathit{a\mhyphen type},~\mathit{e\mhyphen type},~\mathit{e\mhyphen id}\rangle$. The tuple contains the name of the screen ($\mathit{s\mhyphen name}$) where the action was performed, the type ($\mathit{a\mhyphen type}$) of the action, and the type ($\mathit{e\mhyphen type}$) and the identifier ($\mathit{e\mhyphen id}$) of the element exercised by the action. \tech collects this information to uniquely identify an action during the app's execution.
The approach uses the traces to also refine the GUI model. Specifically, \tech translates the information contained in an action (containing screen, action type, and affected element) into corresponding nodes and edges in the GUI model.

\begin{figure*}[t]
	\centering
	\begin{minipage}[t]{.45\linewidth}
	 \begin{Verbatim}[fontsize=\scriptsize, commandchars=\\\{\}]
...
\(\langle{}\)AccountsActivity,CLICK,TextView,menu_add_account\(\rangle{}\)
\(\langle{}\)AccountsActivity,TYPE,EditText,edit_text_account_name\(\rangle{}\)
\(\langle{}\)AccountsActivity,CLICK,TextView,menu_save\(\rangle{}\)
\(\langle{}\)AccountsActivity,CLICK,ImageButton,btn_new_transaction\(\rangle{}\)
\(\langle{}\)TransactionsActivity,TYPE,EditText,input_transaction_name\(\rangle{}\)
\(\langle{}\)TransactionsActivity,TYPE,EditText,input_transaction_amount\(\rangle{}\)
\(\langle{}\)TransactionsActivity,CLICK,TextView,menu_save\(\rangle{}\)
...
	 \end{Verbatim}
	\begin{minipage}{.9\textwidth}
	\vspace{-11pt}
	\caption{Portion of a user trace collected from \textsc{GnuCash} app.}
	\label{fig:usertrace}
	\vspace{10pt}
	\end{minipage}
	\end{minipage}	
	\begin{minipage}[t]{.54\linewidth}
	\begin{Verbatim}[fontsize=\scriptsize, breaklines=true, breakanywhere=true]
... AccountsActivity.CLICK.menu_add_account.TextView AccountsActivity.TYPE.EditText.edit_text_account_name AccountsActivity.CLICK.TextView.menu_save AccountsActivity.CLICK.ImageButton.btn_new_transaction TransactionsActivity.TYPE.EditText.input_transaction_name TransactionsActivity.TYPE.EditText.input_transaction_amount TransactionsActivity.CLICK.TextView.menu_save ...
	 \end{Verbatim}
	\begin{minipage}{.99\textwidth}
	\vspace{4pt}
	\caption{Portion of the GUI action trace computed from the user trace in Fig.~\ref{fig:usertrace}.}
	\label{fig:guiactiontrace}
	\end{minipage}
	\end{minipage}	
	\begin{minipage}[t]{.99\linewidth}
	 \begin{Verbatim}[fontsize=\scriptsize, breaklines=true, breakanywhere=true]
... AccountsActivity.CLICK AccountsActivity.menu_add_account.TextView AccountsActivity.TYPE AccountsActivity.EditText.edit_text_account_name AccountsActivity.CLICK AccountsActivity.TextView.menu_save AccountsActivity.CLICK AccountsActivity.ImageButton.btn_new_transaction TransactionsActivity.TYPE TransactionsActivity.EditText.input_transaction_name TransactionsActivity.TYPE TransactionsActivity.EditText.input_transaction_amount TransactionsActivity.CLICK TransactionsActivity.TextView.menu_save ...
	 \end{Verbatim}
	\begin{minipage}{.9\textwidth}
	\vspace{-7pt}
	\caption{Portion of the GUI element trace computed from the user trace in Fig.~\ref{fig:usertrace}.}
	\label{fig:guielementtrace}
	\end{minipage}
	\end{minipage}
	\vspace{-10pt}
\end{figure*}

Fig.~\ref{fig:usertrace} provides an example of a user trace. The figure reports a portion of a trace collected from the \textsc{GnuCash} app. In this portion, the user created an account and added a transaction to the account. In Fig.~\ref{fig:usertrace}, the fourth tuple represents a click on the ``plus'' button highlighted with a dashed red line in Fig.~\ref{fig:accountsscreen}.
%%@Alex: removed to shorten the paper but I also feel it would be good to have it
%In this tuple, \texttt{AccountsActivity} is the name of the screen ($s\mhyphen name$) where the action was performed, \texttt{CLICK} is the type ($a\mhyphen type$) of the action, \texttt{ImageButton} is the type ($e\mhyphen type$) of the element exercised by the action, and \texttt{btn\_new\_transaction} is the identifier ($e\mhyphen id$) of the element.

\tech computes the two prediction models using two different sets of traces. \tech derives these two sets of traces from the user traces. We call the two sets the \textit{GUI action traces} (GATs) and the \textit{GUI element traces} (GETs). \tech uses the former set to create the GAPM and the latter set to generate the GEPM. \tech creates a GAT and a GET for each user trace. The GAT and the GET are composed of a space-separated sequence of tokens. \tech computes the tokens from the tuples in the corresponding user trace. Fig.~\ref{fig:guiactiontrace} and~\ref{fig:guielementtrace} portray the GAT and the GET, respectively, derived from the user trace depicted in Fig.~\ref{fig:usertrace}. (In both traces, the symbol $\hookrightarrow$ indicates that the sequence included in the trace continues.)
%%@Alex: removed to shorten the paper but I also feel it would be good to have it
%The fourth tuple from Fig.~\ref{fig:usertrace} is converted into the token $\texttt{AccountsActivity}.\texttt{CLICK}.\texttt{ImageButton}.\texttt{btn\_new\_transaction}$ in the GAT and the same tuple is tramslated into the sequence of the two tokens $\texttt{AccountsActivity}.\texttt{CLICK}$ and {\small$\texttt{AccountsActivity}.\texttt{ImageButton}.\texttt{btn\_new\_transaction}$} in GET.

\tech uses \textit{n}-gram-based language modeling~\cite{2009_jurafsky_speech} to build the GAPM and the GEPM from the GATs and the GETs, respectively. At a high level, language models assign probabilities to sequences of words and can be used to estimate the probability of a word given its preceding words (also known as the history of a word)~\cite{2009_jurafsky_speech}. \tech uses this characteristic of language models during its bug reporting phase to predict the occurrence of an S2R (or part thereof) given a sequence of previously occurring S2Rs. Because the GUI action and element traces do not contain sequences of S2Rs but contain sequences of GUI-action-based tokens, in this step, \tech creates token-based prediction models. \tech maps tokens to S2Rs during its bug reporting phase. We chose to leverage n-gram based models over other more complex sequence based prediction models (RNNs and other Neural Language models) as such models typically require large amounts of training data to be effective.

\revision{
\tech's \textit{n}-gram models are able to provide suggestions for sequences that are shorter than the order of the \textit{n}-gram models or for sequences that appear with a slightly different context (\ie arrangement) by leveraging the Kneser-Ney smoothing method~\cite{1998_chen_tr_an}. This method is one of the most commonly used and best performing \textit{n}-gram smoothing methods~\cite{2009_jurafsky_speech}. Using the Kneser-Ney smoothing method, lower-order \textit{n}-gram sequences are encoded in the models. Lower-order \textit{n}-gram sequences are then used to make predictions for sequences shorter than the order of the models and to provide suggestions for sequences appearing with a slightly different context~\cite{2009_jurafsky_speech}.
}

To compute \tech's prediction models, \tech treats trace tokens as words and uses the \text{GATs} and GETs as the training data to build the the \textit{n}-gram-based models characterizing the GAPM and the GEPM.
\tech's \textit{n}-gram models use closed vocabularies that are computed using the tokens of the training traces.
\revision{
Based on this characteristic, when \tech translates the text of an S2R into a prediction model token in the bug reporting phase, and the token is not part of the model vocabulary, the technique does not make any suggestions to the reporter. Because \tech considers \textit{n}-gram  based models and the length of the token sequence used to make the prediction has a fixed length (\ie \textit{n}-1), a token that is not part of the vocabulary would impact (\ie prevent) only \textit{n}-1 predictions. This characteristic does not affect the ability of the models to identify suggestions for sequences that were not observed in the training traces.}

To instantiate the language modeling framework described so far, \tech needs to determine the order of the \textit{n}-grams used by the models. \tech determines the order using the concept of \textit{wasted effort}. We define wasted effort ($we$) as the number of S2R-related suggestions that a user needs to process before finding the suggestion that completes the sequence of already entered S2Rs. (The concept of wasted effort is similarly used in the fault localization literature~\cite{2016_tse_wong} to evaluate fault localization techniques.) To give an example, if a user needs to process two ``unrelated'' suggestions before finding the ``useful'' suggestion, then $we\: \shorteq\: 2$. When considering multiple prediction tasks, we determine the performance of a prediction model using the \textit{wasted effort score} metric:
\begin{equation}
\label{eq:qs}
wes\: \shorteq\: \sum\limits_{i=1}^{k}we_{i} \bigg/ \sum\limits_{i=1}^{k}c_{i}
\end{equation}

In Equation~\ref{eq:qs}, $we_{i}$ is the wasted effort in task $i$ and $c_{i}$ 
is equal to $1$ if the model provided the desired suggestion for task $i$, $0$ otherwise. \revision{To give an example, in a model that provides three suggestions for every prediction task $i$, $c_{i}$ will be equal to zero if none of the three suggestions is the desired suggestion, and the wasted effort $we_{i}$ would be equal to three. If the model provided the desired suggestion in the third position of the suggestion list, $c_{i}$ will be equal to one and $we_{i}$ will be equal to two.}
%We use the wasted effort score so that we are also able to discriminate across models with equal wasted effort.

Because $wes$ is affected by the length of the suggestions provided to the users, we search the best performing model across two dimensions: the \textit{n}-gram order and the length of the suggestions. \tech limits the search space by consider n-grams up to length $10$ and suggestions up to length $10$. (Our experiments detailed in Sec.~\ref{sec:evaluation} show that higher values for the order and the length is highly unlikely to lead to better results). \tech searches the prediction model that minimizes $wes$ and uses that model in the bug reporting phase. (The approach performs the search for each of the two prediction models used by \tech.) \tech leverages the GATs and the GETs to compute the $wes$ of different models. If the relevant app has a database of traces extracted from existing bug reports, \tech also uses these traces to compute the $wes$ of different models. While computing $wes$, \tech maps the concept of wasted effort to trace tokens and uses the token sequences from the traces to check if a token suggestion is a \textit{useful} suggestion. A token suggestion is useful if the suggested token is the same as the next token in the sequence considered. \tech computes $wes$ for a model using leave-one-out (sequence) cross-validation. For each leave-one-out \textit{test} sequence, the approach considers the first token in the sequence and uses the model to provide suggestions for the next token. (For the GEPM, \tech makes predictions only for the element tokens.) \tech repeats this process until it reaches the last token in the sequence. \tech divides the sum of the wasted effort by number of ``useful'' suggestions across all folds to compute $wes$ for the model. Finally, \tech selects the model with the minimum $wes$ to find the model that minimizes the effort needed by the user to find a ``useful'' suggestion. In computing $wes$, we consider that the benefit of finding a ``useful'' suggestion has the same weight as the cost of processing an ``unrelated'' suggestion. We leave as future work the idea of investigating different weights in the computation of $wes$.

In the process of comparing prediction models, we do not need to use an intrinsic evaluation metric such as perplexity~\cite{2009_jurafsky_speech} as our extrinsic evaluation metric $wes$ is fairly inexpensive to compute. Because \tech can determine the \textit{n}-gram order of the prediction models automatically, the approach can update the models as new training traces or bug reports are collected from the users. Although \tech provides a mechanism to automatically determine the \textit{n}-gram order of the prediction models, developers can also specify this value as well as the desired length for the suggestions manually.

When \tech reaches the end of this step, the GAPM and the GEPM are ready for use in the bug reporting phase. %\tech uses these prediction models (and the GUI model) in the next phase of the technique.

\subsection{Bug Reporting Phase}
\label{sec:bug-reporting}

\begin{figure}[t]
	\centering
	\centerline{\includegraphics[width=.99\linewidth]{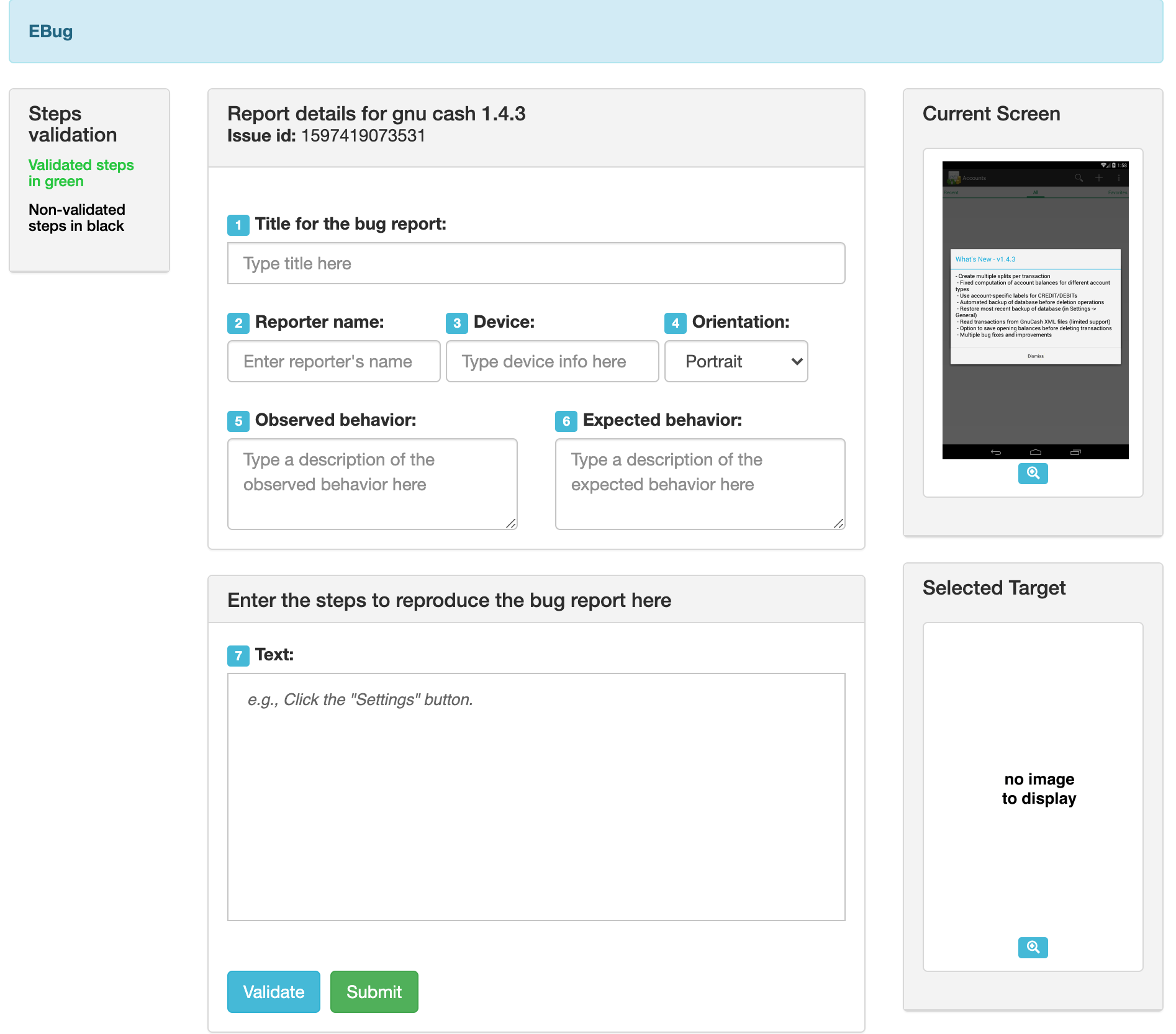}}
	%\vspace{-5pt}
	\caption{\tech's bug reporting interface.} \label{fig:ebuginterface}
	\vspace{0pt}
\end{figure}

In this phase, \tech guides its users in the process of reporting failures so that app developers can easily reproduce submitted bug reports. Specifically, when a user is reporting a failure, the approach analyzes the S2Rs contained in the report and provides S2R-related suggestions so that the user can include high-quality (\ie no missing or ambiguous) S2Rs in the report. Additionally, \tech also provides S2R-related suggestions to accelerate the bug reporting process.

Users can report bugs using the interface depicted in Fig.~\ref{fig:ebuginterface}, where S2Rs are reported in the text box labeled with $\fbox{7}$.
%%@Alex: removed to shorten the paper but I also feel it would be good to have it
%With this interface, users can enter the title of their bug reports (text field labeled $\fbox{1}$), their name ($\fbox{2}$), the device and version of the operating system on which they experienced the bug ($\fbox{3}$), the initial orientation of the device ($\fbox{4}$), the observed behavior ($\fbox{5}$), the expected behavior ($\fbox{6}$), and the S2Rs for the bug ($\fbox{7}$). \tech's bug reporting phase focuses on the information entered in the S2Rs section.
Algorithm~\ref{alg:reporting} describes how \tech guides users in the process of reporting S2Rs. The algorithm takes as inputs the GUI model (\texttt{GM}), the GUI action prediction model (\texttt{GAPM}), and the GUI element prediction model (\texttt{GEPM}).
%The algorithm also processes S2Rs as the user enters them in the report.
The algorithm's output is a bug report (\texttt{bugReport}) containing the S2Rs reported by the user. This output is also the final output of the approach. At a high level, the algorithm uses natural language processing to map the S2Rs to GUI actions in the application, analyzes the sequence of actions to provide suggestions for not-yet-typed S2Rs, and processes partial sentences to provide suggestions on how to complete partially written S2Rs.

\setlength{\algomargin}{10pt}
\setlength{\textfloatsep}{0pt}%
	\begin{algorithm}[t]
		\begin{scriptsize}
			\DontPrintSemicolon
			\SetKwInOut{Input}{Input}
			\SetKwInOut{Output}{Output}
			\caption{S2Rs analysis in the bug reporting phase.}
			\label{alg:reporting}
			\Input{
				$\texttt{GM}$: GUI model for the relevant app\\
				~$\texttt{GAPM}$: GUI action prediction model\\
				~$\texttt{GEPM}$: GUI element prediction model\\
			}
			\Output {
				$\texttt{bugReport}$: bug report describing the failure\\
			}
			\Begin{
				$\texttt{currS2REs}\; \shorteq\; \texttt{[]}$\;\label{alg:reporting:ps2rsinit}
				%$\texttt{tpMapping}\; \shorteq\; \texttt{\{\}}$\;
				$\texttt{currScreen}\; \shorteq\; \texttt{Get-Current-Screen(GM,currS2REs)}$\;\label{alg:reporting:screeninit}
				$\texttt{Display-Screen(currScreen)}$\;
				\While{$\texttt{TRUE}$}{\label{alg:reporting:mainstart}
					\Event($\texttt{S2RS\_TEXT\_CHANGE}\; \textbf{do}$){\label{alg:reporting:textstart}
						$\texttt{currS2RsText}\; \shorteq\; \texttt{Get-S2Rs-Text()}$\;\label{alg:reporting:gets2rstext}
						$\texttt{textEdit}\; \shorteq\; \texttt{Get-Text-Edit()}$\;
						%$\texttt{Update-Mapping(tpMapping,currS2RsText,textEdit,prevS2REs)}$\;\label{alg:reporting:computemapping}
						$\texttt{Check-Updated(currS2RsText,textEdit,currS2REs)}$\;\label{alg:reporting:computemapping}
						$\texttt{prevS2REs}\; \shorteq\; \texttt{currS2REs}$\;		
						$\texttt{textEditOp}\; \shorteq\; \texttt{textEdit.Get-Text-Operation()}$\;				
						\If{$\texttt{textEditOp}\; \shorteq\; \texttt{INSERT}$}{\label{alg:reporting:insertstart}
							$\texttt{newText}\; \shorteq\; \texttt{textEdit.Get-New-Text()}$\;
							\If{$\texttt{textEdit.Get-New-Text()}\; \shorteq\shorteq\; \texttt{ST}$}{\label{alg:reporting:ststart}
								%$\texttt{currS2REs}\; \shorteq\; \texttt{Compute-S2REs(GM,currS2RsText,}$ $\texttt{prevS2REs,tpMapping)}$\;
								$\texttt{currS2REs}\; \shorteq\; \texttt{Compute-S2REs(GM,currS2RsText,prevS2REs)}$\;		
								$\texttt{Display-Validated-S2Rs(currS2REs)}$\;				
								$\texttt{Display-Screen(GM,currS2REs)}$\;					
								$\texttt{S2RSuggestions}\; \shorteq\; \texttt{Get-GA-Suggestions(GAPM,currS2REs)}$\;
								$\texttt{Display-S2Rs-Suggestions(S2RSuggestions)}$\;
								%$\texttt{currScreen}\; \shorteq\; \texttt{Get-Current-Screen(GM,currS2REs)}$\;
							}\label{alg:reporting:stend}
							\ElseIf{$\texttt{newText}\; \shorteq\shorteq\; \texttt{SPACE}$}{\label{alg:reporting:spacestart}
								$\texttt{currS2RText}\; \shorteq\; \texttt{Get-Current-S2R-Text(currS2RsText,textEdit,}$ $\texttt{prevS2REs)}$\;
								$\texttt{suggestionType}\; \shorteq\; \texttt{Analyze-Partial-S2R-Text(currS2RText)}$\;
								\If{$\texttt{suggestionType}\; \shorteq\shorteq\; \texttt{TARGET}$}{\label{alg:reporting:targetstart}
									$\texttt{currS2RAction}\; \shorteq\; \texttt{Get-Current-S2R-Action(currS2RText)}$\;
									$\texttt{precS2REs}\; \shorteq\; \texttt{Get-Prec-S2REs(Get-Prec-Text(currS2RText,}$ $\texttt{currS2RsText),prevS2REs)}$\;
									$\texttt{GESuggestions}\; \shorteq\; \texttt{GEPM.Get-GE-Suggestions(precS2REs,}$ $\texttt{currS2RAction)}$\;
									$\texttt{Display-GE-Suggestions(GESuggestions)}$\;
								}\label{alg:reporting:targetend}
								\ElseIf{$\texttt{suggestionType}\; \shorteq\shorteq\; \texttt{PARAM}$}{\label{alg:reporting:paramstart}
									$\texttt{PSuggestion}\; \shorteq\; \texttt{Get-P-Suggestion(currS2RText)}$\;
									$\texttt{Display-P-Suggestion(PSuggestion)}$\;
								}\label{alg:reporting:paramend}
								\ElseIf{$\texttt{suggestionType}\; \shorteq\shorteq\; \texttt{STRUCTURE}$}{\label{alg:reporting:structurestart}
									$\texttt{SSuggestion}\; \shorteq\; \texttt{Get-S-Suggestion(currS2RText)}$\;
									$\texttt{Display-S-Suggestion(SSuggestion)}$\;
								}\label{alg:reporting:structureend}
							}\label{alg:reporting:spaceend}
						}\label{alg:reporting:insertend}
						\ElseIf{$\texttt{textEditOp}\; \shorteq\; \texttt{DELETE}$}{\label{alg:reporting:deletestart}
%							\If{$\texttt{PS2REs-Change(currS2RsText,prevS2REs,textEditOp},$ $\texttt{GM)}$}{
%								$\texttt{currS2REs}\; \shorteq\; \texttt{Identify-S2REs(currS2RsText,prevS2REs)}$\;
%								%$\texttt{currScreen}\; \shorteq\; \texttt{Get-Current-Screen(GM,currS2REs)}$\;			
%								$\texttt{Display-Screen(GM,currS2REs)}$\;
%							}
							$\texttt{currS2REs}\; \shorteq\; \texttt{Compute-S2REs(GM,currS2RsText,prevS2REs)}$\;	
							$\texttt{Display-Validated-S2Rs(currS2REs)}$\;	
							$\texttt{Display-Screen(GM,currS2REs)}$\;
						}\label{alg:reporting:deleteend}
					}\label{alg:reporting:textend}
				\Event($\texttt{SUBMIT\_BUG\_REPORT}\; \textbf{do}$){\label{alg:reporting:submitstart}
						$\texttt{currS2RsText}\; \shorteq\; \texttt{Get-S2Rs-Text()}$\;
						$\texttt{prevS2REs}\; \shorteq\; \texttt{currS2REs}$\;
						$\texttt{currS2REs}\; \shorteq\; \texttt{Compute-S2REs(GM,currS2RsText,prevS2REs)}$\;	
						$\texttt{bugReport}\; \shorteq\; \texttt{Get-Bug-Report()}$\;
						$\texttt{bugReport.Set-PS2Rs(currS2REs)}$\;
						\Return $\texttt{bugReport}$\;
				}\label{alg:reporting:submitend}
			}\label{alg:reporting:mainend}
		}
		\end{scriptsize}
	\vspace{0pt}
	\end{algorithm}

Throughout the bug reporting phase, Algorithm~\ref{alg:reporting} processes provided S2Rs and encodes them into a list of \textit{S2R entities} (\texttt{currS2REs} in Algorithm~\ref{alg:reporting}). \tech creates an entity for each S2R identifying a GUI action in the relevant app. An S2R entity is a tuple that contains: the text of the S2R ($\mathit{s2r\mhyphen text}$), the \textit{abstract GUI action} (AGA) representing the S2R ($\mathit{a\mhyphen action}$), the GUI action corresponding to the AGA ($\mathit{action}$), the GUI screen on which the action is performed ($\mathit{b\mhyphen screen}$), and the screen displayed by the relevant app after performing the action ($\mathit{a\mhyphen screen}$). 

An AGA is composed of the elements of a sentence that describe a GUI action. More precisely, an AGA is characterized by the type of the action ($\mathit{a\mhyphen type}$), the description of the element exercised by the action ($\mathit{e\mhyphen desc}$), and the description of the parameters associated with the action ($\mathit{p\mhyphen desc}$). All of the AGAs contain the type of the action, but not all of them contain the description of the element or the action, as this information might not apply. We call the action ``abstract'' because the action does not contain executable information. To provide an example, the AGA for the sentence \textit{Enter ``Transaction'' in the ``Description'' text box} (that describes an operation in the app of the motivating example) is $\langle \texttt{TYPE}, \texttt{the }\textit{``}\texttt{Description}\textit{''}\texttt{ text box},\textit{``}\texttt{Transaction}\textit{''} \rangle$. We detail how \tech computes AGAs in Sec.~\ref{sec:nlp-s2rs}.

\begin{table*}[!t]
  \setlength\tabcolsep{1.4pt}
  \renewcommand{\arraystretch}{1.25}
  \caption{Some of \tech's rules to identify a GUI action and its details from a clause's dependency tree.}
  \label{tab:grammar-rules-complete}
  \vspace{-15pt}
  \begin{footnotesize}
    \begin{center}
      \begin{tabular}[h]{|l|l|l|l|}\hline
      			\multicolumn{1}{|c|}{\textit{ID}} &
				\multicolumn{1}{|c|}{\textit{Grammar Rules}} & 
				\multicolumn{1}{c|}{\textit{Clause Example}}  &
				\multicolumn{1}{c|}{\textit{Abstract GUI Action}} \\
				\hline \hline
				$\mathit{A1}$ & \rule{0pt}{10pt}\scriptsize{$(\mathit{n_{1}} \in \mathit{C}) \xrightarrow{\mathit{e_{1}} \texttt{=} \mathit{dobj}} (\mathit{n_{2}})[\mathit{tree}(\mathit{n_{2}})]_{\mathit{e\mhyphen desc}}$} & \scriptsize{\textit{Click the ``Transaction'' element.}} & \scriptsize{$\langle \texttt{CLICK},\textit{the ``Transaction'' element},\rangle$} \\
				$\mathit{A2}$ & \rule{0pt}{10pt}\makecell[l]{\scriptsize{$(\mathit{n_{1}} \in \mathit{C}) \xrightarrow{\mathit{e_{1}} \texttt{=} \mathit{advmod}} (\mathit{n_{2}} \texttt{=} \mathit{long})$} \\ \scriptsize{\rule{35pt}{0pt}$\xrightarrow{\mathit{e_{2}} \texttt{=} \mathit{dobj}} (\mathit{n_{3}})[\mathit{tree}(\mathit{n_{3}})]_{\mathit{e\mhyphen desc}}$}} & \scriptsize{\textit{Long click the ``Transaction'' element.}} & \scriptsize{$\langle \texttt{LONG\_CLICK},\textit{the ``Transaction'' element},\rangle$} \\
				$\mathit{A3}$ & \rule{0pt}{10pt}\scriptsize{$(\mathit{n_{1}} \in \mathit{R})$} & \scriptsize{\textit{Rotate the screen.}} &  \scriptsize{$\langle \texttt{ROTATE},,\rangle$} \\
				$\mathit{A4}$ & \rule{0pt}{10pt}\makecell[l]{\scriptsize{$(\mathit{n_{1}} \in \mathit{S}) \xrightarrow{\mathit{e_{1}} \texttt{=} \mathit{prt}} (\mathit{n_{2}} \in \mathit{D})[n_{2}]_{\mathit{p\mhyphen desc}}$} \\ \scriptsize{\rule{33pt}{0pt}$\xrightarrow{\mathit{e_{2}} \texttt{=} \mathit{prep}} (\mathit{n_{3}}) \xrightarrow{\mathit{e_{3}} \texttt{=} \mathit{pobj}} (\mathit{n_{4}})[\mathit{tree}(\mathit{n_{4}})]_{\mathit{e\mhyphen desc}}$}} & \scriptsize{\textit{Scroll up on the ``Transactions'' list.}} & \scriptsize{$\langle \texttt{SCROLL},\textit{the ``Transaction'' list},\texttt{UP}\rangle$} \\
				$\mathit{A5}$ & \rule{0pt}{10pt}\makecell[l]{\scriptsize{$(\mathit{n_{2}})[\mathit{tree}(\mathit{n_{2}})\setminus\mathit{tree}(\mathit{n_{2}},\mathit{prep})]_{\mathit{p\mhyphen desc}} \xleftarrow{\mathit{e_1} \textit{=} \mathit{dobj}} (\mathit{n_{1}} \in \mathit{T})$} \\ \scriptsize{\rule{15pt}{0pt}$\xrightarrow{\mathit{e_{2}} \texttt{=} \mathit{prep}} (\mathit{n_{3}}) \xrightarrow{\mathit{e_{3}} \texttt{=} \mathit{pobj}} (\mathit{n_{4}})[\mathit{tree}(\mathit{n_{4}})]_{\mathit{e\mhyphen desc}}$}} & \scriptsize{\textit{Enter ``Transaction'' in the ``Description'' text box.}} & \scriptsize{$\langle \texttt{TYPE},\textit{the ``Description'' text box},\textit{``Transaction''} \rangle$} \\
				\hline 
      \end{tabular}
    \end{center}
  \end{footnotesize}
 \vspace{-24pt}
\end{table*}

A GUI action contains the type of the action ($\mathit{a\mhyphen type}$), the element affected by the action ($\mathit{element}$), and the parameters of the action ($\mathit{params}$). An element is composed of the GUI screen of the element ($\mathit{e\mhyphen screen}$), the type of the element ($\mathit{e\mhyphen type}$), the identifier of the element ($\mathit{e\mhyphen id}$), and the text displayed by the element ($\mathit{e\mhyphen text}$). To provide an example, the GUI action for the sentence \textit{Enter ``Transaction'' in the ``Description'' text box} is $\langle \texttt{TYPE},$ $\langle \texttt{TransactionsActivity},\texttt{EditText},\texttt{input\_transaction\_name},$ $\textit{``}\texttt{Description}\textit{''}\rangle,[\textit{``}\texttt{Transaction}\textit{''}]\rangle$. We describe how \tech identifies GUI actions from AGAs in Sec.~\ref{sec:s2rs-proc}.

\begin{table}[!t]
  \setlength\tabcolsep{1.4pt}
  \renewcommand{\arraystretch}{1.25}
  \caption{Some of \tech's rules to identify suggestions for partial clauses.}
  \label{tab:grammar-rules-partial}
    \vspace{-20pt}
  \begin{footnotesize}
    \begin{center}
      \begin{tabular}[h]{|l|l|l|l|}\hline
      			\multicolumn{1}{|c|}{\textit{ID}} &
				\multicolumn{1}{|c|}{\textit{Grammar Rules}} & 
				\multicolumn{1}{c|}{\textit{Clause Example}}  &
				\multicolumn{1}{c|}{\textit{Suggestion}} \\
				\hline \hline
				$\mathit{S1}$ & \rule{0pt}{10pt}\scriptsize{$(\mathit{n_{1}} \in \mathit{C}) \nrightarrow$} & \scriptsize{\textit{Click}} & \scriptsize{\texttt{PARTICLE}}  \\
				$\mathit{S2}$ & \rule{0pt}{10pt}\scriptsize{$(\mathit{n_{1}} \in \mathit{T}) \nrightarrow$} & \scriptsize{\textit{Type}} & \scriptsize{\texttt{PARAM}}  \\
				$\mathit{S3}$ & \rule{0pt}{10pt}\makecell[l]{\scriptsize{$(\mathit{n_{2}}) \xleftarrow{\mathit{e_{1}} \texttt{=} \mathit{dobj}} (\mathit{n_{1}} \in \mathit{T})$} \\ \scriptsize{\rule{10pt}{0pt}$\xrightarrow{\mathit{e_{2}} \texttt{=} \mathit{prep}} (\mathit{n_{3}}) \xrightarrow{\mathit{e_{3}} \texttt{=} \mathit{pobj}} (\mathit{n_{4}} \in \mathit{DET}) \nrightarrow$}} & \makecell[l]{\scriptsize{\textit{Type "Transaction"}} \\  \scriptsize{\textit{in the}}} & \scriptsize{\texttt{TARGET}}  \\
				\hline 
      \end{tabular}
    \end{center}
  \end{footnotesize}
 \vspace{0pt}
\end{table}

Algorithm~\ref{alg:reporting} starts by initializing the current list of S2R entities to be empty (line~\ref{alg:reporting:ps2rsinit}). In this initialization phase, the algorithm also identifies the app's initial screen (line~\ref{alg:reporting:screeninit}) and displays this information to the user while preemptively retrieving the screen information from $\texttt{GM}$. Although \tech displays this information, the user can also report S2Rs starting from a different screen of the apps.
%%@Alex: removed to shorten the paper but I also feel it would be good to have it
%Fig.~\ref{fig:ebuginterface} portrays \tech's interface after the approach performs its initialization steps for the app of the motivating example. The interface displays the initial screen in its ``Current Screen'' section.
After these initialization steps, the algorithm executes its main loop (lines~\ref{alg:reporting:mainstart}-\ref{alg:reporting:mainend}), where \tech helps the user in reporting S2Rs (lines~\ref{alg:reporting:textstart}-\ref{alg:reporting:textend}). The loop terminates when the user decides to submit the bug report (lines~\ref{alg:reporting:submitstart}-\ref{alg:reporting:submitend}). In this last step, the algorithm stores the content of the bug report together with the S2R entities associated with the S2R description.

We now detail \tech's NLP analysis (Sec.~\ref{sec:nlp-s2rs}) and then describe how \tech leverages the analysis to helps users in reporting S2Rs (Sec.~\ref{sec:s2rs-proc}).

\subsubsection{Natural Language Processing of S2Rs}
\label{sec:nlp-s2rs}

\tech processes each sentence in the text of the S2Rs using a two-step approach. First, \tech preprocesses a sentence in the report to simplify its analysis. Specifically, the approach performs noise removal, lexicon normalization, and object standardization~\cite{2012_miner_practical} on the sentence as described in related work~\cite{2018_issta_fazzini_automatically}. 
\revision{For noise removal, \tech discards content within parenthesis, which, in our experience, is generally unnecessary for mapping an S2R to the corresponding GUI action. For lexicon normalization, \tech normalizes non-standard words to their canonical form (\eg \tech replaces the word ``Tap'' with ``Click''). (We use the word replacements identified in related work~\cite{2018_issta_fazzini_automatically} to perform this task.) For object standardization, the approach simplifies the text of a sentence by replacing sequences of words referring to a GUI element in the relevant app with a freshly-created textual identifier. \tech performs this operation using the textual properties saved in the GUI model (\texttt{GM}) and leveraging the fact that the text displayed by an app usually follows a title or sentence-case convention~\cite{2018_saito_medium_case}. This last operation helps in simplifying the analysis of the text of the S2Rs.}
%%@Alex: removed to shorten the paper but I also feel it would be good to have it
%For example, \tech would transform the sentence \textit{Enter ``Transaction'' in the ``Description'' text box} into \textit{Enter ``Transaction'' in the ``0\_element'' text box}. This preprocessing step is particularly useful when the text being replaced contains multiple words.

\begin{figure}[!t]
	\centering
	\vspace{10pt}
	\centerline{\includegraphics[width=.99\linewidth]{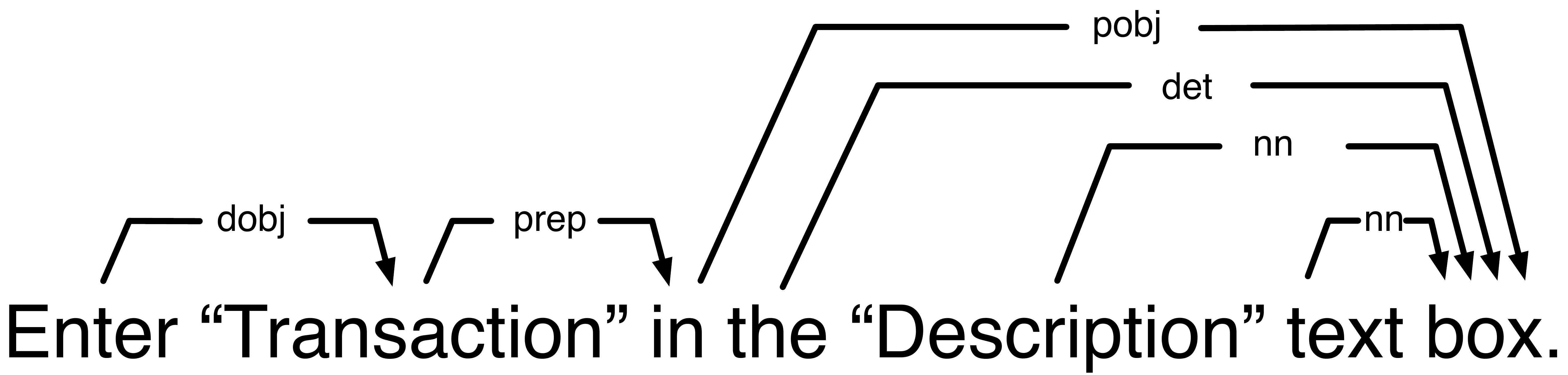}}
	%\vspace{-5pt}
	\caption{Dependency tree for a sentence.} \label{fig:dep-tree}
	\vspace{5pt}
\end{figure}

After its preprocessing step, \tech analyzes the sentence. More precisely, the approach analyzes each clause that appears in a sentence, as each of the clauses could described a different action in the relevant app.
%%@Alex: removed to shorten the paper but I also feel it would be good to have it
%For example, the sentence \textit{Click ``Add'' and enter ``Transaction'' in the ``Description'' text box} contains two clauses connected by a coordinating conjunction, and each clause describes a different action.
To identify the clauses of a sentence, the approach leverages related work~\cite{angeli2015leveraging}, which parses a dependency parse tree recursively and, at each step, predicts whether an edge should yield an independent clause. \tech analyzes clauses using their dependency tree representation~\cite{2009_jurafsky_speech, 2006_lrec_marneffe_generating}.
A dependency tree is a directed graph that captures the syntactic structure of a clause and provides a representation of the grammatical relations between words in the clause. Words are nodes, and relations are edges in the tree. 
Fig.~\ref{fig:dep-tree} provides an example of a dependency tree (extracted from the clause \textit{enter ``Transaction'' in the ``Description'' text box}). (The dependency tree in the figure does not report relations between words and punctuation for readability.)

To identify whether a clause describes a GUI action, \tech analyzes a clause's dependency tree using a rule-based approach as done in related work~\cite{2018_issta_fazzini_automatically, 2019_icse_zhao_recdroid}. Specifically, \tech uses grammar rules to identify whether a tree represents a GUI action and to and extract the relevant information characterizing the action into an AGA. Table~\ref{tab:grammar-rules-complete} reports some of \tech's grammar rules to extract AGAs (For the complete list of grammar rules, please check our online appendix~\cite{appendix}.) Column \textit{Grammar Rule} reports the rule information, column \textit{Clause Example} provides an example of a clause matched by the corresponding rule, and column \textit{Abstract GUI Action} details the AGA extracted from the corresponding clause. For example, the rule {\scriptsize $(\mathit{n_{1}} \in \mathit{C}) \xrightarrow{\mathit{e_{1}} \texttt{=} \mathit{dobj}} (\mathit{n_{2}})[\mathit{tree}(\mathit{n_{2}})]_{\mathit{e\mhyphen desc}}$} matches the clause \textit{Click the ``Transaction'' element.}, and $\langle \texttt{CLICK},\textit{the ``Transaction'' element},\rangle$ is the AGA extracted from clause.

\begin{figure*}[!t]
	\centering
	\centerline{\includegraphics[width=\linewidth]{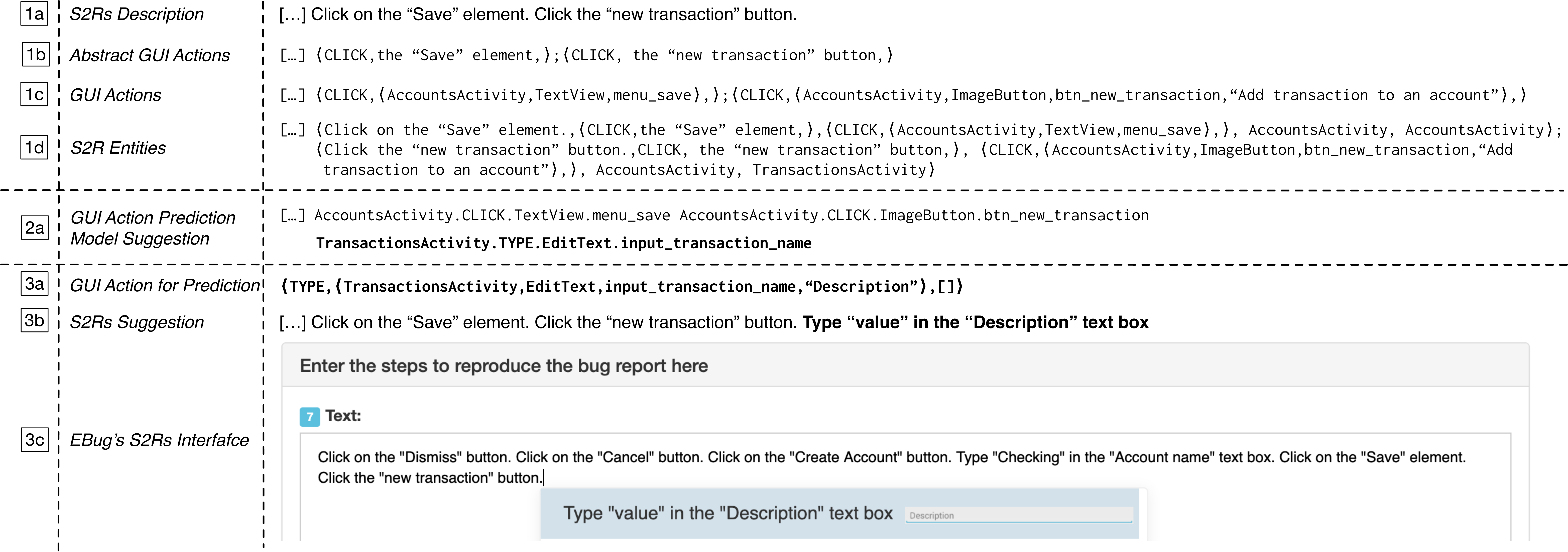}}
	\vspace{-5pt}
	\caption{Example of the information used by \tech to analyze new text in the S2Rs description.} \label{fig:supporting-example}
	\vspace{-10pt}
\end{figure*}

\tech uses grammar rules also to analyze partial clauses.
Table~\ref{tab:grammar-rules-partial} reports some of the grammar rule used for this task. Given a partial clause, \tech uses the grammar rules to identify which suggestion to make. The approach makes three types of suggestions: \texttt{PARTICLE}, \texttt{PARAM}, and \texttt{TARGET}. Suggestions of type \texttt{PARTICLE} (\eg ``the'') have the objective to help users write well-formed clauses, which help \tech's analysis. Suggestions of type \texttt{PARAM} remind the users that they also need to specify a parameter for certain types of action (\eg when users describe that they typed some text in the GUI, they should also specify the text they typed). Suggestions of type \texttt{TARGET} provide users with a list of likely targets for the partial S2R, and this list is computed using the GUI element prediction model. For example, given the partial clause \textit{Type ``Transaction'' in the}, \tech uses the rule $\mathit{S3}$ to identify that this clause requires an action target to define a complete S2R.

\subsubsection{S2Rs Reporting}
\label{sec:s2rs-proc}

This part of \tech is summarized in Algorithm~\ref{alg:reporting} at lines~\ref{alg:reporting:mainstart}-\ref{alg:reporting:mainend}. \tech's analysis performs different operations based on whether the user enters (lines~\ref{alg:reporting:insertstart}-\ref{alg:reporting:insertend}) or removes (lines~\ref{alg:reporting:deletestart}-\ref{alg:reporting:deleteend}) text to/from the S2Rs description. At a high level, when the user enters new text (Case 1), the algorithm helps the user by validating the user-provided S2Rs (\ie informs the user whether the S2Rs can be mapped to GUI actions in the app) and by providing S2R-related suggestions on how to complete the report. When the user removes text from the S2Rs description, \tech re-validates user-provided S2Rs (Case 2). We now describe Case 1 and Case 2 in detail.

\subsubsection*{Case 1: Handling Additions to the S2Rs Description}
\tech processes new text in the S2Rs description only when the user either enters a sentence terminator ($\texttt{ST}$) or a space character ($\texttt{SPACE}$). When the user enters a sentence terminator (Case 1a), \tech assumes that the user finished describing a certain S2R and provides suggestions for possibly following S2Rs. When the users enters a space character (Case 1b), the algorithm considers the description of the certain S2R to be partial, and helps the user in completing the S2R by providing suggestions on its content and structure.

\vspace{5pt}
\noindent\textit{Case 1a: Processing Complete S2Rs}\\
When the user enters a sentence terminator (lines~\ref{alg:reporting:ststart}-\ref{alg:reporting:stend}), the algorithm (i) maps the S2Rs to GUI actions in the app while encoding the GUI actions into S2R entities (\texttt{Compute-S2REs}), (ii) informs the user about validated S2Rs and, if necessary, updates \tech's interface to display the app's screen associated with the sequence of GUI actions extracted from the user-provided S2Rs (\texttt{Display-Screen}), (iii) computes suggestions for not-yet-typed S2Rs (\texttt{Get-GA-Suggestions}), and (iv) provides the suggestions to the user (\texttt{Display-S2Rs-Suggestions}). Fig.~\ref{fig:supporting-example} provides an example of some of the information computed by \tech in this process.
%%@Alex: removed to shorten the paper but I also feel it would be good to have it
%Specifically, the figure reports the information for a portion of an S2Rs description that describes the bug discussed in the motivating example. In the figure, the part labeled with $\fbox{1c}$ illustrates the GUI actions extracted from the S2Rs description, the section labeled with $\fbox{2a}$ reports the suggestion computed by \tech, and the portion labeled with $\fbox{3c}$ shows how \tech's interface displays the suggestions to the user.
We now describe the four steps of \tech that perform these operations.

\begin{figure*}[!t]
{
\footnotesize
\begin{equation*}
R(\mathit{lrs},\mathit{e\mhyphen desc},v)\: \shorteq\:
\begin{cases}
\alpha* \mathit{Max}(\mathit{S}(\mathit{e\mhyphen desc},v.\mathit{e\mhyphen text}),\mathit{S}(\mathit{e\mhyphen desc},v.\mathit{e\mhyphen id})) + (1-\alpha) * \frac{1}{1+D(\mathit{lrs},v)} & \text{if $\mathit{Max}(\mathit{S}(\mathit{e\mhyphen desc},v.\mathit{e\mhyphen text}),\mathit{S}(\mathit{e\mhyphen desc},v.\mathit{e\mhyphen id}))>\beta$}\\
0 &\text{otherwise}
\end{cases}
\end{equation*}
}
\vspace{-10pt}
\caption{Equation to rank relevant GUI elements when mapping abstract GUI actions into GUI actions.}
\label{fig:matching}
\vspace{-10pt}
\end{figure*}

\textit{Step 1: Mapping S2Rs to S2R entities.} After the user completes a specific S2R
% (\ie enters a sentence terminator in the S2Rs description)
, \tech uses the NLP analysis described in Sec.~\ref{sec:nlp-s2rs} to extract a list of abstract GUI actions (AGAs) from the S2Rs description. Given a list of AGAs (\eg the list of tuples labeled with $\fbox{1b}$ in Fig.~\ref{fig:supporting-example}), \tech uses the GUI model ($\texttt{GM}$ in Algorithm~\ref{alg:reporting}) to map the AGAs into a sequence of GUI actions and encode the actions into S2R entities. At a high level, \tech identifies GUI actions based on the description contained in the AGAs. The mapping process operates as follows. \tech uses the list of S2R entities from the previous mapping ($\texttt{prevS2REs}$) to identify the oldest S2R that the approach did not process yet. To do so, \tech processes the newly computed AGAs and finds the longest sequence matching the AGAs in $\texttt{prevS2REs}$. The approach starts processing AGAs from the first one that does not have a match, and also processes all subsequent AGAs. \tech looks at past matches to avoid processing AGAs that did not change. If $\texttt{prevS2REs}$ is empty, \tech starts the mapping task from the first AGA in the list.

For each AGA, \tech identifies the corresponding action using four pieces of information: (i) the AGA ($aga$), (ii) $\texttt{GM}$, (iii) the screen ($\mathit{rs}$) ``reached'' by the GUI action matched with the previously considered AGA, and (iv) the screen ($\mathit{lrs}$) ``reached'' by the last GUI action that was successfully matched to an AGA. For example, based on the GUI model represented in Fig.~\ref{fig:guimodel}, if action $\langle \texttt{CLICK},\langle \texttt{AccountsActivity},\texttt{ImageButton},$ $\texttt{btn\_new\_transaction},\textit{``}\texttt{Add transaction to an account}\textit{''}\rangle,$ $[]\rangle$ was the action matched to the previous AGA, then $\mathit{rs}$ would correspond to $\texttt{TransactionsActivity}$, and $\mathit{lrs}$ would correspond to the same screen. For the first AGA in the list, \tech uses the initial screen of the relevant app as the value for $\mathit{rs}$ and $\mathit{lrs}$. When the approach is not able to find a matching GUI action while analyzing the preceding AGA, \tech assigns the special value ($\ast$) to $\mathit{rs}$, which indicates that the currently reached screen could be any screen in $\texttt{GM}$.

To find the GUI action corresponding to an AGA, the approach performs four operations. First, \tech identifies all the relevant elements $v$ in $\mathit{rs}$. (If $\mathit{rs} \shorteq \ast$, \tech selects all the elements in $\texttt{GM}$. ) Second, \tech refines the list of elements by selecting only the ones that have an outgoing transition edge ($\xrightarrow{t}$) such that the action type of the edge is equal to the action type of the AGA. Third \tech ranks the elements based on how relevant they are to the element description ($\mathit{e\mhyphen desc}$) contained in the AGA and how close they are to $\mathit{lrs}$. The approach uses the equation in Fig.~\ref{fig:matching} to determine how relevant (function $R(\mathit{lrs},\mathit{e\mhyphen desc},v)$ or $R$ in short) an element is. Finally, if the top-ranked element has $R>0$, the approach selects the element as the target of the GUI action and creates the corresponding S2R entity.  For example, considering the AGA $\langle CLICK, \texttt{the }\textit{``}\texttt{new transaction}\textit{``}\texttt{ button},\rangle$, the GUI model represented in Fig.~\ref{fig:guimodel}, and $\mathit{rs} \shorteq \mathit{lrs} \shorteq \texttt{AccountsActivity}$, \tech would find that the top-ranked element is $\langle\texttt{AccountsActivity},\texttt{ImageButton},$ $\texttt{btn\_new\_transaction},\textit{``}\texttt{Add transaction to an account}\textit{''}\rangle$.
When the top-ranked element has $R=0$, \tech creates an S2R entity without a GUI action, indicating that the approach could not find a matching GUI action for the S2R. We now describe the details of the equation in Fig.~\ref{fig:matching}.

The equation is based on three parameters ($\mathit{lrs}$,$\mathit{e\mhyphen desc}$, and $v$). The first part of the equation has two components, {\small $\alpha* \mathit{Max}(\mathit{S}(\mathit{e\mhyphen desc},v.\mathit{e\mhyphen text}),\mathit{S}(\mathit{e\mhyphen desc},v.\mathit{e\mhyphen id}))$} and {\small $(1-\alpha) * \frac{1}{1+D(\mathit{lrs},v)}$}. \tech uses the first component to compute the similarity between $\mathit{e\mhyphen desc}$ and $v$. The approach computes the similarity by taking the maximum (function $\mathit{Max}$) of two values. The first value corresponds to the semantic similarity between $\mathit{e\mhyphen desc}$ and the text of the GUI element ($\mathit{e\mhyphen text}$). The second value represents the semantic similarity between $\mathit{e\mhyphen desc}$ and the identifier of the GUI element ($\mathit{e\mhyphen id}$). \tech uses semantic similarity values, and not string-distance-based metrics (\eg the Levenshtein distance~\cite{levenshtein1966binary}), to account for the fact that users might use, in their bug reports, words that have a different representation from the ones in the developer-defined labels, but these words could be either synonyms or similar in meaning (\eg ``create'' and ``add''). \tech computes semantic similarity values using vectors based on word embeddings extracted from a FastText model~\cite{bojanowski2017enriching}.  Our approach uses FastText as it is has been shown to outperform related work~\cite{mikolov2013efficient, mikolov2013distributed, luong2013better, qiu2014co, soricut2015unsupervised} in different contexts~\cite{bojanowski2017enriching}.

The approach represents each of the three textual properties ($\mathit{e\mhyphen desc}$, $\mathit{e\mhyphen text}$, and $\mathit{e\mhyphen id}$) in Fig.~\ref{fig:matching} as a single vector computed by removing stopwords (as they introduce unnecessary noise) and by averaging the vectors of the remaining words. By taking the average, \tech is able to incorporate the meaning of every word in the vector representation. For the element identifier, \tech also performs an additional preprocessing step in which the approach splits the value of the property into words. Specifically, \tech replaces underscores with spaces and splits apart composite words that follow a camel case convention, both of which are common occurrences in apps~\cite{ostrander2012android}. To determine the similarity of two properties (\eg $\mathit{e\mhyphen desc}$ and $\mathit{e\mhyphen text}$), the approach computes the cosine similarity between the corresponding vectors. The cosine similarity ranges between $[0, 1]$, where $1$ corresponds to the highest similarity. If both $\mathit{S}(\mathit{e\mhyphen desc},v.\mathit{e\mhyphen text})$ and $\mathit{S}(\mathit{e\mhyphen desc},v.\mathit{e\mhyphen id})$ are below a threshold value $\beta$, \tech does not consider the element as being relevant to the AGA and $R$ returns $0$. The approach uses $0.5$ as the default value for $\beta$. \tech uses this value to disregard unrelated matches and the value is based on an empirical analysis done in related work \cite{2018_issta_fazzini_automatically}.

\tech uses the second component ({\small $(1-\alpha) * \frac{1}{1+D(\mathit{lrs},v)}$}) to prioritize matches with elements belonging to screens that are in the proximity of the screen containing the previously matched element. Specifically, $D(\mathit{lrs},v)$ is the distance between $\mathit{lrs}$ and the screen $vs$ containing $v$. The approach measures the distance by finding the shortest path between $\mathit{lrs}$ and $vs$ in $\texttt{GM}$ and counting the number of transition edges in the path. If \tech does not find a path connecting $\mathit{lrs}$ and $vs$, the approach sets $D(\mathit{lrs},v) \shorteq 0$. {\small $\frac{1}{1+D(\mathit{lrs},v)}$} ranges between $[0, 1]$, where $1$ corresponds to the shortest distance (\ie the screens are the same). When $\mathit{lrs} \shorteq \mathit{rs}$ and $\mathit{rs} \neq \ast$, the value of {\small $\frac{1}{1+D(\mathit{lrs},v)}$} is always $1$ as \tech selects elements from $\mathit{rs}$.

Finally, the equation in Fig.~\ref{fig:matching} uses the constant $\alpha$ to define the weight of the two equation's components. \tech uses $0.5$ as the default value for $\alpha$ to assign the same weight to the two components.

\textit{Step 2: Displaying Validated S2Rs and Current Screen.} In this step, \tech provides feedback to the user based on whether the approach was able to map the S2Rs into GUI actions. The goal of this step is to inform the user that certain S2Rs might require further refinement. Specifically, for the S2R entities that \tech could map to $\texttt{GM}$ (\ie entities containing the corresponding GUI action), the approach displays and highlights (in green) the text of the S2Rs in its graphical interface. For S2R entities not mapped to $\texttt{GM}$, the approach reports their S2Rs in its graphical interface but does not highlight them, as we wanted to avoid reporting false negatives (\ie valid actions not mapped to $\texttt{GM}$ due to the possibly partial information encoded in the model). \tech displays the results of the validation process in the section of Fig.~\ref{fig:ebuginterface} labeled with \textit{Steps validation}. During this step, the approach also updates its interface to display the reached screen $\mathit{rs}$ associated with the last reported S2R entity. If $\mathit{rs} \shorteq \ast$, \tech displays a screen informing the user that the last S2R was not matched. The screen is displayed in the part of Fig.~\ref{fig:ebuginterface} labeled with \textit{Current Screen}.

\textit{Step 3: Computing S2R Suggestions.} \tech uses the GUI action prediction model ($\texttt{GAPM}$ in Algorithm~\ref{alg:reporting}) to provide suggestions for not-yet-typed S2Rs. To identify relevant suggestions, the approach first encodes the S2R entities into a sequence of tokens and then uses $\texttt{GAPM}$ to identify the next token (S2R) for the sequence. Specifically, \tech translates the last $n-1$ S2R entities into tokens. (The tokens are in the same order as they appear in the S2Rs description.) \tech uses the GUI actions associated with the S2R entities to compute the sequence tokens. The structure and the content of the tokens processed by $\texttt{GAPM}$ are detailed in Sec.~\ref{sec:pred-mod-gen}.
In Fig.~\ref{fig:supporting-example}, the parts labeled with $\fbox{1d}$ and $\fbox{2a}$ show how \tech translates S2R entities into tokens suitable for $\texttt{GAPM}$. Once the approach retrieves the token suggestions from $\texttt{GAPM}$, \tech uses them to display suggestions to the user.

\textit{Step 4: Providing S2R Suggestions to the User.} Given a list of suggestions, \tech displays them to the user in the same order as provided by $\texttt{GAPM}$. First, for each token, the approach retrieves additional information associated with the token (\eg its text) from $\texttt{GM}$ Then, the approach displays the text of the S2R corresponding to the token, and, when available, a screenshot of the GUI element affected by the action associated with the token. (\tech does not always have a screenshot for a GUI element, as the approach uses static analysis to identify some of the elements in the GUI model). To translate a token into an S2R, \tech uses textual templates. \tech has templates for each GUI action supported by the approach. The approach fills the template with the information associated with the action represented by a token.
%%@Alex: removed to shorten the paper but I also feel it would be good to have it
%For example, for the token $\texttt{TransactionsActivity.TYPE.EditText.input\_transaction\_}$ $\texttt{name}$, the approach would use the template $\texttt{Type }\textit{``}\texttt{value}\textit{''}\texttt{in the ELEMENT ELEMENT\_TYPE}$, and replace $\texttt{ELEMENT}$ with $\textit{``}\texttt{Description}\textit{''}$ and $\texttt{ELEMENT\_TYPE}$ with $\texttt{text box}$.
In Fig.~\ref{fig:supporting-example}, the parts labeled with $\fbox{2a}$, $\fbox{3a}$, $\fbox{3b}$, and $\fbox{3c}$ show how \tech displays suggestions to the user.

\vspace{5pt}
\noindent\textit{Case 1b: Processing Partial S2Rs}\\
When the user enters a space character (lines~\ref{alg:reporting:spacestart}-\ref{alg:reporting:spaceend}), \tech (i) analyzes the text of the S2R to identify whether the S2R needs further refinement ($\texttt{Analyze-Partial-S2R-Text}$), (ii) computes suggestions on how to complete the S2R, and (iii) provides the suggestions to the user.

\textit{Step 1: Analyzing Partial S2Rs.}  \tech uses the NLP analysis described in Sec.~\ref{sec:nlp-s2rs} to analyze the current S2R
%(\ie the S2R containing the space character)
and identify how to help the user in completing the S2R. \tech can (i) suggest likely targets (\ie GUI elements) for the type of GUI action described in the S2R (lines~\ref{alg:reporting:targetstart}-\ref{alg:reporting:targetend}), (ii) inform the user that the S2R should also specify a parameter (\ie the text to be typed in a text box) for the GUI action described in the S2R (lines~\ref{alg:reporting:paramstart}-\ref{alg:reporting:paramend}), and (iii) add grammatical particles to the S2R so that the S2R is easier to analyze (lines~\ref{alg:reporting:structurestart}-\ref{alg:reporting:structureend}).

\textit{Step 2: Computing Partial S2R Suggestions.} The approach leverages the GUI element prediction model ($\texttt{GEPM}$ in Algorithm~\ref{alg:reporting}) to suggest likely targets for a certain S2R. To identify relevant suggestions, the approach extracts the action type associated with the S2R, encodes the action type into a model token, transforms preceding S2R entities ($\texttt{precS2REs}$) into model tokens, and uses the sequence of tokens to identify predictions based on $\texttt{GEPM}$. Similarly to what \tech does in the case of a complete S2R suggestion, if an S2R entity does not have a corresponding GUI action, \tech does not compute partial S2R suggestions for sequences that include the entity. 
%The structure and the content of the tokens processed by $\texttt{GEPM}$ are detailed in Sec.~\ref{sec:pred-mod-gen}.
Once the approach retrieves the token suggestions from $\texttt{GEPM}$, \tech stores them in the same order as they are provided by the model and uses them to display suggestions to the user. \revision{Additionally, the technique is also able to provide suggestions based on partially typed GUI elements. In this case, the suggestions are not based on the GEPM but are based on text matching.} When \tech identifies that the S2R needs a parameter or a suggestion for improving the structure of the S2R, the approach provides the corresponding textual suggestion.

\textit{Step 3: Providing Partial S2R Suggestions to the User.} \tech offers suggestions for GUI elements similarly to how it provides suggestions for complete S2Rs. Suggestions for parameters and the structure of certain S2Rs are automatically added to the bug report and the user can accept them by pressing the $\texttt{tab}$ character.

\subsubsection*{Case 2: Handling Removals from the S2Rs Description}

When the user removes text from the S2Rs description (lines~\ref{alg:reporting:deletestart}-\ref{alg:reporting:deleteend}), the algorithm analyzes whether the sequence of GUI actions associated with the S2Rs has changed, and updates \tech's interface to display the app's screen associated with the action sequence. Specifically, \tech (i) maps S2Rs into S2R entities, (ii) validates the S2Rs, and (iii) displays the screen reached by the last S2R entity in the sequence. The approach performs these steps following the methodology described in \textit{Case 1a-Step 1} and \textit{Case 1a-Step 2}.
\section{Implementation}
\label{sec:implementation}

We implemented \tech in a system that supports bug reporting for Android apps. The system consists of three main modules. The \textit{GUI models generation} module is written in Java and leverages Gator~\cite{2015_ase_yang_static} to perform \tech's static analysis, builds on UiAutomator~\cite{2021_google_uiautomator} to complete the approach's dynamic analysis, and stores the GUI model using the Neo4J~\cite{2021_neo4j_neo4j} graph database. The \textit{prediction models generation} module leverages the Getevent~\cite{2021_google_getevent} and UIAutomator utilities to collect user traces and builds on the MITLM toolkit~\cite{hsu2008iterative} and the MonkeyLab infrastructure~\cite{Linares:MSR15} to generate the n-gram based language models characterizing the GUI-action and GUI-element prediction models. Finally, the \textit{bug reporting} module is implemented as a web application. This web application leverages the Google Cloud Natural Language infrastructure~\cite{2021_google_natural} to perform \tech's natural language processing analysis and builds on the Quill library~\cite{2021_quilljs_quill} to implement the interactive S2R reporting approach. The module uses the FastText model trained on the Common Crawl dataset~\cite{fastTextDataset} to identify the semantic similarity between S2R descriptions and GUI element properties. The web application also provides access to bug reports saved in a MySQL database.

\section{Empirical Evaluation}
\label{sec:evaluation}

To determine the effectiveness and efficiency of \tech, we used the implementation of our approach and performed two user studies. The studies' main focus is to assess the extent to which \tech is able to facilitate the creation of highly reproducible bug reports. In the studies, we also compared the efficiency and effectiveness of \tech with those of \fusion~\cite{Moran:FSE15}. We selected \fusion as a baseline because the technique also aims to improve bug reporting, handles Android apps, and has been shown to create bug reports that are easier to reproduce compared to the ones generated using traditional issue tracking systems~\cite{Moran:FSE15}. The effectiveness of this reporting system~\cite{Moran:FSE15} makes it a strong baseline against which to compare \tech.

In the evaluation, we also compare the effectiveness of \tech's GUI-action and GUI-element prediction models with those of prediction models based on AKOM~\cite{pitkow1999mining} and CPT+~\cite{gueniche2015cptplus}. We selected AKOM as a baseline because the technique generates a model to perform sequence predictions, the technique functions effectively even without a large amount of training data, and the approach has been used to perform sequence predictions in numerous applications and various domains~\cite{pitkow1999mining}. We selected CPT+ as it has similar characteristics as AKOM (\eg the technique functions effectively even without a large amount of training data) and, in some domains, the technique was shown to provide better accuracy than AKOM~\cite{gueniche2015cptplus}.

In the evaluation, we targeted the following research questions (RQs): 

\begin{itemize}
\item \textbf{RQ1}: \textit{Do developers using \tech's reports reproduce more failures compared to when using \fusion's reports?}
\item \textbf{RQ2}: \textit{Is bug reporting with \tech more efficient than \fusion?}
\item \textbf{RQ3}: \textit{Do developers using \tech's reports reproduce failures more efficiently compared to when using \fusion's reports?}
\item \textbf{RQ4}: \textit{Is bug reporting with \tech leading to a better user experience than the one with \fusion?}
\item \textbf{RQ5}: \textit{Do developers using \tech's reports have a better user experience compared to \fusion?}
\item \textbf{RQ6}: \textit{Do \tech's GUI-action and GUI-element prediction models provide a better wasted effort score than prediction models based on AKOM and CPT+?}
\end{itemize}

\subsection{Experimental Context}

In this section, we present the benchmarks used in the evaluation, describe the process used to collect user traces, and detail the user studies' characteristics.

\subsubsection{Benchmarks}

Our empirical evaluation is based on 20 unique, real-world failures from 11 apps. We identified the failures by randomly selecting bug reports from a collection of bug reports used in related work. These bug reports were used to evaluate different aspects of mobile app bug reporting~\cite{Moran:FSE15,Chaparro:FSE'19}. Before including a bug report in our benchmark dataset, we first confirmed that the failure described in the report was still reproducible. To reproduce the failure, we ran the app on a Nexus 7 emulator. The versions of the app and the Android system running on the emulator were the same as those used in related work. (We used a Nexus 7 emulator as \fusion's empirical evaluation also used this emulator.) The first 20 bug reports that we randomly selected were reproducible. Nineteen failures were reproducible on the emulator running Android 4.4. One failure was reproducible on the emulator running Android 5.0. Nine bug reports were used in \fusion's empirical evaluation~\cite{Moran:FSE15}, whereas the remaining eleven bug reports were used in recent work by Chaparro \etal~\cite{Chaparro:FSE'19}. All the 20 bug reports belong to free-text-based issue tracking systems.
\revision{Among the failures described in the 20 bug reports, four failures manifest as crashes and sixteen failures provide an incorrect output to the user. Examples of output errors include displaying an erroneously computed value, incorrectly displaying a GUI element, and bringing the user to the wrong screen.}

\revision{Comparing our dataset with the ones we used to create it, we can report the following. The dataset from the baseline technique we considered~\cite{Moran:FSE15} includes 15 failures from 14 apps. That dataset has three failures that manifest as crashes and 12 failures that manifest as output failures. The second dataset~\cite{Chaparro:FSE'19} contains 24 failures from six apps. Four failures manifest as crashes and 12 manifest as output failures. We believe that the size and diversity between our dataset and these two other datasets are comparable.}

Table~\ref{tab:bug-reports} details some of the characteristics of the apps, bug reports, and failures. Specifically, column \textit{APP ID} provides a unique identifier for the app, \textit{App Name} reports the name of the app, \textit{Category} provides the category of the app on the Google Play store~\cite{2020_googleplay}, \textit{Version} details the app's version used to reproduce the failure, \textit{Report ID} reports the identifier of the bug report describing the failure, \textit{Failure ID} provides a unique identifier for the failure, \textit{Min GUI Actions} details the minimum number of GUI actions necessary to reproduce the failure, and \textit{OIB} labels the type of observable incorrect behavior (\ie crash or output failure). For the \textsc{GnuCash} app, we use two different identifiers (A06a and A06b) because we reproduced the failures on two different versions of the app. The average and the median number of GUI actions necessary to reproduce the failures are $9.7$ and $10$, respectively. These numbers highlight that the failures we considered require multiple steps to reproduce the issue.

%A01 Books
%A02 Comics
%A03 Auto
%A04 Productivity
%A05 Heath
%A06 Finance
%A07 Finance
%A08 Tools
%A09 Books
%A10 Books
%A11 Productivity

\begin{table}[t]
  \setlength\tabcolsep{1.4pt}
  \renewcommand{\arraystretch}{1.1}
  \caption{Apps, bug reports, and failures used in the empirical evaluation. 
%  \textit{App ID} = identifier of a benchmark app,
%  \textit{App Name} = benchmark app name,
%  \textit{Version} = benchmark app version,
%  \textit{Report ID} = bug report identifier on the app's bug tracking system,
%  \textit{Failure ID} = an identifier for the failure described in the bug report,
%  \textit{Min GUI Actions} = minimum number of GUI actions to reproduce the failure,
 \textit{OIB} = observable incorrect behavior of the reports (O = output, C = crash).
  } 
  \label{tab:bug-reports}
  \vspace{-15pt}
  \begin{footnotesize}
    \begin{center}
      \begin{tabular}[h]{|l|l|l|l|c|c|c|c|}\hline
      			\parbox[c][10mm]{3mm}{\centering\shortstack{\textit{App}\\ \textit{ID}}} &
      			\parbox[c][10mm]{19mm}{\centering\textit{App Name}} & 
      			\parbox[c][10mm]{11mm}{\centering\textit{Category}} & 
				\parbox[c][10mm]{9mm}{\centering\textit{Version}} & 
				\parbox[c][10mm]{9mm}{\centering\shortstack{\textit{Report}\\ \textit{ID}}}  &
				\parbox[c][10mm]{8mm}{\centering\shortstack{\textit{Failure}\\ \textit{ID}}}  &
				\parbox[c][10mm]{11mm}{\centering\shortstack{\textit{Min GUI}\\ \textit{Actions}}} &
				\parbox[c][10mm]{5mm}{\centering\shortstack{\textit{OIB}}} \\
				\hline \hline
				A01 & \textsc{Aard} & Books & 1.6.10 &	\href{https://github.com/aarddict/android/issues/104}{\underline{$104$}} & F01 & $4$ & O\\ \hline
				A02 & \textsc{ACV} & Comics & 1.4.1.4 &	\href{https://github.com/robotmedia/droid-comic-viewer/issues/11}{\underline{$11$}} & F02 & $7$ & C\\ \hline
				A03 & \textsc{Car Report} & Auto & 3.4.1 &  \href{https://bitbucket.org/frigus02/car-report/issues/43/infinity-100-km-and-plot-deathlock}{\underline{$43$}} & F03 & $10$ & O\\ \hline
				A04 & \textsc{Doc Viewer} & Prod. & 2.2 &  \href{https://github.com/SufficientlySecure/document-viewer/issues/48}{\underline{$48$}} & F04 & $9$ & O\\ \hline
				\multirow{2}{*}{A05} & \multirow{2}{*}{\textsc{DroidWeight}} & \multirow{2}{*}{Heath} & \multirow{2}{*}{1.5.4} &  \href{https://code.google.com/archive/p/droidweight/issues/38}{\underline{$38$}} & F05 & $5$ & O\\ \cline{5-8} 
				 & & & &\href{https://github.com/sspieser/droidweight/issues/25}{\underline{$25$}} & F06 &  $11$ & O\\ \hline
				 \multirow{2}{*}{A06a} & \multirow{2}{*}{\textsc{GnuCash}} & \multirow{2}{*}{Finance} & \multirow{2}{*}{1.4.3} & \href{https://github.com/codinguser/gnucash-android/issues/247}{\underline{$247$}} & F07 & $16$ & O\\ \cline{5-8} 
				 & & & & \href{https://github.com/codinguser/gnucash-android/issues/256}{\underline{$256$}} & F08 & $20$ & O\\ \hline
				 \multirow{3}{*}{A06b} & \multirow{3}{*}{\textsc{GnuCash}} & \multirow{3}{*}{Finance} & \multirow{3}{*}{2.1.1} & \href{https://github.com/codinguser/gnucash-android/issues/615}{\underline{$615$}} & F09 & $9$ & O\\ \cline{5-8} 
				 & & & & \href{https://github.com/codinguser/gnucash-android/issues/620}{\underline{$620$}} & F10 & $11$ & O\\ \cline{5-8} 
				 & & & & \href{https://github.com/codinguser/gnucash-android/issues/663}{\underline{$663$}} & F11 & $10$ & C\\ \hline
				  \multirow{3}{*}{A07} & \multirow{3}{*}{\textsc{Mileage}} & \multirow{3}{*}{Finance} & \multirow{3}{*}{3.0.8} & \href{https://github.com/evancharlton/android-mileage/issues/49}{\underline{$49$}} & F12 & $15$ & O\\ \cline{5-8}%mileage-1
				 & & & & \href{https://github.com/evancharlton/android-mileage/issues/53}{\underline{$53$}} & F13 & $12$ & O\\ \cline{5-8}%mileage-3
				 & & & & \href{https://github.com/evancharlton/android-mileage/issues/64}{\underline{$64$}} & F14 & $11$ & O\\ \hline%mileage-2
				  A08 & \textsc{Notepad} & Tools & 1.06 &  \href{https://code.google.com/archive/p/banderlabs/issues/23}{\underline{$23$}} & F15 & $6$ & C\\ \hline
				  A09 & \textsc{Olam} & Books & 1 &  \href{https://github.com/vishnus/Olam/issues/2}{\underline{$2$}} & F16 & $2$ & C\\ \hline
				  A10 & \textsc{Schedule} & Books & 1.32.2 &  \href{https://github.com/tuxmobil/CampFahrplan/issues/154}{\underline{$154$}} & F17 & $7$ & O\\ \hline
				 \multirow{3}{*}{A11} & \multirow{3}{*}{\textsc{Time Tracker}} & \multirow{3}{*}{Prod.} & \multirow{3}{*}{0.17} & \href{https://github.com/netmackan/ATimeTracker/issues/24}{\underline{$24$}} & F18 & $11$ & O\\ \cline{5-8}%time-tracker-3
				 & & & & \href{https://github.com/netmackan/ATimeTracker/issues/25}{\underline{$25$}} & F19 & $10$ & O\\ \cline{5-8}%time-tracker-1
				  & & & & \href{https://github.com/netmackan/ATimeTracker/issues/46}{\underline{$46$}} & F20 & $8$ & O\\ \hline%time-tracker-2
      \end{tabular}
    \end{center}
  \end{footnotesize}
 \vspace{0pt}
\end{table}	

\subsubsection{GUI Models and User Traces}

\revision{\tech uses pre-computed GUI models in its bug reporting phase. In the evaluation, we computed one GUI model for each app version considered. The generation of each GUI model is based on static and dynamic app analysis. The static analysis took, on average, 16 seconds to complete. For the dynamic analysis, we did not perform the analysis exhaustively but, instead, used a timeout of 60 minutes, as this timeout was shown to be effective in related work~\cite{Moran:FSE15}.}

In our experiments, we used user traces to generate the prediction models and gathered the traces by designing a data collection activity. In this activity, we invited nine participants to generate the traces. The participants' demographics include three participants without any computer science (CS) background, three undergraduate students in CS, and three graduate students in CS. \revision{In the activity, we assigned four apps to each participant, asked the participants to first get familiar with each app's functionality for five minutes, and then invited them to collect at least three traces for each app while exercising the app's functionality as they would do in a normal use of the app.} In total, the participants collected $152$ traces. The overall number of GUI actions in the traces is $3,580$. We asked participants to collect traces for both A06a and A06b as the apps' GUI differed significantly. Finally, the data collection activity took about two weeks to complete.

Based on this set of traces, we generated the prediction models following the methodology described in Section~\ref{sec:pred-mod-gen}. Table~\ref{tab:prediction-models-performance} reports the n-gram order (column \textit{Order} under \tech headers) and the number of suggestions provided by each prediction (column \textit{SN} under \tech headers) for both the GUI-action and GUI-element prediction models.

\subsubsection{Bug Reporting Study}

To determine the extent to which \tech is able to facilitate the creation of highly reproducible bug reports and how efficiently users can do so, we ran a user study involving ten participants. \revision{The participants' demographics include two undergraduate and eight graduate students in CS, and none of the participants took part in the data collection activity. Additionally, the participants  had been studying CS for an average of 7.65 years (min=2, max=14), had 7.8 years (min=3, max=14) of programming experience, and had an average of self-determined experience level in mobile app programming of 4.7 (min=2, max=8) based on 1-10 assessment scale. None of the participants in this study is a professional developer, but four of the participants have industry exposure through internships as software engineers.}

In the study, we asked the participants to report the failures from Table~\ref{tab:bug-reports}, using both \fusion and \tech. Specifically, each participant reported five failures using \fusion and five failures using \tech. We assigned failures to participants making sure that participants would not report the same failure with both tools and would not report two failures belonging to the same version of the app while using the same tool. We took these two measures to minimize bias in the reporting task. In the study, the subjects used anonymized variants of the tools, and we did not inform them about which tool was ours and which one was the baseline.
%(but they could have discovered the name of the baseline by searching some information about the baseline online).
\revision{It should be noted that we ensured that the order in which each participant was exposed to a tool was set such that half of the participants used \tech first and half used \fusion first to mitigate potential effects of learning bias.}

Before starting the study, we asked participants to watch two tutorial videos (about five-minutes long) explaining how to use \fusion and \tech. (The participant could skip watching the tutorial videos if they chose to do so.) At the beginning of the study, we provided each participant with a document containing the failures the participant should report and the tool the participant needed to use to report a certain failure. We provided information about each failure in the form of a video. The video contained the sequence of GUI actions necessary to reproduce the failure and how the failure manifested itself. We recorded these videos by performing the GUI actions reported in Table~\ref{tab:bug-reports} on a Nexus 7 emulator. In the document, we asked participants to report failures based on what they saw in the videos. \revision{We chose to expose bugs through video because  we are interested in participants experience reporting bugs \textit{assuming that an end-user has a good understanding of the failure that they are reporting}. As such, while exposing users to annotated videos of bugs may not represent a real-world scenario, it provided us with an avenue to clearly illustrate app failures in a manner in which we could be reasonably certain that the participant understood the bug.} The document listed failures so that half of the participants would start reporting failures using \fusion, and the remaining half would start with \tech. We randomized the order of the remaining failure-tool pairs in each document. The document also instructed participants to report the failures in the order they appeared in the document. During the study, we asked the participants to record the time it took them to report each failure and record their screen as they were reporting the failures. (We used the screen recordings to check whether the information entered by participants was the same as the one saved in the two tools' database. After watching the screen recordings, we could confirm that the information was the same.) Finally, during the study, the participants did not have access to the original bug reports of the failures.

At the end of the study, we asked the participants to complete a survey. In the survey, participants entered demographics information, recorded the time it took them to report each failure, and answered ten user experience (UX) questions for both \fusion and \tech. Among the ten questions, six were usability questions, and four were user preference questions. Table~\ref{tab:report-questions} reports the usability (R-UN questions) and user-preference questions (R-PN questions) we asked to the participants. We formulated the usability questions based on the system usability scale (SUS) by John Brooke~\cite{Brooke:96}, as this scale is often used to compare the usability between systems~\cite{Brooke:96}. The answers to these usability questions are structured using a five-level Likert scale that ranges from ``Strongly disagree'' to ``Strongly agree''. We formulated the user preference questions based on the user experience honeycomb developed by Peter Morville~\cite{Morville:04} and asked the participants to answer the questions using free-form text fields. Finally, the user experience questions were also used in the original evaluation of \fusion~\cite{Moran:FSE15}.

\begin{table}[t]
  \setlength\tabcolsep{1.4pt}
  \renewcommand{\arraystretch}{1.1}
 \caption{User experience questions asked to participants in the bug reporting study.  When asking questions about \fusion, we replaced \texttt{<sys>} with \textit{System A}.  For \tech, we replaced \texttt{<sys>} with \textit{System B}.
%  \textit{QID} = an identifier for the question,
%  \textit{Question} = user experience question.
 }
 \label{tab:report-questions}
  \vspace{-10pt}
  \begin{scriptsize}
    \begin{center}
      \begin{tabular}[h]{|c|p{8cm}|}\hline
      			{\footnotesize \textit{QID}} & \multicolumn{1}{c|}{{\footnotesize \textit{Question}}} \\
				\hline \hline
				R-U1 & I think that I would like to use \texttt{<sys>} frequently.\\
				R-U2 & I found  \texttt{<sys>} very cumbersome to use.\\
				R-U3 & I found the various functions in \texttt{<sys>} were well integrated.\\
				R-U4 & I thought \texttt{<sys>} was easy to use.\\
				R-U5 & I found \texttt{<sys>} unnecessarily complex.\\
				R-U6 & I thought that \texttt{<sys>} was very useful for reporting a bug.\\
				\hline
				\hline
				R-P1 & What functionality did you find useful when reporting bugs with \texttt{<sys>}?\\
				R-P2 & What information (if any) were you not able to report with \texttt{<sys>}?\\
				R-P3 & What elements do you like the most from \texttt{<sys>}\\
				R-P4 & What elements do you like the least from \texttt{<sys>}\\
				\hline
      \end{tabular}
    \end{center}
  \end{scriptsize}
 \vspace{0pt}
\end{table}	

We ran the study over a three-week time frame and obtained a dataset of $80$ bug reports. The dataset includes $40$ bug reports generated by each tool and has two bug reports for each failure-tool pair.

\subsubsection{Bug Reproduction Study}

To determine whether the bug reporting study participants created reproducible bug reports, we ran a user study involving ten professional developers with experience in mobile app development from ten different companies. \revision{The developers had different levels of experience in mobile app programming and bug reporting management. Specifically, the developers had an average programming experience of 9.4 years (min 2, max=12), had an average of self-determined experience level in mobile app programming of 5.4 (min=2, max=9) based on 1-10 assessment scale, and had an average of self-determined experience level in bug reporting and management of 6.8 (min=1, max=9) based on 1-10 assessment scale.} \revision{Although the professional developers involved in the study have experience in app development, these developers were not the developers of the apps considered in the study. We opted not to involve the developers of the benchmark apps as we wanted to avoid including developers that could have been already aware of the reported bugs, which could have impacted the reproduction part of the study.} In the study, we asked the developers to reproduce the bug reports gathered in the bug reporting study. In this study, we also asked the developers to reproduce the original bug reports (\ie the bug reports listed and linked in Table~\ref{tab:bug-reports}). We assigned ten bug reports to each developer. Four bug reports were generated using \fusion, four were created using \tech, and two were from the set of original bug reports. We assigned bug reports to developers making sure that they would not try to reproduce the same report. \revision{Similar to the bug reporting study, it should be noted that we ensured that the order in which each participant was exposed to each type of report was set such that one-third of the participants used \tech first and one-third used \fusion first, and one-third used the original reports first to mitigate potential effects of learning bias.}

During the study, the developers could access the bug reports using the anonymized variants of the tools. Like in the bug reporting study, we did not inform the subjects about which tool was ours and which one was the baseline. For the original bug reports, developers had access to a screenshot of the original bug reports. We took this measure so that the developers could not access any follow-up discussion in the bug tracking system of the original bug report. In the study instructions, we asked developers to reproduce the bug reports using a set of Nexus 7 emulators. These emulators contained the apps associated with the reports that the developers needed to reproduce. We also informed the developers that they needed to record the time it took them to reproduce each of the bug reports. We asked the developers to collect evidence (in the form of a screenshot or files stored in the emulator) that they reproduced a bug report (without including the time it took to perform this operation in the time they logged to reproduce a bug report). We used the evidence to verify that the subjects succefully reproduced the bug reports. Finally, we informed the developers that they should stop trying to reproduce a report if the report took more than $10$ minutes to reproduce. \revision{To select the $10$ minutes value, we relied on our experience from related work~\cite{Moran:FSE15} and used the same value as in that work. We used the same value to provide a consistent evaluation of the techniques, and in related work, we also observed that the value had mitigated participant fatigue. Although setting the limit could provide a partial view of the total number of bug reports that could be reproduced, developers generally need to decide how much time they would like to invest in reproducing a report, and our results could also be interpreted as that time being set to 10 minutes.}

Like in the bug reporting study, we asked participants to complete a survey at the end of the study. In the survey, developers entered demographics information, recorded the time it took them to reproduce each bug report, and answered nine user experience (UX) questions about \fusion, \tech, and the original bug reports. Among the nine questions, five were usability questions, and four were user preference questions. Table~\ref{tab:repro-questions} reports the usability (D-UN questions) and user-preference questions (D-PN questions). We formulated the questions following the same principles adopted in the bug reporting study.

\begin{table}[t]
  \setlength\tabcolsep{1.4pt}
  \renewcommand{\arraystretch}{1.1}
 \caption{User experience questions asked in the bug reproduction study.
  \textit{QID} = an identifier for the question,
  \textit{Question} = user experience question. 
 }
 \label{tab:repro-questions}
  \vspace{-10pt}
  \begin{scriptsize}
    \begin{center}
      \begin{tabular}[h]{|c|p{8cm}|}\hline
      			{\footnotesize \textit{QID}} & \multicolumn{1}{c|}{{\footnotesize \textit{Question}}} \\
				\hline \hline
				D-U1 & I think that I would like to use this type of bug report frequently.\\
				D-U2 & I found this type of bug report unnecessarily complex.\\
				D-U3 & I thought this bug report was easy to read/understand.\\
				D-U4 & I found this type of bug report very cumbersome to read.\\
				D-U5 & I thought that this type of report was very useful for reproducing the bug.\\
				\hline
				\hline
				D-P1 & What feature(s) did you find useful when reproducing the bugs with using this type of bug report?\\
				D-P2 & What additional information (if any) would you like to see in this type of bug report?\\
				D-P3 & What elements do you like the most from this type of bug report?\\
				D-P4 & What elements do you like the least from this type of bug report?\\
				\hline
      \end{tabular}
    \end{center}
  \end{scriptsize}
 \vspace{0pt}
\end{table}	

\subsection{Results}

\subsubsection{RQ1: Effectiveness}

To answer RQ1, we identified how many failures developers could reproduce using the reports generated in the bug reporting study. Table~\ref{tab:study-data} provides this information. Specifically, column \textit{Repro} under the \tech header reports which failures developers could reproduce using the reports generated with \tech. Column \textit{Repo} under the \fusion header reports the results for \fusion.
%The symbol \cmark indicates that a developer could reproduce the failure FN using the bug report associated with the failure, while symbol \xmark expresses the opposite.
%Table~\ref{tab:study-data} reports two symbols for each failure because, in the bug reporting study, reporters gathered two bug reports for each failure.
Table~\ref{tab:study-data} also reports the identifier of the developer that tried to reproduce a certain report (columns \textit{Dev ID}).

\begin{table*}[t!]
  \setlength\tabcolsep{1.4pt}
  \renewcommand{\arraystretch}{1.1}
  \caption{Bug reporting and bug reproduction studies results.
  \textit{App ID} = identifier of the benchmark app,
  \textit{Fail. ID} = identifier for the failure in the benchmark app,
  \textit{Rep ID} = participant identifier in the bug reporting study,
  \textit{S2Rs} = number of steps to reproduce in the bug report submitted by the participant,
  \textit{$T_{\mathit{report}}$} = time taken by the participant to submit the bug report,
  \textit{Dev ID} = developer identifier in the bug reproduction study,
  \textit{$T_{\mathit{repro}}$} = time taken by the participant to submit the bug report,
  \textit{Repro} = \cmark if a developer reproduced the bug report (\xmark otherwise),
  \textit{Val S2Rs} = number of S2Rs validated by \tech,
  \revision{
  \textit{GUI Action Sugg.} = GUI action suggestions (GAS),
  \textit{GAS P} = number of GUI Action suggestions provided to the participants of the bug reporting study,
  \textit{GAS A} = number of GUI Action suggestions accepted by the participants of the bug reporting study,
  \textit{GAS A\Arrow{.15cm}\hspace{-2pt}E} = number of GUI Action suggestions accepted and then edited by the participants of the bug reporting study,
  \textit{GAS A\Arrow{.15cm}\hspace{-2pt}D} = number of GUI Action suggestions accepted and then deleted by the participants of the bug reporting study,
  \textit{GUI Element Sugg.} = GUI element suggestions (GES),
  \textit{GES P} = number of GUI Element suggestions provided to the participants of the bug reporting study,
  \textit{GES A} = number of GUI Element suggestions accepted by the participants of the bug reporting study,
  \textit{GES A\Arrow{.15cm}\hspace{-2pt}E} = number of GUI Element suggestions accepted and then edited by the participants of the bug reporting study,
  \textit{GES A\Arrow{.15cm}\hspace{-2pt}D} = number of GUI Element suggestions accepted and then deleted by the participants of the bug reporting study,
  \textit{PS A} = number of particle suggestions accepted by the participants of the bug reporting study.
  }
  }
  \label{tab:study-data}
  \vspace{-15pt}
  \begin{footnotesize}
    \begin{center}
      \begin{tabular}[h]{|l|c||c|c|c|c|r|c||c|c|c|c|c|c|c|c|c|c|c|c|r|c|r|c|}
      			\hhline{|-|-||------||----------------|}
      			\multirow{3}{*}{\parbox[c][15mm]{5mm}{\centering\shortstack{\textit{App}\\ \textit{ID}}}} &
      			\multirow{3}{*}{\parbox[c][15mm]{5mm}{\centering\shortstack{\textit{Fail.}\\ \textit{ID}}}} &
      			\multicolumn{6}{c||}{\textit{\fusion}} & \multicolumn{16}{c|}{\textit{\tech}}\\
      			\hhline{|~|~||------||----------------|}
      			& & \multicolumn{3}{c|}{\textit{Reporting}} & \multicolumn{3}{c||}{\textit{Reproducing}} & \multicolumn{13}{c|}{\textit{Reporting}} & \multicolumn{3}{c|}{\textit{Reproducing}}\\
      			\hhline{|~|~||------||----------------|}
      			& &  \multirow{2}{*}{\parbox[c]{4mm}{\centering\shortstack{\textit{Rep}\\ \textit{ID}}}} & \multirow{2}{*}{\textit{S2Rs}} & \multirow{2}{*}{$T_{\mathit{report}}$} & \multirow{2}{*}{\parbox[c]{5mm}{\centering\shortstack{\textit{Dev}\\ \textit{ID}}}} & \multirow{2}{*}{$T_{\mathit{repro}}$} & \multirow{2}{*}{\rotatebox[origin=c]{90}{\textit{Repro}}}
      			& \multirow{2}{*}{\parbox[c]{5mm}{\centering\shortstack{\textit{Rep}\\ \textit{ID}}}}
      			& \multirow{2}{*}{\parbox[c]{6mm}{\centering\shortstack{\textit{S2Rs}}}}  & \multirow{2}{*}{\parbox[c]{6mm}{\centering\shortstack{\textit{Val}\\ \textit{S2Rs}}}} & \multicolumn{4}{c|}{\textit{\parbox[c]{24mm}{\centering\shortstack{\textit{GUI Action Sugg.}}}}}	& \multicolumn{4}{c|}{\textit{\parbox[c]{24mm}{\centering\shortstack{\textit{GUI Element Sugg.}}}}} & \multirow{2}{*}{\parbox[c]{6mm}{\centering\shortstack{\textit{PS$_{A}$}}}} 
      			& \multirow{2}{*}{$T_{\mathit{report}}$} & \multirow{2}{*}{\parbox[c]{5mm}{\centering\shortstack{\textit{Dev}\\ \textit{ID}}}} & \multirow{2}{*}{$T_{\mathit{repro}}$} & \multirow{2}{*}{\rotatebox[origin=c]{90}{\textit{Repro}}}  \\
      			\hhline{|~|~||~~~~~~||~~~--------~~~~~|}
      			& &  & & & & & & & & & \parbox[c]{5mm}{\centering\shortstack{\textit{P}}} & \parbox[c]{5mm}{\centering\shortstack{\textit{A}}} & \parbox[c]{6mm}{\centering\shortstack{\textit{A\Arrow{.15cm}\hspace{-2pt}E}}} & \parbox[c]{6mm}{\centering\shortstack{\textit{A\Arrow{.15cm}\hspace{-2pt}D}}} & \parbox[c]{5mm}{\centering\shortstack{\textit{P}}} & \parbox[c]{5mm}{\centering\shortstack{\textit{A}}} & \parbox[c]{6mm}{\centering\shortstack{\textit{A\Arrow{.15cm}\hspace{-2pt}E}}} & \parbox[c]{6mm}{\centering\shortstack{\textit{A\Arrow{.15cm}\hspace{-2pt}D}}} & & & & & \\
				\hhline{|-|-||------||----------------|}
				\noalign{\vspace{1.5pt}}
				\hhline{|-|-||------||----------------|}
				\multirow{2}{*}{A01} & \multirow{2}{*}{F01} & R03 & $4$ & $9$m$10$s & D10 & $1$m$56$s & \cmark & R02 & $5$ & $3$ & $5$ & $1$ & $1$ & $1$ & $1$ & $1$ & $0$ & $0$ & $0$ & $4$m$58$s & D02 & $2$m$36$s & \cmark\\
				                                                                        &  & R08 & $4$ & $9$m$14$s & D03 & $4$m$00$s & \cmark & R10 & $4$ & $3$ & $3$ & $0$ & $1$ & $0$ & $3$ & $1$ & $0$ & $0$ & $0$ & $2$m$30$s & D07 & $1$m$13$s & \cmark\\
				\hhline{|-|-||------||----------------|}
				\multirow{2}{*}{A02} & \multirow{2}{*}{F02} & R04 & $4$ & $7$m$18$s & D09 & $4$m$30$s & \cmark & R02 & $6$ & $4$ & $1$ & $0$ & $0$ & $0$ & $5$ & $4$ & $1$ & $0$ & $0$ & $5$m$17$s & D01 & $1$m$45$s & \cmark\\
			                                                                             &  & R08 & $6$ & $6$m$20$s & D04 & $2$m$37$s & \cmark & R09 & $6$ & $4$ & $7$ & $1$ & $0$ & $5$ & $15$ & $3$ & $0$ & $11$ & $0$ & $9$m$42$s & D06 & $59$s & \cmark\\
				\hhline{|-|-||------||----------------|}
				\multirow{2}{*}{A03} & \multirow{2}{*}{F03} & R05 & $8$ & $7$m$48$s & D09 & $4$m$10$s & \cmark & R02 & $10$ & $10$ & $9$ & $0$ & $3$ & $0$ & $7$ & $7$ & $0$ & $0$ & $2$ & $4$m$55$s & D05 & $1$m$04$s & \cmark\\
			                                                                             &  & R10 & $10$ & $5$m$40$s & D01 & $9$m$55$s & \cmark & R09 & $16$ & $16$ & $14$ & $3$ & $5$ & $0$ & $6$ & $5$ & $0$ & $0$ & $0$ & $6$m$42$s & D07 & $1$m$47$s & \cmark\\
				\hhline{|-|-||------||----------------|}
				\multirow{2}{*}{A04} & \multirow{2}{*}{F04} & R05 & $7$ & $6$m$34$s & D07 & $10$m$00$s & \xmark & R07 & $8$ & $4$ & $0$ & $0$ & $0$ & $0$ & $12$ & $6$ & $1$ & $3$ & $1$ & $7$m$29$s & D09 & $5$m$40$s & \cmark\\
			                                                                             &  & R06 & $4$ & $11$m$57$s & D02 & $3$m$33$s & \cmark & R01 & $10$ & $7$ & $6$ & $1$ & $0$ & $0$ & $4$ & $2$ & $0$ & $2$ & $0$ & $6$m$27$s & D01 & $2$m$30$s & \cmark\\
				\hhline{|-|-||------||----------------|}
				\multirow{4}{*}{A05} & \multirow{2}{*}{F05} & R03 & $5$ & $7$m$36$s & D07 & $1$m$40$s & \cmark & R07 & $5$ & $5$ & $1$ & $0$ & $0$ & $0$ & $5$ & $4$ & $0$ & $1$ & $1$ & $3$m$42$s & D10 & $45$s & \cmark\\
			                                                                             &  & R08 & $3$ & $4$m$36$s & D01 & $10$m$00$s & \xmark & R01 & $5$ & $4$ & $2$ & $0$ & $0$ & $0$ & $3$ & $2$ & $0$ & $0$ & $0$ & $4$m$07$s & D05 & $1$m$13$s & \cmark\\
			                                                                             \hhline{|~|-||------||----------------|}
			                                          & \multirow{2}{*}{F06} & R09 & $7$ & $10$m$38$s & D02 & $4$m$18$s & \cmark & R06 & $5$ & $5$ & $2$ & $0$ & $0$ & $0$ & $5$ & $5$ & $0$ & $0$ & $1$ & $8$m$02$s & D08 & $10$m$00$s & \xmark\\
			                                                                             &  & R02 & $5$ & $2$m$50$s & D06 & $10$m$00$s & \xmark & R04 & $5$ & $5$ & $0$ & $0$ & $0$ & $0$ & $3$ & $3$ & $0$ & $0$ & $1$ & $3$m$15$s & D03 & $1$m$20$s & \cmark\\
				\hhline{|-|-||------||----------------|}
				\multirow{4}{*}{A06a} & \multirow{2}{*}{F07} & R03 & $13$ & $16$m$51$s & D09 & $3$m$30$s & \cmark & R02 & $12$ & $11$ & $15$ & $5$ & $2$ & $1$ & $6$ & $4$ & $0$ & $0$ & $0$ & $10$m$30$s & D02 & $2$m$26$s & \cmark\\
			                                                                             &  & R07 & $12$ & $21$m$29$s & D04 & $3$m$10$s & \cmark & R09 & $9$ & $9$ & $31$ & $6$ & $3$ & $11$ & $5$ & $1$ & $0$ & $1$ & $0$ & $20$m$11$s & D07 & $1$m$07$s & \cmark\\
			                                                                             \hhline{|~|-||------||----------------|}
			                                          & \multirow{2}{*}{F08} & R05 & $13$ & $19$m$01$s & D07 & $10$m$00$s & \xmark & R04 & $15$ & $11$ & $14$ & $2$ & $0$ & $4$ & $20$ & $6$ & $0$ & $10$ & $4$ & $18$m$40$s & D04 & $7$m$36$s & \cmark\\
			                                                                             &  & R10 & $19$ & $17$m$20$s & D02 & $6$m$19$s & \cmark & R08 & $13$ & $12$ & $5$ & $0$ & $2$ & $0$ & $13$ & $10$ & $0$ & $0$ & $0$ & $8$m$11$s & D09 & $10$m$00$s & \xmark\\
				\hhline{|-|-||------||----------------|}
				\multirow{6}{*}{A06b} & \multirow{2}{*}{F09} & R03 & $9$ & $10$m$33$s & D06 & $10$m$00$s & \xmark & R10 & $9$ & $8$ & $12$ & $4$ & $0$ & $2$ & $6$ & $4$ & $0$ & $0$ & $0$ & $3$m$35$s & D08 & $3$m$18$s & \cmark\\
			                                                                             &  & R09 & $9$ & $14$m$12$s & D03 & $3$m$30$s & \cmark & R01 & $7$ & $6$ & $7$ & $2$ & $0$ & $1$ & $4$ & $1$ & $0$ & $3$ & $0$ & $4$m$07$s & D01 & $3$m$00$s & \cmark\\
			                                                                             \hhline{|~|-||------||----------------|}
			                                          & \multirow{2}{*}{F10} & R02 & $8$ & $12$m$29$s & D08 & $8$m$10$s & \cmark & R05 & $9$ & $9$ & $2$ & $0$ & $0$ & $0$ & $13$ & $8$ & $0$ & $3$ & $0$ & $8$m$14$s & D05 & $3$m$35$s & \cmark\\
			                                                                             &  & R06 & $10$ & $21$m$15$s & D01 & $3$m$15$s & \cmark & R09 & $7$ & $7$ & $12$ & $1$ & $0$ & $3$ & $12$ & $4$ & $0$ & $5$ & $0$ & $10$m$04$s & D10 & $2$m$55$s & \cmark\\
			                                                                              \hhline{|~|-||------||----------------|}
			                                          & \multirow{2}{*}{F11} & R04 & $6$ & $11$m$52$s & D10 & $1$m$08$s & \cmark & R08 & $10$ & $9$ & $3$ & $0$ & $0$ & $0$ & $9$ & $8$ & $0$ & $1$ & $0$ & $5$m$03$s & D06 & $5$m$10$s & \cmark\\
			                                                                             &  & R07 & $15$ & $12$m$36$s & D05 & $2$m$33$s & \cmark & R03 & $11$ & $7$ & $9$ & $4$ & $0$ & $2$ & $2$ & $1$ & $0$ & $0$ & $0$ & $9$m$49$s & D03 & $3$m$30$s & \cmark\\
				\hhline{|-|-||------||----------------|}
				\multirow{6}{*}{A07} & \multirow{2}{*}{F12} & R08 & $13$ & $10$m$03$s & D01 & $10$m$00$s & \xmark & R07 & $22$ & $20$ & $5$ & $0$ & $0$ & $0$ & $16$ & $14$ & $0$ & $0$ & $7$ & $7$m$32$s & D06 & $3$m$00$s & \cmark\\
			                                                                             &  & R05 & $11$ & $4$m$50$s & D09 & $8$m$50$s & \cmark & R01 & $21$ & $12$ & $2$ & $0$ & $0$ & $0$ & $3$ & $3$ & $0$ & $0$ & $0$ & $6$m$23$s & D03 & $4$m$00$s & \cmark\\
			                                                                             \hhline{|~|-||------||----------------|}
			                                          & \multirow{2}{*}{F13} & R02 & $16$ & $7$m$55$s & D10 & $1$m$58$s & \cmark & R03 & $17$ & $16$ & $15$ & $7$ & $6$ & $0$ & $2$ & $2$ & $0$ & $0$ & $0$ & $6$m$22$s & D02 & $2$m$18$s & \cmark\\
			                                                                             &  & R06 & $6$ & $8$m$32$s & D05 & $3$m$01$s & \cmark & R08 & $10$ & $9$ & $0$ & $0$ & $0$ & $0$ & $7$ & $7$ & $0$ & $0$ & $0$ & $5$m$01$s & D07 & $1$m$22$s & \cmark\\
			                                                                              \hhline{|~|-||------||----------------|}
			                                          & \multirow{2}{*}{F14} & R04 & $11$ & $5$m$37$s & D08 & $3$m$26$s & \cmark & R05 & $12$ & $10$ & $1$ & $0$ & $0$ & $0$ & $12$ & $11$ & $0$ & $1$ & $1$ & $5$m$13$s & D04 & $2$m$50$s & \cmark\\
			                                                                             &  & R09 & $16$ & $10$m$08$s & D03 & $7$m$00$s & \cmark & R10 & $10$ & $10$ & $8$ & $2$ & $4$ & $1$ & $5$ & $4$ & $0$ & $0$ & $0$ & $3$m$33$s & D09 & $3$m$15$s & \cmark\\
				\hhline{|-|-||------||----------------|}
				\multirow{2}{*}{A08} & \multirow{2}{*}{F15} & R10 & $6$ & $4$m$03$s & D05 & $1$m$14$s & \cmark & R04 & $5$ & $5$ & $3$ & $1$ & $0$ & $0$ & $3$ & $3$ & $0$ & $0$ & $1$ & $2$m$48$s & D02 & $1$m$22$s & \cmark\\
			                                                                             &  & R01 & $7$ & $5$m$00$s & D08 & $1$m$01$s & \cmark & R06 & $6$ & $6$ & $7$ & $4$ & $1$ & $0$ & $3$ & $1$ & $0$ & $1$ & $0$ & $4$m$58$s & D10 & $25$s & \cmark\\
				\hhline{|-|-||------||----------------|}
				\multirow{2}{*}{A09} & \multirow{2}{*}{F16} & R07 & $3$ & $2$m$40$s & D05 & $3$m$27$s & \cmark & R03 & $2$ & $2$ & $3$ & $1$ & $0$ & $0$ & $1$ & $1$ & $0$ & $0$ & $0$ & $2$m$03$s & D04 & $1$m$00$s & \cmark\\
			                                                                             &  & R01 & $2$ & $1$m$49$s & D06 & $16$s & \cmark & R06 & $2$ & $2$ & $1$ & $0$ & $0$ & $0$ & $2$ & $2$ & $0$ & $0$ & $1$ & $2$m$47$s & D10 & $29$s & \cmark\\
				\hhline{|-|-||------||----------------|}
				\multirow{2}{*}{A10} & \multirow{2}{*}{F17} & R06 & $7$ & $9$m$11$s & D04 & $10$m$00$s & \xmark & R07 & $7$ & $7$ & $0$ & $0$ & $0$ & $0$ & $7$ & $5$ & $0$ & $0$ & $0$ & $5$m$38$s & D08 & $35$s & \cmark\\
			                                                                             &  & R01 & $7$ & $6$m$19$s & D06 & $2$m$11$s & \cmark & R05 & $7$ & $7$ & $3$ & $0$ & $0$ & $0$ & $10$ & $7$ & $0$ & $1$ & $0$ & $3$m$04$s & D03 & $5$m$00$s & \cmark\\
				\hhline{|-|-||------||----------------|}
				\multirow{6}{*}{A11} & \multirow{2}{*}{F18} & R01 & $4$ & $2$m$57$s & D07 & $5$m$51$s & \cmark & R10 & $10$ & $9$ & $0$ & $0$ & $0$ & $0$ & $9$ & $8$ & $0$ & $1$ & $0$ & $2$m$41$s & D09 & $4$m$10$s & \cmark\\
			                                                                             &  & R07 & $9$ & $7$m$41$s & D03 & $10$m$00$s & \xmark & R04 & $7$ & $6$ & $0$ & $0$ & $0$ & $0$ & $2$ & $0$ & $0$ & $2$ & $0$ & $5$m$28$s & D01 & $10$m$00$s & \xmark\\
			                                                                             \hhline{|~|-||------||----------------|}
			                                          & \multirow{2}{*}{F19} & R02 & $8$ & $13$m$45$s & D08 & $2$m$00$s & \cmark & R05 & $7$ & $5$ & $0$ & $0$ & $0$ & $0$ & $11$ & $5$ & $0$ & $0$ & $0$ & $5$m$45$s & D05 & $2$m$34$s & \cmark\\
			                                                                             &  & R10 & $11$ & $5$m$55$s & D04 & $7$m$22$s & \cmark & R06 & $8$ & $6$ & $0$ & $0$ & $0$ & $0$ & $10$ & $6$ & $0$ & $0$ & $1$ & $7$m$40$s & D06 & $48$s & \cmark\\
			                                                                              \hhline{|~|-||------||----------------|}
			                                          & \multirow{2}{*}{F20} & R04 & $10$ & $4$m$27$s & D10 & $1$m$40$s & \cmark & R08 & $12$ & $12$ & $0$ & $0$ & $0$ & $0$ & $9$ & $9$ & $0$ & $0$ & $0$ & $3$m$31$s & D08 & $45$s & \cmark\\
			                                                                             &  & R09 & $11$ & $7$m$02$s & D02 & $3$m$42$s & \cmark & R03 & $12$ & $12$ & $1$ & $1$ & $0$ & $0$ & $7$ & $7$ & $0$ & $0$ & $0$ & $5$m$50$s & D04 & $2$m$58$s & \cmark\\
				\hhline{|-|-||------||----------------|}
				\multicolumn{1}{c}{} & \multicolumn{1}{c}{} & \multicolumn{1}{|c|}{-} & $339$ & $365$m$13$s & - & $201$m$13$s & 32 &- & $364$ & $315$ & $209$ & $46$ & $28$ & $31$ & $278$ & $185$ & $2$ & $46$ & $21$ & $251$m$47$s & - & $120$m$20$s & 37\\
				\hhline{~~||------||----------------|}
      \end{tabular}
    \end{center}
  \end{footnotesize}
  \vspace{-18pt}
\end{table*}	

Overall, developers using \tech's reports could successfully reproduce the report's failure in 37 out of 40 cases ($92.5$\% success rate). Developers using \fusion's reports could do so in 32 out of 40 cases ($80$\% success rate). After inspecting the bug reports, we explain this result with the fact that bug report participants created more complete bug reports while using \tech. This characteristic can be seen in Table~\ref{tab:study-data} by comparing the \textit{S2Rs} columns in the \tech and \fusion sections of the table. Specifically, participants reported a total of $364$ S2Rs using \tech and $339$ using \fusion. Additionally, the result can also be related to the fact that $315$ out of the $364$ S2Rs reported using \tech, were also validated by the technique. As another explanation of the result, the clarity of the the S2Rs provided by \tech also appeared as a comment in the survey answers of seven out the $10$ developers.
Finally, developers using \tech's bug reports could reproduce all of the $20$ failures we selected for the empirical evaluation.

\revision{We manually analyzed the three reports submitted using \tech that could not be reproduced by developers and identified that all three reports had missing S2Rs. The first report (F11 submitted by R04) had two missing S2Rs, the second report (F06 submitted by R06) had six missing S2Rs, and the third report (F08 submitted by R08) had seven missing S2Rs. Because of the missing S2Rs, two of the reports (F06 submitted by R06 and F08 submitted by R08) had some S2Rs that could not be validated by \tech. In the third report (F11 submitted by R04), all S2Rs could be validated by \tech as the missing S2Rs corresponded to GUI actions in the screens as already validated S2Rs, and therefore \tech did not identify issues with the sequence of S2Rs. Although the developers did not provide the reasons for which they could not reproduce these reports in their exit surveys, we believe that these missing S2Rs might have affected the bug reproduction outcome. We believe that automated bug reproduction techniques could be combined with \tech to provide feedback on whether the reports could be reproduced upon submission, and this feedback might incentivize users to refine their reports.}

In the bug reproduction study, we also asked developers to reproduce the failures using the original bug reports. Developers were able to do so in 15 out of 20 cases ($75$\% success rate). These results show that \tech is more effective than \fusion and the original bug tracking systems in providing reproducible bug reports to developers.

In summary, we believe that the results show that \tech can be effective in providing reproducible bug reports and that \tech is more effective than \fusion.

\subsubsection{RQ2: Bug Reporting Efficiency}

To answer RQ2, we compare the amount of time the participants took to submit a bug report using \fusion and \tech. For each bug reporting task, Table~\ref{tab:study-data} reports the time (columns {\small $T_{\mathit{report}}$}) participants took to report a specific failure.

In total, when using \tech, the participants submitted the $40$ bug reports in $251$m$47$s (about $6$m$18$s per report). When using \fusion instead, the participants took $365$m$13$s to report the same failures (about $9$m$8$s per report). These results show a decrease of $31.06$\% in the bug reporting time when using \tech. If we only considered successfully reproduced bug reports, the average bug reporting time for \tech is $6$m$14$s and for \fusion is $9$m$13$s. These numbers indicate that the decrease in bug reporting time not only applies to all bug reports but, in particular, also applies to successfully reproduced ones. 

\revision{The bug reporting time of \tech is characterized by a number of suggestions provided by \tech and accepted by the participants of the study. For both GUI action and GUI element suggestions (\textit{GUI Action Sugg.} and \textit{GUI Element Sugg.} headers), Table~\ref{tab:study-data} reports provided (\textit{P}), accepted (\textit{A}), accepted and then edited (\textit{A\Arrow{.15cm}\hspace{-2pt}E}), and accepted and then deleted (\textit{A\Arrow{.15cm}\hspace{-2pt}D}) suggestions. In total, \tech provided 209 GUI Action suggestions, and participants accepted 46 suggestions (without any modification). The participants also accepted 28 additional suggestions and then edited the suggestions. In the modifications, the participants refined the content of the suggestion by providing additional information in the S2R. The participants also accepted 31 suggestions that they then removed from the bug report.  One participant (R09) rewrote part of two bug reports multiple times before submitting the results, and this contributed to 12 suggestions being deleted. Although deleted suggestion could represent suggestions that were not ultimetly helpful for participants, deleted suggestions might be leveraged by automated bug reproduction techniques~\cite{2018_issta_fazzini_automatically,2019_icse_zhao_recdroid} to guide the exploration performed by the techniques to reproduce bug reports. The number of GUI element suggestions is 278. The participants accepted 185 suggestions, accepted and then edited two suggestions, and accepted and then deleted 46 suggestions. The GUI element suggestions reported in Table~\ref{tab:study-data} only include suggestions provided by the GUI element prediction model. As a side note, the participants also accepted 42 suggestions created for partially written GUI elements. (These suggestions are based on text matching and not the GUI element prediction models). The participants also accepted $21$ particle suggestions (\textit{PS$_{A}$}). The small number of accepted particle suggestions is due to the significant number of accepted GUI action and element suggestions. The usefulness of the suggestions provided by \tech appeared as a comment in the survey answers of nine out the ten participants and the comments highlighted that the suggestions made reporting faster.}

\revision{Analyzing failures individually, we identified two cases (F02 and F16) in which the average time to report the failure (the study had two submitted bug reports per failure and tool) was lower in \textsc{Fusion} as compared to \tech. The first failure (F02) is an output failure and the second one (F16) is a crash. In the first case, the average reporting time with \tech was higher because one of the two users (R09) had a significantly higher reporting time. Although we do not know the exact reason behind this higher reporting time, we observed that the user rewrote four S2Rs three times because the subsequent S2R could not be validated by \tech. This result shows that improving the accuracy of \tech's GUI model, which is used to validate S2Rs, can further improve \tech's efficiency. The second failure (F16) is the failure caused by the shortest number of GUI actions (two). In this case, the average reporting time with \tech ($2$m$25$s) is slightly slower than \textsc{Fusion} ($2$m$15$s). We believe that \tech does not perform better than \textsc{Fusion} in this case because users cannot take full advantage of \tech's suggestions.}

Give the results of our bug reporting study, we can conclude that \tech allows users to submit bug reports more efficiently than \fusion.

\subsubsection{RQ3: Bug Reproduction Efficiency}

To answer RQ3, we compare the time developers took to reproduce \fusion's and \tech's bug reports. Table~\ref{tab:study-data} reports the time (columns $T_{\mathit{repro}}$) developers took to reproduce the failure described in the bug reports.

Overall, developers took $120$m$20$s to reproduce \tech's reports and $201$m$13$s to reproduce \fusion's reports. These numbers identify a $40.2$\% decrease in the time to reproduce bug reports when \tech's reports. (Note that this number is affected by the 10m time limit set for the bug reproduction task.) As in RQ1, we explain this improvement due to the fact that \tech provides more complete bug reports with respect to \fusion. If we only consider bug reports successfully reproduced by developers, developers could reproduce \tech's bug reports in $2$m$26$s and \fusion's bug reports in $3$m$46$s on average. This result highlights that developers can reproduce \tech's bug reports faster even without considering the cases where the bug reproduction task is not successful. Finally, developers could also reproduce \tech's bug reports faster than the original bug reports, which took $3$m$5$s on average in the successful cases.

\revision{Considering failures for which developers could reproduce both participant-submitted bug reports (the study had two submitted bug reports per failure and tool), developers using bug reports submitted with \tech were slower in only one case (F11). To report this failure, users had to provide a value from a dropdown list that contained a large number of values (225). Because \tech also allows users to report this information just by using text instead of specifically selecting a value from a suggestion, reporters using \tech provided this information using a textual description without going through the technique's suggestions. The description provided by both users was not complete, and we believe that this aspect might have caused the higher reproduction time with \tech, as developers had to identify the right value among a large number of values. This result highlights a tradeoff between flexibility in the bug reporting process and efficiency in bug reproduction time. We believe that this tradeoff is worth investigating in future work through additional user studies.}

\begin{figure}[t!]
	\centering
	\centerline{\includegraphics[width=0.95\linewidth]{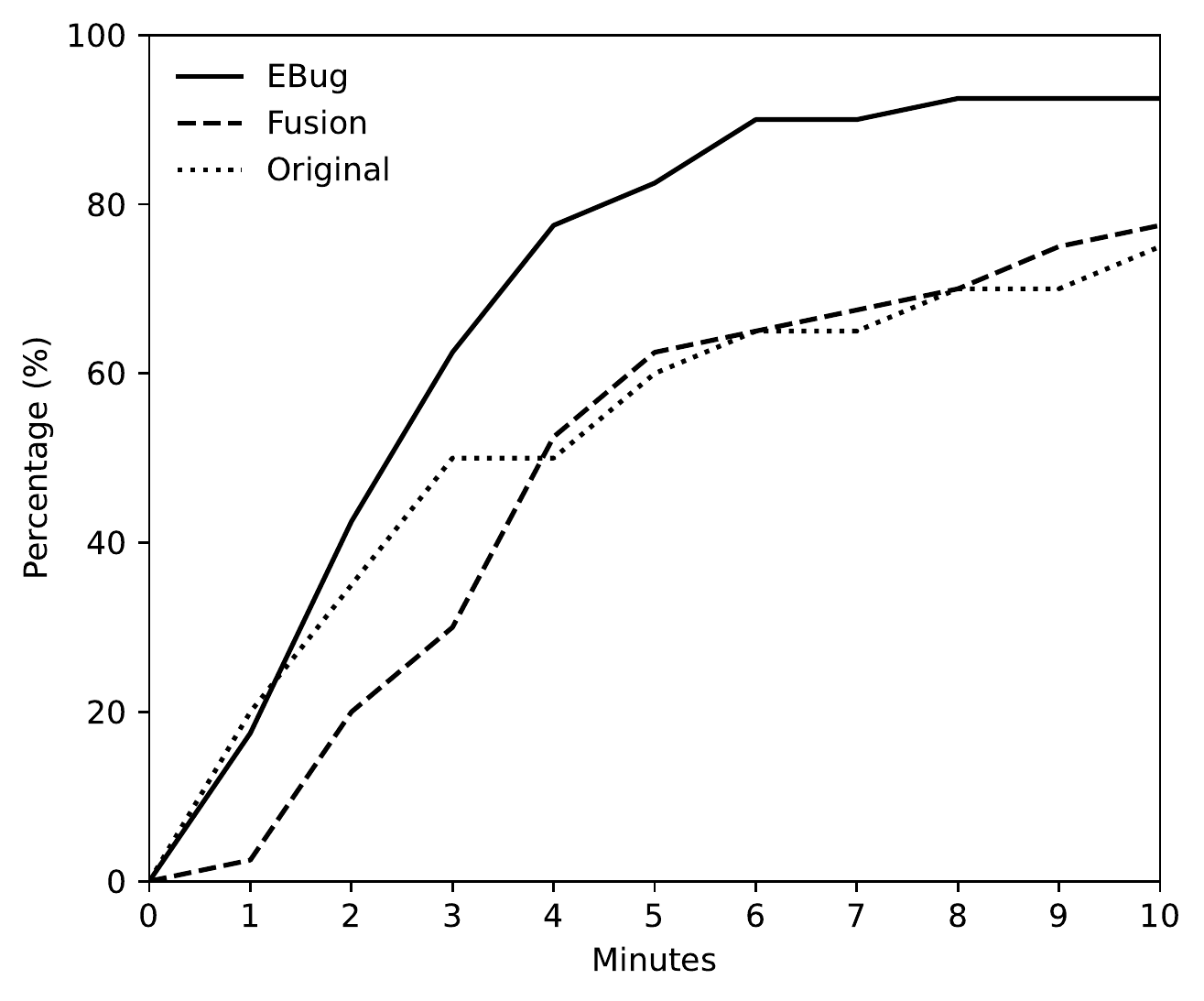}}
	\vspace{-15pt}
	\caption{\revision{\revisiontwo{Percentage of bug reports reproduced by developers over time.}}}
	\label{fig:repro-graph}
	\vspace{5pt}
\end{figure}

\revision{Fig.~\ref{fig:repro-graph} reports the percentage of bug reports reproduced within $n$ minutes (where $n$ ranges from $0$ to $10$). The graph confirms that developers could reproduce more bug reports and faster using \tech's reports as compared to \textsc{Fusion}'s reports and the original reports.
%The graph also shows that developers could generally reproduce more bug reports while being faster using \tech's reports as compared to \textsc{Fusion}'s reports and the original reports.
The only case where developers were not the fastest using \tech's reports is within the first minute, where 20\% of the original reports were reproduced as opposed to 17.5\% of \tech's reports. We believe that this last result can be explained by the fact that some of the original reports were slightly quicker to process as they contained less information, but in the following minutes, when relevant S2Rs were necessary to reproduce the failures, \tech outperformed both \textsc{Fusion} and the original reports.}

\revision{Because reports for the same failure were submitted by different participants and reproduced by different developers, it is not possible for us to compare results individually and identify exactly why it took longer for a certain failure to be reproduced using reports from \tech. However, analyzing the content of submitted reports, we believe that missing S2Rs and the developers' experience could be factors that affected the failure reproduction time.}

In summary, we can conclude that \tech allows developers to reproduce bug reports more efficiently than \fusion and the issue tracking systems of the original bug reports.

\subsubsection{RQ4: Bug Reporting User Experience}

\begin{figure}[t!]
	\centering
	\centerline{\includegraphics[width=0.95\linewidth]{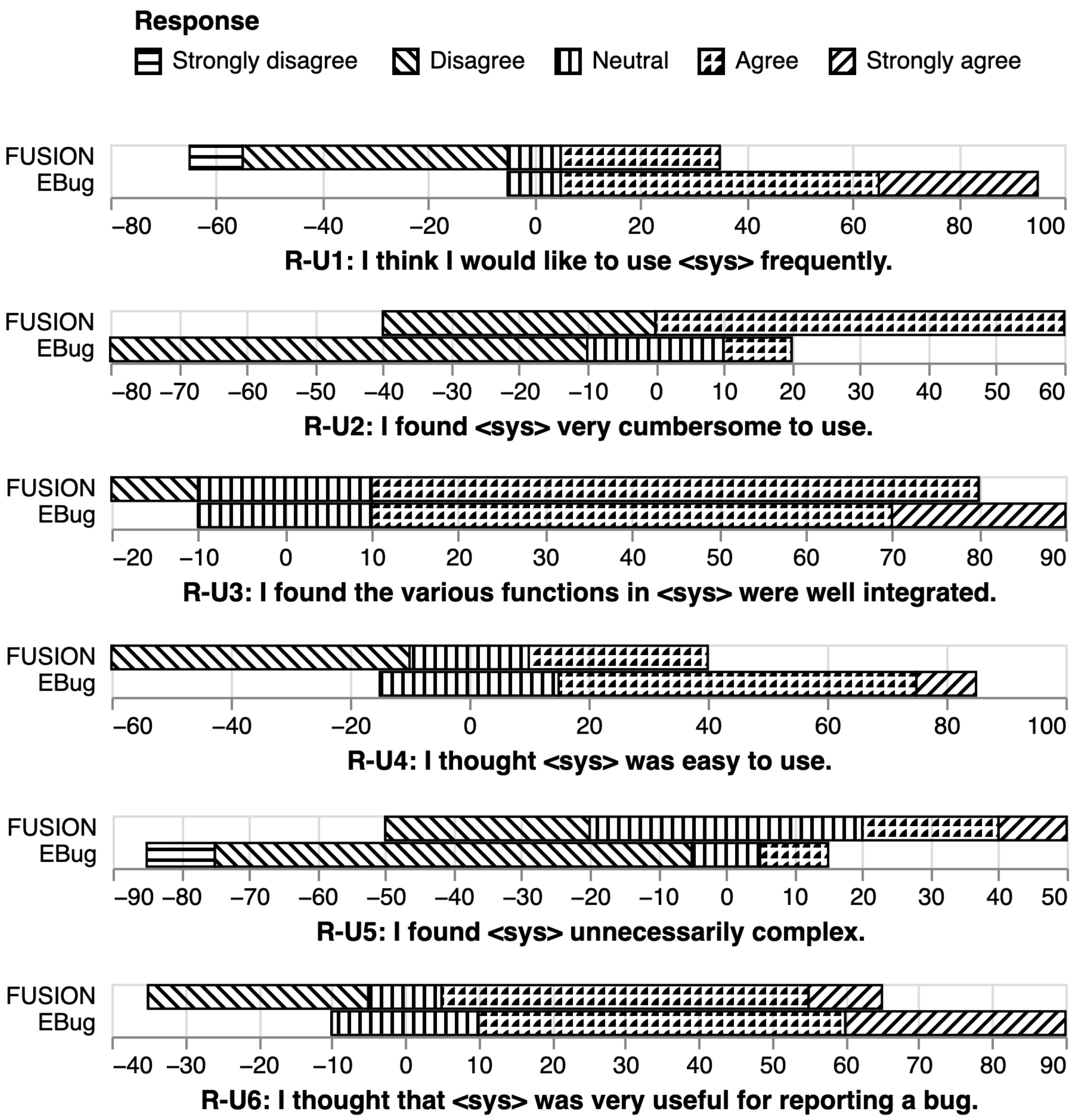}}
	\vspace{-8pt}
	\caption{\revisiontwo{Usability answers in the survey of the bug reporting study.}}
	\label{fig:reporting-charts}
	\vspace{5pt}
\end{figure}

To answer RQ4, we collected and analyzed the participants' answers to the exit survey questions of the bug reporting study. We report the answer to the usability questions (R-UN questions in Table~\ref{tab:report-questions}) in Figure~\ref{fig:reporting-charts} using centered stacked bar charts. For questions R-U1, R-U3, R-U4, and R-U6 agreement is better, while for questions R-U2 and R-U5 disagreement is better.

Overall, the participants provided very positive answers to the usability questions related to \tech. Nine of the ten participants either agreed ($6$) or strongly agreed ($3$) with the statement ``\textit{I think I would like to use \tech frequently}'' (R-U1). Participants also provided positive answers ($5$ agree and $3$ strongly agree) in relation to the statement ``\textit{I thought that \tech was very useful for reporting a bug}'' (R-U6). Among all answers provided by all of the participants, only one was negative. Specifically, one participant agreed with the statement ``\textit{I found \tech very cumbersome to use}'' (R-U2). The same participant had a neutral answer in relation to the statement ``\textit{I thought \tech was easy to use}'' (R-U4). We believe that this answer was related to the fact that the participant experienced networking issues during the study, and our system relies on a network connection to operate.

In relation to user experience with \tech, we also want to report a few excerpts from the free-text answers we received from the participants when they answered the user preference questions (R-PN questions in Table~\ref{tab:report-questions}).
%These answers echo some of our claims about \tech's usefulness.
Some answers indicate that the suggestions provided by \tech make the bug reporting process more efficient: ``\textit{[...] Autocomplete really speeds up the process [...]}'' (participant R02), ``\textit{[...] The context-aware list of components helps finding the one I'm thinking of [...] Templates of step description reduce a lot of typing}'' (R03), and ``\textit{[...] I like the action and GUI components' suggestions. That functionality is very useful for saving time [...]}'' (R6). Finally, some answers indicate that \tech is useful and a natural way to submit bug reports: ``\textit{I liked that I was basically typing a paragraph of instructions that happened to automatically link with images.}'' (R01), ``\textit{It feels natural while reporting bugs with [\tech] [...]}'' (R05), and ``\textit{I felt that this tool was very useful and would prove very easy to use, even if you had no coding experience [...]}'' (R07). 

\revision{The free-text answers also provided useful feedback for improving \tech and bug reporting through future work. A comment (from participant R03) mentioned that ``\textit{If the app contains same-looking components on multiple screens, the components list shows me multiple options, when there should be only one valid option if the tool considers the current state of the app}''. This situation can appear when \tech is not able to validate previously typed S2R. In this case, \tech provides a suggestion that contains all available GUI actions in the app (ordered by their distance in the GUI model with respect to the last validated step). This feedback highlights the opportunity to improve \tech in two ways. First, future research could define techniques to create an even more accurate GUI model. Second, future research could focus on user studies in which users receive a limited number of options in the suggestions based on a pre-defined GUI model distance value (and not include all available GUI actions as the options). A second comment (from R05) highlighted that ``\textit{[...] external controls such as power button, or emulator controls for changing orientation were not accessible.}. Power button support is currently not provided by \tech, but it could be added by representing the operation in the GUI model. For both types of actions, a user might not receive such suggestions also because the prediction model did not contain such actions. Having a more complete prediction model could be achieved by collecting additional user traces, but there is always the possibility of not having some actions in the model. Considering the results obtained while assessing \tech's bug reporting efficiency, we believe that the technique has good results even if some actions are not present in the prediction model. Another comment (from R06) reported that ``\textit{The selected target screen is unnecessary. At least in my case, I never used it at all.}''. The selected target screen is the one in the lower-right part of Fig.~\ref{fig:ebuginterface}. Future user studies related to \tech's usage could focus on assessing the usefulness of each of the technique's interface components.}

\revisiontwo{In summary, these preliminary results demonstrate that the participants had a positive user experience when submitting reports with \tech, and this experience was better than the one with \fusion.}

\subsubsection{RQ5: Bug Reproduction User Experience}

\begin{figure}[t!]
	\centering
	\centerline{\includegraphics[width=0.95\linewidth]{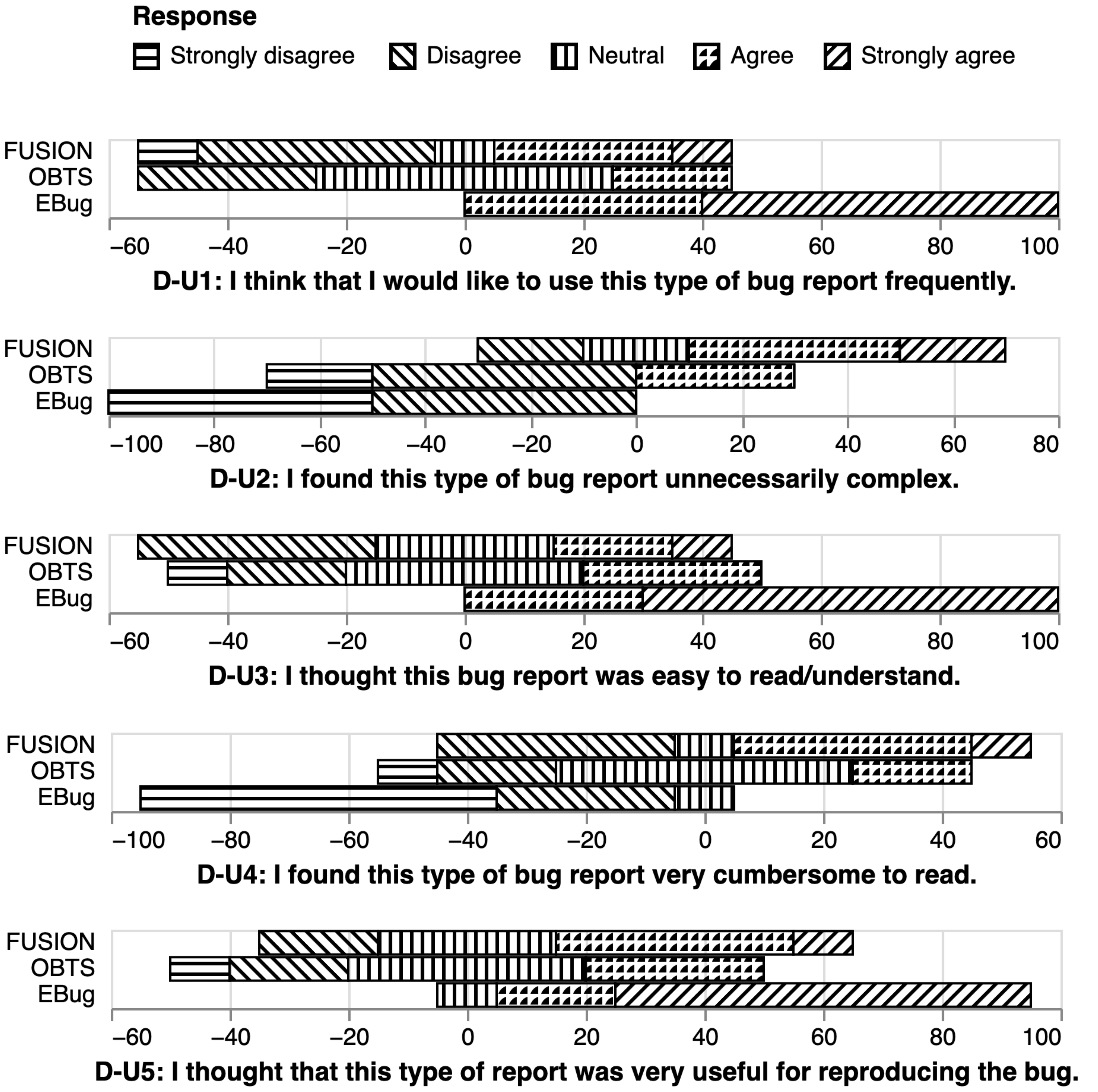}}
	\vspace{-8pt}
	\caption{\revisiontwo{Usability answers in the survey of the bug reproduction study. \textit{OBTS} is the original bug tracking system.}} \label{fig:repro-charts}
	%\vspace{-8pt}
\end{figure}

We answer RQ5 by analyzing the developers' answers to the exit survey of the bug reproduction study. We report the answer to the usability questions (D-UN questions in Table~\ref{tab:repro-questions}) in Figure~\ref{fig:repro-charts} using centered stacked bar charts. For questions D-U1, D-U3, and D-U5 agreement is better, while for questions D-U2 and D-U4 disagreement is better.

Also for RQ5, the subjects provided very positive answers to the usability questions related to \tech. All developers either agreed (four) or strongly agreed (six) with the statement ``\textit{I think that I would like to use this type of bug report frequently}'' (D-U1). All developers also agreed (three) or strongly agreed (seven) with the statement ``\textit{I thought this bug report was easy to read/understand}''' (D-U3). Additionally, developers provided positive answers (two agree and seven strongly agree) in relation to the statement ``\textit{I thought that this type of report was very useful for reproducing the bug}'' (D-U5). If we consider the participants' answers to all the questions, we found that all the answers were either neutral (two answers) or positive ($48$ answers). Developers did not provide as positive answers for \fusion's usability as the ones provided for \tech. In fact, some participants disagreed with statement D-U1 (four disagree and one strongly disagree) and D-U5 (two disagree). Figure~\ref{fig:repro-charts} also shows that the answers about the original bug tracking systems are not as positives as the ones provided for \tech.

\begin{table*}[t!]
  \setlength\tabcolsep{1.4pt}
  \renewcommand{\arraystretch}{1.1}
  \caption{Prediction models comparison.
  \textit{App ID} = app identifier,
  \textit{Traces} = number of traces ,
  \textit{Predictions} = number of predictions,
  \textit{Order} = order of AKOM's and \tech's models,
  \textit{SN} = number of suggestions provided by the models,
  \textit{wes} = wasted effort score,
  \textit{PL} = sequence prefix length for CPT+.  
  }
  \label{tab:prediction-models-performance}
  \vspace{-15pt}
  \begin{footnotesize}
    \begin{center}
      \begin{tabular}[h]{|l|c|c||c|c|c|c|c|c|c|c|c||c|c|c|c|c|c|c|c|c|}
      			\hhline{|-|-|-||---------||---------}
      			\multirow{3}{*}{\textit{App ID}} &
				\multirow{3}{*}{\textit{Traces}} & 
				\multirow{3}{*}{\textit{Predictions}} &
				\multicolumn{9}{c||}{\textit{GUI Action Models}} & 
				\multicolumn{9}{c|}{\textit{GUI Element Models}}\\
				\hhline{|~|~|~||---------||---------}
				& & & \multicolumn{3}{c|}{\textit{AKOM}} & \multicolumn{3}{c|}{\textit{CPT+}} & \multicolumn{3}{c||}{\textit{\tech}} & \multicolumn{3}{c|}{\textit{AKOM}} & \multicolumn{3}{c|}{\textit{CPT+}} & \multicolumn{3}{c|}{\textit{\tech}}\\
				\hhline{|~|~|~||---------||---------}
				& & & \textit{Order} & \textit{SN} & $wes$ & \textit{PL} & \textit{SN} & $wes$ & \textit{Order} & \textit{SN} & $wes$ & \textit{Order} & \textit{SN} & $wes$ & \textit{PL} & \textit{SN} & $wes$ & \textit{Order} & \textit{SN} & $wes$ \\
				\hhline{|---||---------||---------}
				\noalign{\vspace{1.5pt}}
				\hhline{|---||---------||---------}
				A01 & $12$ & $285$ & $4$ & $1$ & $1.159$ & $4$ & $1$ & $1.568$ & $3$ & $1$ & $\textbf{1.021}$ & $4$ & $1$ & $\textbf{0.727}$ & $3$ & $1$ & $2.314$ & $5$ & $1$ & $\textbf{0.727}$\\ %data-3
				A02 & $15$ & $283$ & $7$ & $1$ & $\textbf{0.85}$ & $3$ & $1$ & $1.55$ & $5$ & $1$ & $0.952$ & $6$ & $1$ & $\textbf{0.626}$ & $3$ & $1$ & $2.291$  & $9$ & $1$ & $0.675$\\ %data-4
				A03 & $14$ & $458$ & $4$ & $1$ & $\textbf{0.877}$ & $2$ & $1$ & $1.018$ & $4$ & $1$ & $0.885$ & $6$ & $1$ & $\textbf{0.69}$ & $3$ & $1$ & $3.362$ & $8$ & $1$ & $0.703$\\ %data-5
				A04 & $15$ & $284$ & $2$ & $1$ & $\textbf{2.595}$ & $2$ & $1$ & $3.239$ & $3$ & $1$ & $3.239$ & $4$ & $1$ & $\textbf{1.427}$ & $2$ & $1$ & $8.793$ & $5$ & $1$ & $1.513$\\ %data-6
				A05 & $12$ & $211$ & $1$ & $1$ & $\textbf{1.542}$ & $3$ & $1$ & $1.931$ & $3$ & $1$ & $1.638$ & $2$ & $1$ & $1.542$ & $3$ & $1$ & $4.024$ & $4$ & $1$ & $\textbf{1.269}$\\ %data-26
				A06a & $15$ & $464$ & $2$ & $1$ & $1.549$ & $2$ & $1$ & $1.698$ & $7$ & $1$ & $\textbf{1.275}$ & $4$ & $1$ & $1.468$ & $3$ & $1$ & $8.098$ & $9$ & $1$ & $\textbf{1.275}$\\ %data-25
				A06b & $14$ & $398$ & $1$ & $1$ & $1.355$ & $2$ & $1$ & $1.745$ & $6$ & $1$ & $\textbf{1.151}$ & $4$ & $1$ & $0.98$ & $3$ & $1$ & $4.169$ & $9$ & $1$ & $\textbf{0.961}$\\ %data-29
				A07 & $13$ & $285$ & $2$ & $1$ & $0.717$ & $5$ & $1$ & $0.759$ & $3$ & $1$ & $\textbf{0.686}$ & $4$ & $1$ & $0.647$ & $3$ & $1$ & $2.132$ & $5$ & $1$ & $\textbf{0.566}$\\ %data-10
				A08 & $12$ & $250$ & $3$ & $1$ & $0.799$ & $2$ & $1$ & $0.88$ & $4$ & $1$ & $\textbf{0.603}$ & $8$ & $1$ & $0.563$ & $3$ & $1$ & $0.908$ & $8$ & $1$ & $\textbf{0.429}$\\ %data-12
				A09 & $9$ & $126$ & $2$ & $1$ & $0.636$ & $2$ & $1$ & $0.909$ & $2$ & $1$ & $\textbf{0.575}$ & $4$ & $1$ & $\textbf{0.313}$ & $3$ & $1$ & $1.1$ & $5$ & $1$ & $0.355$\\ %data-16
				A10 & $11$ & $265$ & $2$ & $1$ & $2.193$ & $2$ & $1$ & $2.581$ & $2$ & $1$ & $\textbf{2.118}$ & $3$ & $1$ & $1.325$ & $2$ & $1$ & $4.096$ & $4$ & $1$ & $\textbf{1.284}$\\ %data-28
				A11 & $10$ & $271$ & $3$ & $1$ & $1.221$ & $2$ & $1$ & $1.398$ & $5$ & $1$ & $\textbf{1.117}$ & $6$ & $1$ & $0.936$ & $2$ & $1$ & $4.42$ & $5$ & $1$ & $\textbf{0.726}$\\ %data-15
				\hhline{|---||---------||---------}
      \end{tabular}
    \end{center}
  \end{footnotesize}
 \vspace{-15pt}
\end{table*}

The developers' answers to the user preference questions (D-PN questions in Table~\ref{tab:repro-questions}) suggest that \tech's reports are actionable and easy to understand: ``\textit{Simple step details, ordered and simple}'' (developer D03), ``\textit{list of operation is very easy to understand}'' (D04), ``\textit{I found both the summary of the bug on the left side, with expected vs observed behavior to be useful to provide context. [...] I think the list of steps with screenshot on the right is what makes this type of bug report system particularly effective }'' (D06), ``\textit{I liked that it was clear and with the right amount of information to reproduce the steps. The list of actions/screenshots is the most useful.}'' (D07), and ``\textit{The screenshots were very useful in finding the bug [...]} (D10).

\revision{The free-text answers from the bug reproduction study also provided useful insights for improving \tech and bug reproduction in future work. A comment (from developer D01) suggested that an ``\textit{[i]mage of the final (buggy) state of the app}'' would be a good addition to \tech. Currently, \tech does not focus on improving the bug reporting process of the observed behavior. However, we believe that automated, real-time, and iterative bug reproduction techniques could help users report a screenshot of the observed behavior and believe that this is an interesting direction for future work. A second comment (from D02) mentioned that ``\textit{I think the description field doesn't have relevant information if it's a copy of the steps, but could be useful to hold additional information}''. The description field mentioned by the developer is the description typed by the reporter. The developer found this description redundant as compared to the list of S2Rs validated by \tech. The approach displays the information to the developer in case S2Rs were not suitably validated, and, in this way, the developer could always refer to the description originally provided by the user. We believe that additional user studies could further improve \tech's interface, and these studies could possibly display the user description on an on-demand basis. Finally, another comment (from D04) highlighted that images associated with S2Rs are not always present, and this increases the difficulty in understanding the reported bug: ``\textit{[n]ot always image[s] are present, this increase[s] [the] difficult[y] to understand}''. This situation can appear when \tech is not able to validate an S2R. Devising techniques to improve \tech's GUI model would mitigate this type of situation.}

\revisiontwo{The answers provided in the study offer some preliminary results on the user experience with \tech and reveal that participants had a positive user experience when reproducing bug reports with \tech. Additionally, the results provide initial evidence that this experience was better than the one with \fusion's reports or the original bug reports.}

\subsubsection{RQ6: Prediction Models' Effectiveness}

To answer RQ6, we leveraged the user traces collected for \tech's evaluation and compared \tech's prediction models with prediction models generated using AKOM~\cite{pitkow1999mining} and CPT+~\cite{gueniche2015cptplus}. Specifically, we compared the prediction models of \tech used in the bug reporting study with the best performing models generated using AKOM and CPT+. To compute AKOM and CPT+ prediction models, we follow a similar methodology as the one described in Section~\ref{sec:pred-mod-gen}. For AKOM models, we identified the best performing model for each app by varying the model order and doing leave-one-out (sequence) cross-validation. We varied the sequence prefix length to identify the best performed CPT+ models. \revision{We use the wasted effort score (\textit{wes}) metric to find the best performing models for a specific technique and compare models across approaches. The wasted effort score is an extrinsic evaluation metric and we used this metric as extrinsic evaluations are the most suitable way to evaluate models in sequence prediction tasks~\cite{2009_jurafsky_speech}.}

Table~\ref{tab:prediction-models-performance} reports the performance of the models. For each app, the table reports the best performing models for predicting GUI actions and GUI elements using AKOM, CPT+, and \tech.
Table~\ref{tab:prediction-models-performance} is divided into three sections. The first section reports the identifier (column \textit{App ID}) of the benchmark apps considered in the evaluation, the number of traces (\textit{Traces}) to evaluate the models, and the number of predictions used to evaluate the models (\textit{Predictions}). The GUI action and GUI element models were evaluated using the same number of predictions because when we evaluated GUI element models, we only assessed predictions on the tokens representing GUI elements (see Section~\ref{sec:bug-reporting} for more details). The second section reports the best performing models for making GUI Action suggestions and the third section for making GUI element suggestions. Columns \textit{Order} provide the order of the models, columns \textit{SN} provide the number of suggestions provided by the models for each prediction task, columns \textit{wes} provide the wasted effort score for the models, and columns \textit{PL} provide the sequence prefix length for the CPT+ models. Table~\ref{tab:prediction-models-performance} reports in bold the wasted effort score of the best performing model across all models for a certain app.

Overall, \tech provides the best performing models in $16$ out of the $24$ cases considered (one of the cases is a tie with a model based on AKOM). AKOM offers the best performing model in nine out of the $24$ (including the tie). CPT+ never provides the best performing model. If we considered the wasted effort score of the GUI action prediction models across all apps, \tech's \textit{wes} is $1.127$ and AKOM's \textit{wes} is $1.198$. For GUI element prediction models, \tech's \textit{wes} is $0.846$ and AKOM's \textit{wes} is $0.902$.

\revision{Although the overall difference between AKOM models and \tech's $n$-gram models is not big, the $n$-gram models performed better than the AKOM models both in terms of wasted effort and correctly made predictions. Across all the GUI action prediction tasks, \tech's $n$-gram models produced 1,683 correct predictions and a wasted effort of 1,897. AKOM models led to 1,629 correct predictions and a wasted effort of 1,951. For GUI element prediction tasks, \tech's $n$-gram models produced 1,939 correct predictions and a wasted effort of 1,641. AKOM models led to 1,882 correct predictions and a wasted effort of 1,698. Even if these differences are not big, when considering that the same failure might be reported by multiple users, the differences can become more significant. Considering these results, \tech's $n$-gram models can help reporters by providing more useful predictions while lowering the wasted effort. However, additional studies are also needed to confirm and extend these results and we suggest those as possible future work.}

Based on these results, we can conclude that \tech provides the best performing prediction models in the majority of the cases, and when we consider all models of all apps, \tech has the best wasted effort score as compared with baseline approaches.

\subsection{Threats to Validity}

As it is the case for most empirical evaluations, there are both external and construct threats to validity associated with our results. In the bug reporting study, the participants were CS students. These participants might have a different background compared to other bug reporters. To mitigate this threat to validity, the study included participants from three institutions and at a different level of their studies. \revision{Additionally, past work has illustrated that performance of CS students and professional developers do not differ dramatically in some software engineering tasks~\cite{Salman:ICSE'15}}.
\revision{The results of the bug reporting study might also be affected by the user traces collected in the data collection activity. To mitigate this threat, we invited participants with different backgrounds (some of which did not have a CS background) and asked them to use the benchmark apps as they would do in a normal use of the apps.}
The results of the bug reproduction study might also not generalize to other developers. To mitigate this threat to validity, the study included developers from ten different companies at different career stages.
\revision{In the bug reproduction study, we used $10$ minutes as a time limit for developers to reproduce bug reports. This value could provide a partial view of the total number of reproduced bug reports as more bug reports could be reproduced after the time limit. We used this value to be consistent with related work~\cite{Moran:FSE15}. We also decide to use the value, as, in related work, we also observed that the value had mitigated participant fatigue.}

Another threat to external validity is that our results might not generalize to other failures or apps. In particular, we only considered $20$ failures in our empirical evaluation. This limitation is an artifact of the complexity of our evaluation. To mitigate this threat, we used randomly selected real-world failures of varying types and complexity. These failures belong to real-world apps from different categories and functions. These apps include apps such as \textsc{GnuCash}, \textsc{Mileage}, and \textsc{Doc Viewer}, which have complex functionality, hundreds of widgets, and hundreds of thousands of users. We believe that, given the complexity of the failures and apps considered in the studies, \tech should also apply to other types of failures and apps.

Finally, in terms of construct validity, there might be errors in the implementation of our approach or our experimental infrastructure. To mitigate this threat, we extensively inspected the results of the evaluation manually.

\section{Discussion}
\label{sec:discussion}

\revision{
Through our empirical evaluation, we found that \tech is more efficient and effective than the baseline. Based on the evaluation results and the feedback received from the participants of the bug reporting study, we believe that the results can be attributed to the S2R suggestions and the validation operations performed by \tech. For one of the failures considered, bug reporting with \tech was not more efficient. In this case, one of the bug reporters rewrote four S2Rs three times. We believe that this situation appeared because \tech could not validate a subsequent S2R in the report. This result reveals that improving the accuracy of the GUI models generated by \tech, which are used to validate the S2Rs, can further improve \tech's efficiency. Developers using \tech's bug reports were generally more efficient in reproducing bug reports as compared to developers using the baseline's bug reports. Developers using bug reports submitted with \tech were slower only in the case of one specific failure. To report this failure, the reporters had to provide a value from a dropdown list that contained a large number of values, and the reporters decided to use free text instead of inspecting a list of options provided by \tech. The description provided by both users was not complete, and we believe that this aspect might have caused the higher reproduction time with \tech. This result shows a tradeoff between flexibility in the bug reporting process and efficiency in bug reproduction time. We believe that this tradeoff is an interesting aspect related to bug reporting and should be the target of future user studies.

During our evaluation, we focused on examining how useful \tech is for reporting bugs, however, it is also important to comment on the general scalability and applicability of our technique. Given that a majority of the ``heavy'' processing required by \tech is performed \textit{a-priori}, that is before the reporting process commences, and the ``online'' portions of \tech generally scale well, we expect \tech to scale reasonably well for a variety of complex applications. The one potential limitation with scaling may be with \tech's GUI-component matching procedure. That is, if there is a \textit{very} large number of components present within an app, searching for the corresponding component may introduce a short delay in the reporting interface. However, while we cannot confirm \tech's performance outside the context of our study, we speculate that, based on the authors' experience of the general size of most Android applications, \tech would generally scale to a majority of apps available today. 

Another important point of investigation to understand the reasons for \tech general performance in our empirical evaluation is \textit{the extent to which static and dynamic GUI models contribute to the accuracy of the suggestion engine.} To identify the contributions of the static and dynamic GUI models (which together form the GUI model used by \tech), we analyzed the S2Rs that were entered by the participants of the bug reporting study and that were also validated by \tech. As reported in Table \ref{tab:study-data}, the technique validated 315 S2Rs across all apps. One of those S2Rs had the corresponding GUI action that was present only in the static GUI model, 140 S2Rs had corresponding GUI actions appearing in both the static and the dynamic GUI model, and 174 S2Rs had corresponding GUI actions present only in the dynamic GUI model. This result highlights that the dynamic GUI is of primary importance for \tech. If we look at all the GUI actions (1,255) encoded in the GUI models of all the apps considered, 155 GUI actions are present only in the static GUI models, 274 appear in both the static and the dynamic GUI models, and 826 are present only in the dynamic GUI models. Although a large number of GUI actions are part of the dynamic GUI models, the number of GUI actions provided by the static GUI model is not negligible, making them potentially useful for validating S2Rs.

The predictions accepted during bug reporting are another important aspect of \tech. In our study we found that participants accepted a large number of both GUI action and GUI element suggestions. The wasted effort score during bug reporting was higher for GUI action predictions and lower for GUI element predictions as compared to when models were built (\ie when evaluating the models on user traces). Although the wasted effort scores associated with GUI action prediction models are higher during bug reporting, the number of accepted suggestions, the study participant's feedback, and the overall lower bug reporting time of \tech suggests that the predictions can be helpful during bug reporting. We believe that performing additional studies on wasted effort during bug reporting is an interesting venue for future work.

Another important aspect of \tech's practical applicability is concerned with the collection of the app usage models. One of the key facets of \tech is the utilization of ``real-world'' usage traces of a given application to help prioritize potential predicted S2Rs. There are several ways in which such traces could be collected. For example, as developers release beta versions of an application to crowd-testers, traces could be collected from crowd-testers~\cite{Mao:ASE'17}. 
Additionally, traces could be collected from developer written test cases, or even usages collected by developers themselves during the testing process of each version of their app. \tech uses traces to gather an understanding of commonly performed GUI action sequences and provides suggestions when those GUI action sequences are part of a sequence that reveals a bug. For this reason, \tech does \textit{\textbf{not}} need to collect traces that expose bugs to provide suggestions.

There are many potential future directions of work for improving \tech. For instance, currently our execution traces are tied to a specific application, meaning that data collection must be done on a per-app basis. However, recent work from the Ubicomp community has illustrated the benefits of modeling user behaviors across applications~\cite{Chen:IMWUT19}. This suggests that future work could look to learn a generalized abstract usage model across many applications that is supplemented with a smaller amount of app-specific data from a diverse set of users. This may help to improve the need of collecting usage data for \tech. We also believe future work could examine different types of reporting interfaces, such as on-device reporting or chatbot-based systems, given the success of prior programming tools that use these interfaces~\cite{Ko2008}. Finally, given the additional information that is captured via \tech, we believe future work could examine potential benefits to downstream bug reporting tasks such as bug triaging or localization.}

\section{Limitations}
\label{sec:limitations}

\revision{\tech computes a GUI model of the relevant app in its models generation phase. GUI model generation is based on static and dynamic analysis. In the evaluation, we used a timeout of 60 minutes for the dynamic analysis. Although the dynamic analysis might take longer if the analysis is run exhaustively (or a higher timeout is used), our evaluation indicates that even with a timeout of 60 minutes \tech can achieve good results. Furthermore, the analysis needs to be run only once for every app release. For this reason, we believe that the cost associated with this part of \tech should have a limited impact on \tech. The analyses performed by \tech might not fully capture all the transitions and screens in the relevant app. \tech's evaluation results give us confidence that generated models are effective, but this aspect of the technique could be improved in future work by leveraging ongoing research on GUI model abstraction and representation (\eg~\cite{2019_ase_duling_goal}).}

Our current prototype tool, in its static analysis part of the GUI model generation phase, does not currently create different screens for the different fragments in an activity of a certain app. This characteristic can lead the prototype tool to create an overapproximated GUI model of the app. \revision{The GUI model is overapproximated as it might contain transitions that do not actually exist in the app due to overapproximations (\eg context-insensitive analysis) made in the static analysis used by \tech.} Although this aspect might lead to false positives in the S2Rs validation process, we did not encounter significant challenges related to this aspect in the studies performed to evaluate \tech.
%Additionally, it is our plan for future work to support fragments in the GUI model in the static analysis of the GUI model generation phase.

\revision{The main assumption for \tech to work effectively is that reported S2Rs describe GUI actions in the app. The description of a S2R needs to be provided through text but does not need to follow an imposed structure. \tech is able to process S2R descriptions having different structures by leveraging the dependency trees associated with the descriptions~\cite{2009_jurafsky_speech,2006_lrec_marneffe_generating}. To take full advantage of \tech, the text of a S2R should describe the components of the corresponding GUI action, that is, the type of action, the GUI element affected by the action, and the properties of the action (when applicable). \tech is also able to operate even if one or more S2R descriptions do not provide the desired information. Although the technique will not be able to map those GUI descriptions into GUI actions, it can map subsequent descriptions leveraging the GUI model and feedback from the user (\ie the user can correct the provided mapping ). This part of \tech also allows the technique to support incomplete or missing S2R descriptions.}

\revision{Our approach does not currently offer support for actions based on multi-touch gestures (e.g., pinch in). If an app uses that type of action, \tech would not be able to validate or provide suggestions for that type of action but would still be able to support other actions in the app. We are currently investigating ways to support that type of action, and, based on our experience, that type is not the predominant one in a large number of apps.}

\revision{
Our approach binds app interactions to GUI elements. Certain Android apps, however, rely on bitmapped (rather than GUI) elements. Hence, the approach cannot currently handle such apps. Luckily, the vast majority of these apps are games, whereas other types of apps tend to rely on standard GUI elements. We also believe that those apps could be supported by identifying interactable elements using approaches based on computer vision techniques (\eg edge detection).
}

Finally, our predictive models require training data. Although it might not be possible to collect such data from users in some cases, we believe that developers can mitigate this problem by performing this activity in-house as described also in Sec.~\ref{sec:discussion}.
\vspace{-5pt}
\section{Related Work}
\label{sec:rel-work}

\subsection{Automation for Bug Reporting Systems}

\textsc{Fusion} is a bug reporting system developed by Moran \etal~\cite{Moran:FSE15} that has the goal of facilitating the reporting process by providing pre-populated structured fields that reporters can use to compose the reproduction steps of their reports. While this approach led to the construction of bug reports that were more reproducible than a baseline issue tracking system, it came at the cost of longer reporting times. Our work in developing the \tech approach differs in two important aspects. First, we are the first to explore how \textit{predictive modeling} can enhance the bug reporting process by proactively providing suggestions to users about yet-to-be-completed information. Second, our interface blends a predictive free-form text field in conjunction with visual information (\eg screenshots), whereas \textsc{Fusion} used more structured menus. We compared \tech against \textsc{Fusion} in our comprehensive evaluation and found that reporters could create bug reports faster with \tech and these reports were more reproducible in far less time, illustrating the advantages of \tech.

Other approaches from related work focus on the analysis of bug reports after they are created.  Specifically these techniques analyze completed bug reports and (i) detect missing information~\cite{Chaparro:FSE'17}, (ii) assess the quality of the reproduction steps~\cite{Chaparro:FSE'19}, and (iii) automatically reproduce bug reports~\cite{2018_issta_fazzini_automatically,2019_icse_zhao_recdroid, li2020automated}. While such research is related to ours, we view our proposed approach as \textit{complimentary}, as it aims to increase bug report quality \textit{at the time of reporting} which in turn, can help to improve the performance of the techniques in this area of research by providing higher quality bug information to analyze. \revision{Addtionally, automated reproduction techniques of bug reports~\cite{2018_issta_fazzini_automatically,2019_icse_zhao_recdroid} and app reviews~\cite{li2020automated} could be integrated with bug reporting systems like \tech so that the bug reporting process would have an iterative feedback loop in which the results of automated bug reproduction are provided to the users as they are reporting bugs to help them refine the reports.}

\revision{Finally, a wealth of work has been conducted on automating downstream SE tasks that are tied to bug reporting, such as (i) triaging, (ii) fault localization, and (iii) duplicate detectio. We now provide a summary of the work in this area of research but refer the reader to surveys on this line of research for a more thorough discussion~\cite{zhang2015survey,2016_tse_wong,uddin2017surveynew}. Related work on triaging improved assignment of bug reports to developers~\cite{7Shokripour:MSR13,10Naguib:MSR13,45Park:AAAI2011, Linares-Vasquez:ICSM2012}, reduced bug report tossing~\cite{40Jeong:FSE2009}, and prioritized bug reports~\cite{51Haihao:ICST2011}. Work on bug reports and fault localization used information retrieval to identify files that likely need fixing~\cite{8Zhou:ICSE12}, leveraged multi-step recommendation models based on information retrieval~\cite{44Kim:TOSE2013}, used version control history information to identify relevant buggy files~\cite{13Wang:ICPC14}, utilized stack traces to identify files leading to crashing faults~\cite{42Wu:ISSTA2014,moreno2014use}. Related work on duplicate bug report detection used information retrieval~\cite{14Nguyen:ASE12}, leveraged natural language processing and execution information~\cite{17Wang:ICSE08}, and adopted a learning to rank approach~\cite{21Zhou:CIKM12}. We view \tech as complimentary to the techniques in this line of research, as its main goal is to improve the quality of bug reports \textit{before} they are submitted, which can ultimately benefit all of the downstream approaches by providing reports of higher quality.}

\subsection{Bug Reporting Studies}

A large body of work focused on understanding the various aspects of the bug reporting processes, and we draw inspiration from several of the findings of this research area. Bettenburg \etal~\cite{3Bettenburg:FSE08} performed a study with developers of open source software systems and found that (i) steps to reproduce, (ii) stack traces, and (iii) test cases were the most important pieces of information to be included within a bug report. However, these pieces of information were also the most difficult for reporters to provide. This helps to illustrate the importance of \tech's real-time understanding and auto-completion of S2Rs as these features can help reporters in quickly writing effective S2Rs.  
Bettenburg \etal~\cite{32Bettenburg:ICSM08} also examined developer's perceptions of duplicate bug reports, and found that many developers did not necessarily view them as harmful, but rather, as providing important information. While \tech does not focus on duplicate reports, we hope that the accurate reporting mechanism provided by our approach could aid future automated analysis tools that aim to find related or duplicate bug reports for developers. The same authors also proposed a tool that is capable of extracting structural information from bug reports~\cite{11Bettenburg:MSR08}. Similarly, Song \& Chaparro studied and proposed techniques for automatically identifying different types of lexical information from bug reports~\cite{Chaparro:FSE'17,Song:FSE20}. \tech is complementary to this line of research as the approach introduces a new form of real-time analysis of S2Rs that can identify various components in the S2Rs.

\subsection{Field Failure Reproduction}

The research area of field failure reproduction shares similar goals with our work, and with bug reporting systems in general, in that they attempt to capture and relay fault information to developers.
\revision{Related work on field failure reproduction proposed techniques to reproduce failures using genetic programming~\cite{Soltani:ICSE'17,Kifetew:ICST'14}, crowdsourcing~\cite{Gomez:MobileSoft'16}, symbolic execution~\cite{Chen:TSE'15,Wei:TOSEM'15,Wei:ICSE'12,Cao:ASE'14}, static analysis combined with symbolic execution~\cite{Zamfir:EuroSys'10}, model checking~\cite{Nayrolles:JSS'17}, dynamic exploration~\cite{White:ICPC15}, sequential pattern minin on user traces~\cite{Roehm:SANER'15}, and test case mutaiton~\cite{Xuan:FSE'15}.}

However, we view \tech, and other user-facing bug reporting systems as largely complimentary to these field failure reproduction approaches. This is mainly due to the fact that in-field techniques require a known oracle, such as a crash, to identify that a failure has occurred. Conversely, user-facing bug reporting systems can be used for failures discovered by end-users or other developers. These bugs have been shown to account for more than half of the reported bugs in open source systems~\cite{Tan:EMSE'14}. 

\revision{\subsection{User Support Systems and Interactive Debugging}}

\revision{Previous work in the SE research community has focused on creating tools that emulate a chatbot-like interface for debugging, question answering, and user support. For instance, the work by Ko \etal focused on helping programmers ask questions about program behavior~\cite{Ko2008}. Parmit \etal introduced the LemonAid system that provided contextual assistance on websites through crowdsourcing techniques. There are also a number of commercial tools such as BugClipper~\cite{bugclipper} or BugSee~\cite{bugsee} provide features for screen recording, and bug report annotation. In contrast to these tools, \tech offers proactive suggestions to help speed up and improve the quality of information included in the bug reporting process. While \tech does not make use of a chatbot interface, we believe this is an interesting direction of future work to explore for the domain of bug reporting.}

\subsection{Automated GUI Exploration for Mobile Apps}

\revision{The modeling techniques employed by \tech share certain similarities with automated GUI exploration techniques for mobile apps. Automated input generation approaches for mobile apps perform \textit{random-based} input generation \cite{android-monkey,Machiry:FSE13,Sasnauskas:WODA14,Ravindranath:Mobisys2014}, do \textit{systematic} input generation \cite{Amalfitano:ASE12,Anand:FSE12,Azim:OOPSLA13,Moran:ICST16,google-robo-test}, employ \textit{model-based} input generation \cite{Amalfitano:IEEE14,Azim:OOPSLA13,Choi:OOPSLA13,Zaeem:ICST2014,Yang:FASE13,Linares:MSR15,Zhang:ICSE17}, use a \textit{search-based} approach~\cite{Mahmood:FSE14,Mao:ISSTA16}, and perform \textit{symbolic} input generation \cite{Jensen:ISSTA13, Mirzaei:ISSRE15}. Related work, also investigated how to integrate user feedback to improve mobile app testing~\cite{grano2018exploring,pelloni2018becloma}.}

The two most highly related techniques to those proposed in \tech are the MonkeyLab approach proposed by Linares-Vasquez \etal~\cite{Linares:MSR15}, and the Polariz approach proposed by Mao \etal~\cite{Mao:2017}. Both of these approaches attempt to incorporate app usages gathered from crowd workers in order to generate a more diverse set of actions to guide automated input generation. \tech's input generation approach and prediction modeling function in a similar manner, but with a markedly different end goal: aid users in predicting reproduction steps during the bug reporting process. As such, \tech's approach uses an n-gram model to predict likely reproduction steps and target GUI elements for reporters, rather than using this information to guide an input generation technique.

%By combining both the dynamically and statically extracted GUI-states, \tech is able to build a comprehensive GUI model while using the user information to determine how often various states are visited. 

%\KEVIN{We could also talk about bug reporting studies and in-field failure reproduction, but I am not sure that it is necessary.} 

%\KEVIN{Here, talk about all of the automated exploration techniques for Android, take information from previous survey paper. Our main differentiation here is that none of these prior techniques has combined crowdsourced usage models with app models to predict reproduction steps in bug reports.}

%\subsection{GUI-based Analysis of Mobile Apps}

%There is an emerging body of research that seeks to automatically the GUIs of mobile applications in order to support a variety of SE tasks. Such tasks include mobile app testing \cite{Jones:2014,Moran:ICST16,Hu:FSE18,Bernal-Cardenas:ICSE'20}, developer/user behavior modeling \cite{Caetano:02,Bao:ICSE15,Frisson:CHI16}, GUI reverse engineering and code generation \cite{Dixon:11,Nguyen:ASE15,Beltramelli:EICS18,Chen:ICSE18,Moran:TSE18,Chen:FSE20}, analysis of programming videos \cite{MacLeod:ICPC'15,Lasecki:ACM'15,Yadid:2016,Ponzanelliz:TSE'19,Alahmadi:EMSE20,Zhao:ICSE19}, and GUI understanding and verification \cite{Chang:ACM'11,Zhao:ICSE20}. None of these works deal with finding duplicate video-based bug reports, which is our focus.

\section{Conclusion}
\label{sec:conclusion}

When users experience a software failure, they have the option of submitting a bug report to provide information about the failure. Developers rely on submitted bug reports to identify and remedy the faults in their software. Unfortunately, the quality of manually constructed bug reports can vary widely due to the effort required to include essential information, such as a detailed description of the S2Rs. This issue affects and complicates the developers' task of reproducing the failures experienced by the users. To help users in writing bug reports that are easier to reproduce, we propose the \tech bug reporting approach. \tech understands S2Rs written in natural language and, as the user reports them, uses a novel predictive model to suggest additional S2Rs for helping the user complete the bug report. We implemented \tech to support bug reporting in the context of Android apps and empirically evaluated it in two user studies. The studies involved ten participants who submitted ten bug reports each, and ten developers who reproduced the submitted bug reports. Our results show that reporters were able to write bug reports 31\% faster with \tech as compared to a state-of-the-art bug reporting system used as a baseline. The results also show that developers could reproduce more failures using \tech's reports than using those generated using the baseline.

We foresee a number of venues for future work. First, we will investigate ways to help users in defining observed and expected behavior. Second, we will study how to automatically extract more domain knowledge from the relevant app so that \tech can understand and provide suggestions for macro-S2Rs (\ie S2Rs that should be interpreted as sequences of multiple actions on the GUI). Third, we will evaluate the application of the approach in the context of web and GUI-based desktop software. \revision{Fourth, we also think that \tech could be extended by evaluating it in additional studies. Specifically, \tech could be evaluated in studies where developers use \tech to report bugs and studies where \tech is used by the developers of the relevant apps upon receiving newly identified bugs. Fifth, future studies could also analyze the tradeoffs between closed and open vocabularies in \tech's prediction models.}  Finally, and more on the engineering side, we will extend \tech to model fragments as independent screens in the relevant app and add support for multi-touch gestures, which will allow us to handle reports involving these kinds of interactions.

\vspace{-5pt}

\section*{Acknowledgment}
W\&M and GMU co-authors have been supported in part by the NSF CCF-1955853 and CCF-2007246 grants. Any opinions, findings, and conclusions expressed herein are the authors' and do not necessarily reflect those of the sponsors.

\begin{comment}
\appendices
\section{Proof of the First Zonklar Equation}
Appendix one text goes here.

\section{}
Appendix two text goes here.

\ifCLASSOPTIONcompsoc
  \section*{Acknowledgments}
\else
  \section*{Acknowledgment}
\fi

The authors would like to thank...
\end{comment}

\ifCLASSOPTIONcaptionsoff
  \newpage
\fi



%\bibliographystyle{IEEEtran}
\bibliographystyle{IEEEtran}
\bibliography{IEEEabrv,main}

\begin{thebibliography}{100}
\providecommand{\url}[1]{#1}
\csname url@samestyle\endcsname
\providecommand{\newblock}{\relax}
\providecommand{\bibinfo}[2]{#2}
\providecommand{\BIBentrySTDinterwordspacing}{\spaceskip=0pt\relax}
\providecommand{\BIBentryALTinterwordstretchfactor}{4}
\providecommand{\BIBentryALTinterwordspacing}{\spaceskip=\fontdimen2\font plus
\BIBentryALTinterwordstretchfactor\fontdimen3\font minus
  \fontdimen4\font\relax}
\providecommand{\BIBforeignlanguage}[2]{{%
\expandafter\ifx\csname l@#1\endcsname\relax
\typeout{** WARNING: IEEEtran.bst: No hyphenation pattern has been}%
\typeout{** loaded for the language `#1'. Using the pattern for}%
\typeout{** the default language instead.}%
\else
\language=\csname l@#1\endcsname
\fi
#2}}
\providecommand{\BIBdecl}{\relax}
\BIBdecl

\bibitem{25Tassey:NIST}
G.~Tassey, ``The economic impacts of inadequate infrastructure for software
  testing,'' National Institute of Standards and Technology, Tech. Rep., 2002.

\bibitem{Tan:EMSE'14}
L.~Tan, C.~Liu, Z.~Li, X.~Wang, Y.~Zhou, and C.~Zhai, ``Bug characteristics in
  open source software,'' \emph{Empirical Software Engineering}, vol.~19,
  no.~6, pp. 1665--1705, 2014.

\bibitem{3Bettenburg:FSE08}
N.~Bettenburg, S.~Just, A.~Schr\"{o}ter, C.~Weiss, R.~Premraj, and
  T.~Zimmermann, ``What makes a good bug report?'' in \emph{Proceedings of the
  16th ACM SIGSOFT International Symposium on Foundations of Software
  Engineering}.\hskip 1em plus 0.5em minus 0.4em\relax New York, NY, USA: ACM,
  2008.

\bibitem{Moran:FSE15}
K.~Moran, M.~Linares-V\'{a}squez, C.~Bernal-C\'{a}rdenas, and D.~Poshyvanyk,
  ``Auto-completing bug reports for android applications,'' in
  \emph{Proceedings of the 2015 10th Joint Meeting on Foundations of Software
  Engineering}.\hskip 1em plus 0.5em minus 0.4em\relax New York, NY, USA: ACM,
  2015.

\bibitem{smart-compose}
\BIBentryALTinterwordspacing
(2021) Google smart compose. [Online]. Available:
  \url{https://www.blog.google/products/gmail/subject-write-emails-faster-smart-compose-gmail}
\BIBentrySTDinterwordspacing

\bibitem{appendix}
\BIBentryALTinterwordspacing
M.~Fazzini, K.~Moran, C.~Bernal-Cardenas, T.~Wendland, A.~Orso, and
  D.~Poshyvanyk. (2021) \tech's online appendix. [Online]. Available:
  \url{https://www-users.cs.umn.edu/~mfazzini/ebug.html}
\BIBentrySTDinterwordspacing

\bibitem{2020_github_deposit}
\BIBentryALTinterwordspacing
(2021) Deposit/withdrawal change existing entry. [Online]. Available:
  \url{https://github.com/codinguser/gnucash-android/issues/247}
\BIBentrySTDinterwordspacing

\bibitem{2020_github_gnucash}
\BIBentryALTinterwordspacing
(2021) Gnucash github. [Online]. Available:
  \url{https://github.com/codinguser/gnucash-android}
\BIBentrySTDinterwordspacing

\bibitem{2020_googleplay_gnucash}
\BIBentryALTinterwordspacing
(2021) Gnucash. [Online]. Available:
  \url{https://play.google.com/store/apps/details?id=org.gnucash.android}
\BIBentrySTDinterwordspacing

\bibitem{Choi:OOPSLA13}
W.~Choi, G.~Necula, and K.~Sen, ``Guided gui testing of android apps with
  minimal restart and approximate learning,'' in \emph{Proceedings of the 2013
  ACM SIGPLAN International Conference on Object Oriented Programming Systems
  Languages \& Applications}.\hskip 1em plus 0.5em minus 0.4em\relax New York,
  NY, USA: ACM, 2013, pp. 623--640.

\bibitem{2018_issta_fazzini_automatically}
M.~Fazzini, M.~Prammer, M.~d'Amorim, and A.~Orso, ``Automatically translating
  bug reports into test cases for mobile apps,'' in \emph{Proceedings of the
  27th ACM SIGSOFT International Symposium on Software Testing and
  Analysis}.\hskip 1em plus 0.5em minus 0.4em\relax New York, NY, USA: ACM,
  2018, pp. 141--152.

\bibitem{2015_ase_yang_static}
S.~{Yang}, H.~{Zhang}, H.~{Wu}, Y.~{Wang}, D.~{Yan}, and A.~{Rountev}, ``Static
  window transition graphs for android (t),'' in \emph{Proceedings of the 2015
  30th IEEE/ACM International Conference on Automated Software
  Engineering}.\hskip 1em plus 0.5em minus 0.4em\relax Piscataway, NJ, USA:
  IEEE Press, 2015.

\bibitem{2020_ast_wnwarang_testing}
T.~Wanwarang, N.~P. Borges, L.~Bettscheider, and A.~Zeller, ``Testing apps with
  real-world inputs,'' in \emph{Proceedings of the IEEE/ACM 1st International
  Conference on Automation of Software Test}, ser. AST '20.\hskip 1em plus
  0.5em minus 0.4em\relax New York, NY, USA: Association for Computing
  Machinery, 2020, p. 1–10.

\bibitem{2020_github_tesseract}
\BIBentryALTinterwordspacing
(2021) Tesseract ocr. [Online]. Available:
  \url{https://github.com/tesseract-ocr/tesseract}
\BIBentrySTDinterwordspacing

\bibitem{1969_harary_graph}
F.~Harary, \emph{Graph Theory}.\hskip 1em plus 0.5em minus 0.4em\relax USA:
  Avalon Publishing, 1969.

\bibitem{2009_jurafsky_speech}
D.~Jurafsky and J.~H. Martin, \emph{Speech and Language Processing (2nd
  Edition)}.\hskip 1em plus 0.5em minus 0.4em\relax USA: Prentice-Hall, Inc.,
  2009.

\bibitem{1998_chen_tr_an}
S.~F. Chen and J.~Goodman, ``An empirical study of smoothing techniques for
  language modeling,'' Computer Science Group, Harvard University, Tech. Rep.
  TR-10-98, 1998.

\bibitem{2016_tse_wong}
W.~E. {Wong}, R.~{Gao}, Y.~{Li}, R.~{Abreu}, and F.~{Wotawa}, ``A survey on
  software fault localization,'' \emph{IEEE Transactions on Software
  Engineering}, vol.~42, no.~8, pp. 707--740, 2016.

\bibitem{2012_miner_practical}
G.~Miner, J.~Elder, T.~Hill, R.~Nisbet, D.~Delen, and A.~Fast, \emph{Practical
  Text Mining and Statistical Analysis for Non-structured Text Data
  Applications}.\hskip 1em plus 0.5em minus 0.4em\relax Orlando, FL, USA:
  Academic Press, 2012.

\bibitem{2018_saito_medium_case}
\BIBentryALTinterwordspacing
J.~Saito. (2016, Jul.) Making a case for letter case. [Online]. Available:
  \url{https://medium.com/@jsaito/making-a-case-for-letter-case-19d09f653c98}
\BIBentrySTDinterwordspacing

\bibitem{angeli2015leveraging}
G.~Angeli, M.~J.~J. Premkumar, and C.~D. Manning, ``Leveraging linguistic
  structure for open domain information extraction,'' in \emph{Proceedings of
  the 53rd Annual Meeting of the Association for Computational Linguistics and
  the 7th International Joint Conference on Natural Language Processing}.\hskip
  1em plus 0.5em minus 0.4em\relax Stroudsburg, PA, USA: The Association for
  Computer Linguistics, 2015, pp. 344--354.

\bibitem{2006_lrec_marneffe_generating}
M.~de~Marneffe, B.~MacCartney, and C.~D. Manning, ``Generating typed dependency
  parses from phrase structure parses,'' in \emph{Proceedings of the Fifth
  International Conference on Language Resources and Evaluation}.\hskip 1em
  plus 0.5em minus 0.4em\relax Bern, Switzerland: European Language Resources
  Association {(ELRA)}, 2006, pp. 449--454.

\bibitem{2019_icse_zhao_recdroid}
Y.~Zhao, T.~Yu, T.~Su, Y.~Liu, W.~Zheng, J.~Zhang, and W.~G.~J. Halfond,
  ``Recdroid: Automatically reproducing android application crashes from bug
  reports,'' in \emph{Proceedings of the 41st International Conference on
  Software Engineering}.\hskip 1em plus 0.5em minus 0.4em\relax Piscataway, NJ,
  USA: IEEE Press, 2019, p. 128–139.

\bibitem{levenshtein1966binary}
V.~I. Levenshtein, ``Binary codes capable of correcting deletions, insertions,
  and reversals,'' in \emph{Soviet physics doklady}, 1966.

\bibitem{bojanowski2017enriching}
P.~Bojanowski, E.~Grave, A.~Joulin, and T.~Mikolov, ``Enriching word vectors
  with subword information,'' \emph{Transactions of the Association for
  Computational Linguistics}, pp. 135--146, 2017.

\bibitem{mikolov2013efficient}
T.~Mikolov, K.~Chen, G.~Corrado, and J.~Dean, ``Efficient estimation of word
  representations in vector space,'' 2013.

\bibitem{mikolov2013distributed}
T.~Mikolov, I.~Sutskever, K.~Chen, G.~Corrado, and J.~Dean, ``Distributed
  representations of words and phrases and their compositionality,'' in
  \emph{Proceedings of the 26th International Conference on Neural Information
  Processing Systems}.\hskip 1em plus 0.5em minus 0.4em\relax Red Hook, NY,
  USA: Curran Associates Inc., 2013, p. 3111–3119.

\bibitem{luong2013better}
M.-T. Luong, R.~Socher, and C.~D. Manning, ``Better word representations with
  recursive neural networks for morphology,'' in \emph{Proceedings of the
  Seventeenth Conference on Computational Natural Language Learning}, 2013, pp.
  104--113.

\bibitem{qiu2014co}
S.~Qiu, Q.~Cui, J.~Bian, B.~Gao, and T.-Y. Liu, ``Co-learning of word
  representations and morpheme representations,'' in \emph{Proceedings of the
  25th International Conference on Computational Linguistics: Technical
  Papers}, 2014, pp. 141--150.

\bibitem{soricut2015unsupervised}
R.~Soricut and F.~J. Och, ``Unsupervised morphology induction using word
  embeddings,'' in \emph{Proceedings of the 2015 Conference of the North
  American Chapter of the Association for Computational Linguistics: Human
  Language Technologies}, 2015, pp. 1627--1637.

\bibitem{ostrander2012android}
J.~Ostrander, \emph{Android UI Fundamentals: Develop and Design}.\hskip 1em
  plus 0.5em minus 0.4em\relax Berkeley, CA, USA: Peachpit Press, 2012.

\bibitem{2021_google_uiautomator}
\BIBentryALTinterwordspacing
(2021) {UI} automator. [Online]. Available:
  \url{https://developer.android.com/training/testing/ui-automator}
\BIBentrySTDinterwordspacing

\bibitem{2021_neo4j_neo4j}
\BIBentryALTinterwordspacing
(2021) Neo4j. [Online]. Available: \url{https://neo4j.com}
\BIBentrySTDinterwordspacing

\bibitem{2021_google_getevent}
\BIBentryALTinterwordspacing
(2021) Getevent. [Online]. Available:
  \url{https://source.android.com/devices/input/getevent}
\BIBentrySTDinterwordspacing

\bibitem{hsu2008iterative}
B.-J. Hsu and J.~Glass, ``Iterative language model estimation: efficient data
  structure \& algorithms,'' in \emph{Ninth Annual Conference of the
  International Speech Communication Association}, 2008.

\bibitem{Linares:MSR15}
M.~Linares-V\'{a}squez, M.~White, C.~Bernal-C\'{a}rdenas, K.~Moran, and
  D.~Poshyvanyk, ``Mining android app usages for generating actionable
  gui-based execution scenarios,'' in \emph{Proceedings of the 12th Working
  Conference on Mining Software Repositories}.\hskip 1em plus 0.5em minus
  0.4em\relax Piscataway, NJ, USA: IEEE Press, 2015, p. 111–122.

\bibitem{2021_google_natural}
\BIBentryALTinterwordspacing
(2021) Google natural language. [Online]. Available:
  \url{https://cloud.google.com/natural-language}
\BIBentrySTDinterwordspacing

\bibitem{2021_quilljs_quill}
\BIBentryALTinterwordspacing
(2021) Quill. [Online]. Available: \url{https://quilljs.com}
\BIBentrySTDinterwordspacing

\bibitem{fastTextDataset}
\BIBentryALTinterwordspacing
(2021) English word vectors. [Online]. Available:
  \url{https://fasttext.cc/docs/en/english-vectors.html}
\BIBentrySTDinterwordspacing

\bibitem{pitkow1999mining}
J.~Pitkow and P.~Pirolli, ``Mining longest repeating subsequences to predict
  world wide web surfing,'' in \emph{Proceedings of the 2nd Conference on
  USENIX Symposium on Internet Technologies and Systems}.\hskip 1em plus 0.5em
  minus 0.4em\relax USA: USENIX Association, 1999, p.~13.

\bibitem{gueniche2015cptplus}
T.~Gueniche, P.~Fournier-Viger, R.~Raman, and V.~S. Tseng, ``Cpt+: Decreasing
  the time/space complexity of the compact prediction tree,'' in
  \emph{Proceedings of the 19th Pacific-Asia Conference on Knowledge Discovery
  and Data Mining}.\hskip 1em plus 0.5em minus 0.4em\relax Berlin, Germany:
  Springer, 2015.

\bibitem{Chaparro:FSE'19}
O.~Chaparro, C.~Bernal-C\'{a}rdenas, J.~Lu, K.~Moran, A.~Marcus, M.~Di~Penta,
  D.~Poshyvanyk, and V.~Ng, ``Assessing the quality of the steps to reproduce
  in bug reports,'' in \emph{Proceedings of the 2019 27th ACM Joint Meeting on
  European Software Engineering Conference and Symposium on the Foundations of
  Software Engineering}.\hskip 1em plus 0.5em minus 0.4em\relax New York, NY,
  USA: ACM, 2019, p. 86–96.

\bibitem{2020_googleplay}
\BIBentryALTinterwordspacing
(2021) Gnucash. [Online]. Available: \url{https://play.google.com}
\BIBentrySTDinterwordspacing

\bibitem{Brooke:96}
J.~Brooke, ``{SUS}: A quick and dirty usability scale,'' in \emph{Usability
  evaluation in industry}.\hskip 1em plus 0.5em minus 0.4em\relax London:
  Taylor and Francis, 1996.

\bibitem{Morville:04}
\BIBentryALTinterwordspacing
P.~Morville. (2021) User experience design. [Online]. Available:
  \url{http://semanticstudios.com/user_experience_design}
\BIBentrySTDinterwordspacing

\bibitem{Salman:ICSE'15}
I.~Salman, A.~T. Misirli, and N.~Juristo, ``Are students representatives of
  professionals in software engineering experiments?'' in \emph{Proceedings of
  the 37th International Conference on Software Engineering - Volume 1}, ser.
  ICSE '15.\hskip 1em plus 0.5em minus 0.4em\relax IEEE Press, 2015, p.
  666–676.

\bibitem{Mao:ASE'17}
K.~Mao, M.~Harman, and Y.~Jia, ``Crowd intelligence enhances automated mobile
  testing,'' in \emph{Proceedings of the 32nd IEEE/ACM International Conference
  on Automated Software Engineering}, ser. ASE 2017.\hskip 1em plus 0.5em minus
  0.4em\relax IEEE Press, 2017, p. 16–26.

\bibitem{Chen:IMWUT19}
\BIBentryALTinterwordspacing
X.~Chen, Y.~Wang, J.~He, S.~Pan, Y.~Li, and P.~Zhang, ``Cap: Context-aware app
  usage prediction with heterogeneous graph embedding,'' \emph{Proc. ACM
  Interact. Mob. Wearable Ubiquitous Technol.}, vol.~3, no.~1, mar 2019.
  [Online]. Available: \url{https://doi.org/10.1145/3314391}
\BIBentrySTDinterwordspacing

\bibitem{Ko2008}
A.~J. Ko and B.~A. Myers, ``Debugging reinvented: Asking and answering why and
  why not questions about program behavior,'' in \emph{Proceedings of the 30th
  International Conference on Software Engineering (ICSE'08)}, 2008, p.
  301–310.

\bibitem{2019_ase_duling_goal}
D.~Lai and J.~Rubin, ``Goal-driven exploration for android applications,'' in
  \emph{2019 34th IEEE/ACM International Conference on Automated Software
  Engineering (ASE)}, 2019, pp. 115--127.

\bibitem{Chaparro:FSE'17}
O.~Chaparro, J.~Lu, F.~Zampetti, L.~Moreno, M.~Di~Penta, A.~Marcus, G.~Bavota,
  and V.~Ng, ``Detecting missing information in bug descriptions,'' in
  \emph{Proceedings of the 2017 11th Joint Meeting on Foundations of Software
  Engineering}.\hskip 1em plus 0.5em minus 0.4em\relax New York, NY, USA: ACM,
  2017, p. 396–407.

\bibitem{li2020automated}
S.~Li, J.~Guo, M.~Fan, J.-G. Lou, Q.~Zheng, and T.~Liu, ``Automated bug
  reproduction from user reviews for android applications,'' in \emph{2020
  IEEE/ACM 42nd International Conference on Software Engineering: Software
  Engineering in Practice (ICSE-SEIP)}.\hskip 1em plus 0.5em minus 0.4em\relax
  USA: IEEE, 2020, pp. 51--60.

\bibitem{zhang2015survey}
J.~Zhang, X.~Wang, D.~Hao, B.~Xie, L.~Zhang, and H.~Mei, ``A survey on
  bug-report analysis,'' \emph{Science China Information Sciences}, vol.~58,
  no.~2, pp. 1--24, 2015.

\bibitem{uddin2017surveynew}
J.~Uddin, R.~Ghazali, M.~M. Deris, R.~Naseem, and H.~Shah, ``A survey on bug
  prioritization,'' \emph{Artificial Intelligence Review}, vol.~47, no.~2, pp.
  145--180, 2017.

\bibitem{7Shokripour:MSR13}
R.~Shokripour, J.~Anvik, Z.~M. Kasirun, and S.~Zamani, ``Why so complicated?
  simple term filtering and weighting for location-based bug report assignment
  recommendation,'' in \emph{Proceedings of the 10th Working Conference on
  Mining Software Repositories}.\hskip 1em plus 0.5em minus 0.4em\relax
  Piscataway, NJ, USA: IEEE Press, 2013, pp. 2--11.

\bibitem{10Naguib:MSR13}
H.~Naguib, N.~Narayan, B.~Br\"{u}gge, and D.~Helal, ``Bug report assignee
  recommendation using activity profiles,'' in \emph{Proceedings of the 10th
  Working Conference on Mining Software Repositories}.\hskip 1em plus 0.5em
  minus 0.4em\relax Piscataway, NJ, USA: IEEE Press, 2013, pp. 22--30.

\bibitem{45Park:AAAI2011}
J.~woo Park, M.-W. Lee, J.~Kim, S.~won Hwang, and S.~Kim, ``Costriage: A
  cost-aware triage algorithm for bug reporting systems,'' 2011.

\bibitem{Linares-Vasquez:ICSM2012}
M.~Linares-Vasquez, K.~Hossen, H.~Dang, H.~Kagdi, M.~Gethers, and
  D.~Poshyvanyk, ``Triaging incoming change requests: Bug or commit history, or
  code authorship?'' in \emph{Proceedings of the 28th IEEE International
  Conference on Software Maintenance}.\hskip 1em plus 0.5em minus 0.4em\relax
  Piscataway, NJ, USA: IEEE Press, 2012, pp. 451--460.

\bibitem{40Jeong:FSE2009}
G.~Jeong, S.~Kim, and T.~Zimmermann, ``Improving bug triage with bug tossing
  graphs,'' in \emph{Proceedings of the the 7th Joint Meeting of the European
  Software Engineering Conference and the ACM SIGSOFT Symposium on The
  Foundations of Software Engineering}.\hskip 1em plus 0.5em minus 0.4em\relax
  New York, NY, USA: ACM, 2009, pp. 111--120.

\bibitem{51Haihao:ICST2011}
H.~Shen, J.~Fang, and J.~Zhao, ``Efindbugs: Effective error ranking for
  findbugs,'' in \emph{Proceedings of the 2011 IEEE Fourth International
  Conference on Software Testing, Verification and Validation}.\hskip 1em plus
  0.5em minus 0.4em\relax Piscataway, NJ, USA: IEEE Press, 2011, pp. 299--308.

\bibitem{8Zhou:ICSE12}
J.~Zhou, H.~Zhang, and D.~Lo, ``Where should the bugs be fixed? - more accurate
  information retrieval-based bug localization based on bug reports,'' in
  \emph{Proceedings of the 34th International Conference on Software
  Engineering}.\hskip 1em plus 0.5em minus 0.4em\relax Piscataway, NJ, USA:
  IEEE Press, 2012.

\bibitem{44Kim:TOSE2013}
D.~Kim, Y.~Tao, S.~Kim, and A.~Zeller, ``Where should we fix this bug? a
  two-phase recommendation model,'' \emph{IEEE Transactions on Software
  Engineering}, vol.~39, no.~11, pp. 1597--1610, 2013.

\bibitem{13Wang:ICPC14}
S.~Wang and D.~Lo, ``Version history, similar report, and structure: Putting
  them together for improved bug localization,'' in \emph{Proceedings of the
  22Nd International Conference on Program Comprehension}.\hskip 1em plus 0.5em
  minus 0.4em\relax New York, NY, USA: ACM, 2014, pp. 53--63.

\bibitem{42Wu:ISSTA2014}
R.~Wu, H.~Zhang, S.-C. Cheung, and S.~Kim, ``Crashlocator: Locating crashing
  faults based on crash stacks,'' in \emph{Proceedings of the 2014
  International Symposium on Software Testing and Analysis}.\hskip 1em plus
  0.5em minus 0.4em\relax New York, NY, USA: ACM, 2014, pp. 204--214.

\bibitem{moreno2014use}
L.~Moreno, J.~J. Treadway, A.~Marcus, and W.~Shen, ``On the use of stack traces
  to improve text retrieval-based bug localization,'' in \emph{2014 IEEE
  International Conference on Software Maintenance and Evolution}.\hskip 1em
  plus 0.5em minus 0.4em\relax USA: IEEE, 2014, pp. 151--160.

\bibitem{14Nguyen:ASE12}
A.~T. Nguyen, T.~T. Nguyen, T.~N. Nguyen, D.~Lo, and C.~Sun, ``Duplicate bug
  report detection with a combination of information retrieval and topic
  modeling,'' in \emph{Proceedings of the 27th IEEE/ACM International
  Conference on Automated Software Engineering}.\hskip 1em plus 0.5em minus
  0.4em\relax New York, NY, USA: ACM, pp. 70--79.

\bibitem{17Wang:ICSE08}
X.~Wang, L.~Zhang, T.~Xie, J.~Anvik, and J.~Sun, ``An approach to detecting
  duplicate bug reports using natural language and execution information,'' in
  \emph{Proceedings of the 30th International Conference on Software
  Engineering}.\hskip 1em plus 0.5em minus 0.4em\relax New York, NY, USA: ACM,
  2008, pp. 461--470.

\bibitem{21Zhou:CIKM12}
J.~Zhou and H.~Zhang, ``Learning to rank duplicate bug reports,'' in
  \emph{Proceedings of the 21st ACM International Conference on Information and
  Knowledge Management}.\hskip 1em plus 0.5em minus 0.4em\relax New York, NY,
  USA: ACM, 2012, pp. 852--861.

\bibitem{32Bettenburg:ICSM08}
N.~Bettenburg, R.~Premraj, T.~Zimmermann, and S.~Kim, ``Duplicate bug reports
  considered harmful... really?'' in \emph{Proceedings of the 2008 IEEE
  International Conference on Software Maintenance}.\hskip 1em plus 0.5em minus
  0.4em\relax Piscataway, NJ, USA: IEEE Press, 2008, pp. 337--345.

\bibitem{11Bettenburg:MSR08}
------, ``Extracting structural information from bug reports,'' in
  \emph{Proceedings of the 2008 International Working Conference on Mining
  Software Repositories}.\hskip 1em plus 0.5em minus 0.4em\relax New York, NY,
  USA: ACM, 2008, pp. 27--30.

\bibitem{Song:FSE20}
Y.~Song and O.~Chaparro, \emph{BEE: A Tool for Structuring and Analyzing Bug
  Reports}.\hskip 1em plus 0.5em minus 0.4em\relax New York, NY, USA: ACM,
  2020, p. 1551–1555.

\bibitem{Soltani:ICSE'17}
M.~Soltani, A.~Panichella, and A.~van Deursen, ``A guided genetic algorithm for
  automated crash reproduction,'' in \emph{Proceedings of the 39th
  International Conference on Software Engineering}.\hskip 1em plus 0.5em minus
  0.4em\relax Piscataway, NJ, USA: IEEE Press, 2017, pp. 209--220.

\bibitem{Kifetew:ICST'14}
F.~M. {Kifetew}, W.~{Jin}, R.~{Tiella}, A.~{Orso}, and P.~{Tonella},
  ``Reproducing field failures for programs with complex grammar-based input,''
  in \emph{Proceedings of the 2014 IEEE Seventh International Conference on
  Software Testing, Verification and Validation}.\hskip 1em plus 0.5em minus
  0.4em\relax Piscataway, NJ, USA: IEEE Press, 2014, pp. 163--172.

\bibitem{Gomez:MobileSoft'16}
M.~G\'{o}mez, R.~Rouvoy, B.~Adams, and L.~Seinturier, ``Reproducing
  context-sensitive crashes of mobile apps using crowdsourced monitoring,'' in
  \emph{Proceedings of the International Conference on Mobile Software
  Engineering and Systems}.\hskip 1em plus 0.5em minus 0.4em\relax New York,
  NY, USA: ACM, 2016, pp. 88--99.

\bibitem{Chen:TSE'15}
N.~{Chen} and S.~{Kim}, ``Star: Stack trace based automatic crash reproduction
  via symbolic execution,'' \emph{IEEE Transactions on Software Engineering},
  vol.~41, no.~2, pp. 198--220, 2015.

\bibitem{Wei:TOSEM'15}
W.~Jin and A.~Orso, ``Automated support for reproducing and debugging field
  failures,'' \emph{ACM Trans. Softw. Eng. Methodol.}, vol.~24, no.~4, 2015.

\bibitem{Wei:ICSE'12}
------, ``Bugredux: Reproducing field failures for in-house debugging,'' in
  \emph{Proceedings of the 34th International Conference on Software
  Engineering}.\hskip 1em plus 0.5em minus 0.4em\relax Piscataway, NJ, USA:
  IEEE Press, 2012, pp. 474--484.

\bibitem{Cao:ASE'14}
Y.~Cao, H.~Zhang, and S.~Ding, ``Symcrash: Selective recording for reproducing
  crashes,'' in \emph{Proceedings of the 29th ACM/IEEE International Conference
  on Automated Software Engineering}.\hskip 1em plus 0.5em minus 0.4em\relax
  New York, NY, USA: ACM, 2014, pp. 791--802.

\bibitem{Zamfir:EuroSys'10}
C.~Zamfir and G.~Candea, ``Execution synthesis: A technique for automated
  software debugging,'' in \emph{Proceedings of the 5th European Conference on
  Computer Systems}.\hskip 1em plus 0.5em minus 0.4em\relax New York, NY, USA:
  ACM, 2010, pp. 321--334.

\bibitem{Nayrolles:JSS'17}
M.~Nayrolles, A.~Hamou-Lhadj, S.~Tahar, and A.~Larsson, ``A bug reproduction
  approach based on directed model checking and crash traces,'' \emph{Journal
  of Software: Evolution and Process}, vol.~29, no.~3, 2017.

\bibitem{White:ICPC15}
M.~White, M.~Linares-V\'{a}squez, P.~Johnson, C.~Bernal-C\'{a}rdenas, and
  D.~Poshyvanyk, ``Generating reproducible and replayable bug reports from
  android application crashes,'' in \emph{Proceedings of the 2015 IEEE 23rd
  International Conference on Program Comprehension}.\hskip 1em plus 0.5em
  minus 0.4em\relax Piscataway, NJ, USA: IEEE Press, 2015, p. 48–59.

\bibitem{Roehm:SANER'15}
T.~{Roehm}, S.~{Nosovic}, and B.~{Bruegge}, ``Automated extraction of failure
  reproduction steps from user interaction traces,'' in \emph{Proceedings of
  the 2015 IEEE 22nd International Conference on Software Analysis, Evolution,
  and Reengineering}.\hskip 1em plus 0.5em minus 0.4em\relax Piscataway, NJ,
  USA: IEEE Press, 2015, pp. 121--130.

\bibitem{Xuan:FSE'15}
J.~Xuan, X.~Xie, and M.~Monperrus, ``Crash reproduction via test case mutation:
  Let existing test cases help,'' in \emph{Proceedings of the 2015 10th Joint
  Meeting on Foundations of Software Engineering}.\hskip 1em plus 0.5em minus
  0.4em\relax New York, NY, USA: ACM, 2015, pp. 910--913.

\bibitem{bugclipper}
``Bugclipper \url{http://bugclipper.com},'' 2021.

\bibitem{bugsee}
``\url{https://www.bugsee.com},'' 2021.

\bibitem{android-monkey}
\BIBentryALTinterwordspacing
(2021) Android ui/application exerciser monkey. [Online]. Available:
  \url{http://developer.android.com/tools/help/monkey.html}
\BIBentrySTDinterwordspacing

\bibitem{Machiry:FSE13}
A.~Machiry, R.~Tahiliani, and M.~Naik, ``Dynodroid: An input generation system
  for android apps,'' in \emph{Proceedings of the 2013 9th Joint Meeting on
  Foundations of Software Engineering}.\hskip 1em plus 0.5em minus 0.4em\relax
  New York, NY, USA: ACM, 2013, pp. 224--234.

\bibitem{Sasnauskas:WODA14}
R.~Sasnauskas and J.~Regehr, ``Intent fuzzer: Crafting intents of death,'' in
  \emph{Proceedings of the 2014 Joint International Workshop on Dynamic
  Analysis and Software and System Performance Testing, Debugging, and
  Analytics}.\hskip 1em plus 0.5em minus 0.4em\relax New York, NY, USA: ACM,
  2014.

\bibitem{Ravindranath:Mobisys2014}
L.~Ravindranath, S.~Nath, J.~Padhye, and H.~Balakrishnan, ``Automatic and
  scalable fault detection for mobile applications,'' in \emph{Proceedings of
  the 12th Annual International Conference on Mobile Systems, Applications, and
  Services}.\hskip 1em plus 0.5em minus 0.4em\relax New York, NY, USA: ACM,
  2014, pp. 190--203.

\bibitem{Amalfitano:ASE12}
D.~Amalfitano, A.~R. Fasolino, P.~Tramontana, S.~De~Carmine, and A.~M. Memon,
  ``Using gui ripping for automated testing of android applications,'' in
  \emph{Proceedings of the 27th IEEE/ACM International Conference on Automated
  Software Engineering}.\hskip 1em plus 0.5em minus 0.4em\relax New York, NY,
  USA: ACM, 2012, pp. 258--261.

\bibitem{Anand:FSE12}
S.~Anand, M.~Naik, M.~J. Harrold, and H.~Yang, ``Automated concolic testing of
  smartphone apps,'' in \emph{Proceedings of the ACM SIGSOFT 20th International
  Symposium on the Foundations of Software Engineering}.\hskip 1em plus 0.5em
  minus 0.4em\relax New York, NY, USA: ACM, 2012, pp. 59--69.

\bibitem{Azim:OOPSLA13}
T.~Azim and I.~Neamtiu, ``Targeted and depth-first exploration for systematic
  testing of android apps,'' in \emph{Proceedings of the 2013 ACM SIGPLAN
  International Conference on Object Oriented Programming Systems Languages \&
  Applications}.\hskip 1em plus 0.5em minus 0.4em\relax New York, NY, USA: ACM,
  2013, pp. 641--660.

\bibitem{Moran:ICST16}
K.~Moran, M.~Linares-V{\'a}squez, C.~Bernal-C{\'a}rdenas, C.~Vendome, and
  D.~Poshyvanyk, ``Automatically discovering, reporting and reproducing android
  application crashes,'' in \emph{Proceedings of the IEEE International
  Conference on Software Testing, Verification and Validation}.\hskip 1em plus
  0.5em minus 0.4em\relax Piscataway, NJ, USA: IEEE Press, 2016, pp. 33--44.

\bibitem{google-robo-test}
\BIBentryALTinterwordspacing
(2021) Google firebase test lab robo test. [Online]. Available:
  \url{https://firebase.google.com/docs/test-lab/robo-ux-test}
\BIBentrySTDinterwordspacing

\bibitem{Amalfitano:IEEE14}
D.~{Amalfitano}, A.~R. {Fasolino}, P.~{Tramontana}, B.~D. {Ta}, and A.~M.
  {Memon}, ``Mobiguitar: Automated model-based testing of mobile apps,''
  \emph{IEEE Software}, vol.~32, no.~5, pp. 53--59, 2015.

\bibitem{Zaeem:ICST2014}
R.~N. Zaeem, M.~R. Prasad, and S.~Khurshid, ``Automated generation of oracles
  for testing user-interaction features of mobile apps,'' in \emph{Proceedings
  of the 2014 IEEE International Conference on Software Testing, Verification,
  and Validation}.\hskip 1em plus 0.5em minus 0.4em\relax Washington, DC, USA:
  IEEE Computer Society, 2014, pp. 183--192.

\bibitem{Yang:FASE13}
W.~Yang, M.~R. Prasad, and T.~Xie, ``A grey-box approach for automated
  gui-model generation of mobile applications,'' in \emph{Proceedings of the
  16th International Conference on Fundamental Approaches to Software
  Engineering}.\hskip 1em plus 0.5em minus 0.4em\relax Berlin, Heidelberg:
  Springer-Verlag, 2013, pp. 250--265.

\bibitem{Zhang:ICSE17}
H.~Zhang and A.~Rountev, ``Analysis and testing of notifications in android
  wear applications,'' in \emph{Proceedings of the 2017 IEEE/ACM 39th
  International Conference on Software Engineering}.\hskip 1em plus 0.5em minus
  0.4em\relax Piscataway, NJ, USA: IEEE Press, 2017, pp. 347--357.

\bibitem{Mahmood:FSE14}
R.~Mahmood, N.~Mirzaei, and S.~Malek, ``Evodroid: Segmented evolutionary
  testing of android apps,'' in \emph{Proceedings of the 22Nd ACM SIGSOFT
  International Symposium on Foundations of Software Engineering}.\hskip 1em
  plus 0.5em minus 0.4em\relax New York, NY, USA: ACM, 2014, pp. 599--609.

\bibitem{Mao:ISSTA16}
K.~Mao, M.~Harman, and Y.~Jia, ``Sapienz: Multi-objective automated testing for
  android applications,'' in \emph{Proceedings of the 25th International
  Symposium on Software Testing and Analysis}.\hskip 1em plus 0.5em minus
  0.4em\relax New York, NY, USA: ACM, 2016, pp. 94--105.

\bibitem{Jensen:ISSTA13}
C.~S. Jensen, M.~R. Prasad, and A.~M\o{}ller, ``Automated testing with targeted
  event sequence generation,'' in \emph{Proceedings of the 2013 International
  Symposium on Software Testing and Analysis}.\hskip 1em plus 0.5em minus
  0.4em\relax New York, NY, USA: ACM, 2013, p. 67–77.

\bibitem{Mirzaei:ISSRE15}
N.~Mirzaei, H.~Bagheri, R.~Mahmood, and S.~Malek, ``Sig-droid: Automated system
  input generation for android applications,'' in \emph{Proceedings of the 2015
  IEEE 26th International Symposium on Software Reliability Engineering}.\hskip
  1em plus 0.5em minus 0.4em\relax USA: IEEE Computer Society, 2015, p.
  461–471.

\bibitem{grano2018exploring}
G.~Grano, A.~Ciurumelea, S.~Panichella, F.~Palomba, and H.~C. Gall, ``Exploring
  the integration of user feedback in automated testing of android
  applications,'' in \emph{2018 IEEE 25Th international conference on software
  analysis, evolution and reengineering (SANER)}.\hskip 1em plus 0.5em minus
  0.4em\relax USA: IEEE, 2018, pp. 72--83.

\bibitem{pelloni2018becloma}
L.~Pelloni, G.~Grano, A.~Ciurumelea, S.~Panichella, F.~Palomba, and H.~C. Gall,
  ``Becloma: Augmenting stack traces with user review information,'' in
  \emph{2018 IEEE 25th International Conference on Software Analysis, Evolution
  and Reengineering (SANER)}.\hskip 1em plus 0.5em minus 0.4em\relax USA: IEEE,
  2018, pp. 522--526.

\bibitem{Mao:2017}
K.~Mao, M.~Harman, and Y.~Jia, ``Crowd intelligence enhances automated mobile
  testing,'' in \emph{Proceedings of the 32nd IEEE/ACM International Conference
  on Automated Software Engineering}.\hskip 1em plus 0.5em minus 0.4em\relax
  Piscataway, NJ, USA: IEEE Press, 2017, p. 16–26.

\end{thebibliography}


%\vfill


\begin{IEEEbiography}[{\includegraphics[width=1in,height=1.25in,clip,keepaspectratio]{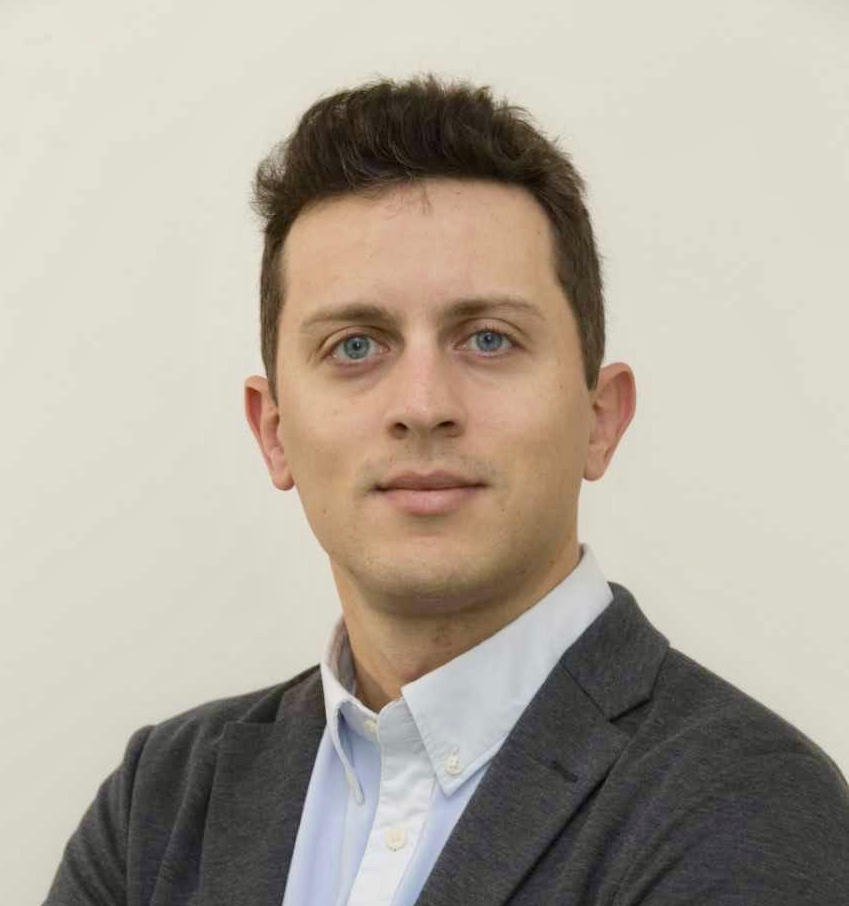}}]{Mattia Fazzini} is an Assistant Professor in the Department of Computer Science \& Engineering at the University of Minnesota. He completed his Ph.D. in Computer Science from the Georgia Institute of Technology in 2019. His research interests lie primarily in the area of software engineering, with emphasis on techniques for improving software quality. The central theme of his research is the development of approaches for testing and maintenance of mobile apps.  He has published in several top peer-reviewed software engineering venues including: ASE, ICSE, ICST, and ISSTA. He received the Facebook Testing and Verification award in 2019 for his research on mobile app testing. More information is available at \url{https://www-users.cs.umn.edu/~mfazzini/}.
\end{IEEEbiography}

\begin{IEEEbiography}[{\includegraphics[width=1in,height=1.25in,clip,keepaspectratio]{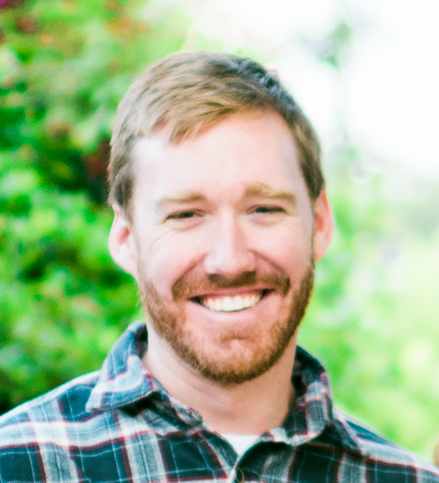}}]{Kevin Moran} is currently an Assistant Professor at George Mason University where he directs the SAGE research group. %He graduated with a B.A. in Physics from the College of the Holy Cross in 2013 and an M.S. degree from William \& Mary in August of 2015.
He received a Ph.D. degree from William \& Mary in August 2018. His main research interest involves facilitating the processes of software engineering, maintenance, and evolution with a focus on mobile platforms. He has published in several top peer-reviewed software engineering venues including: ICSE, ESEC/FSE, TSE, USENIX, ICST, ICSME, and MSR. His research has been recognized with ACM SIGSOFT distinguished paper awards at ESEC/FSE 2019 and ICSE 2020.
%, and a Best Paper Award at CODASPY'19. 
He is a member of the ACM and IEEE. More information is available at \url{http://www.kpmoran.com}
\end{IEEEbiography}

\begin{IEEEbiography}[{\includegraphics[width=1in,height=1.25in,clip,keepaspectratio]{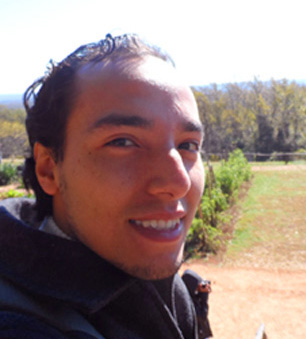}}]{Carlos Bernal-C\'ardenas}
%received the B.S. degree in systems engineering from the Universidad Nacional de Colombia in 2012 and his M.E. in Systems and Computing Engineering in 2015. He 
is currently Ph.D. candidate in Computer Science at the College of William \& Mary as a member of the SEMERU research group advised by Dr Denys Poshyvanyk. His research interests include software engineering, software evolution and maintenance, information retrieval, software reuse, mining software repositories, mobile applications development, and user experience. He has published in several top peer-reviewed software engineering venues including: ICSE, ESEC/FSE, ICST, and MSR.  He has also received the ACM SigSoft Distinguished paper award at ESEC/FSE'15 \& '19 and ICSE'20. 
%Bernal-C\'ardenas is a student member of IEEE and ACM and has served as an external reviewer for ICSE, ICSME, FSE, APSEC, and SCAM.
More information is available at \url{http://www.cs.wm.edu/~cebernal/}.
\end{IEEEbiography}

\vskip -2.5\baselineskip plus -1fil
\begin{IEEEbiography}[{\includegraphics[width=1in,height=1.25in,clip,keepaspectratio]{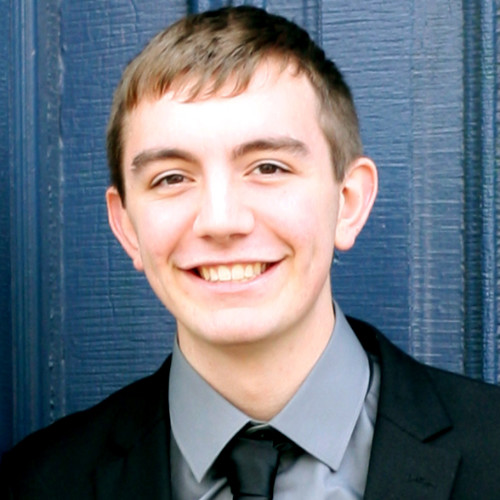}}]{Tyler Wendland} is a graduate student at the University of Minnesota. He graduated with a B.S. in Computer Science from the University of Minnesota in 2020.  His research focuses on devising automated software engineering techniques to improve bug reporting processes. 
\end{IEEEbiography}

\vskip -2.5\baselineskip plus -1fil

\begin{IEEEbiography}[{\includegraphics[width=1in,height=1.25in,clip,keepaspectratio]{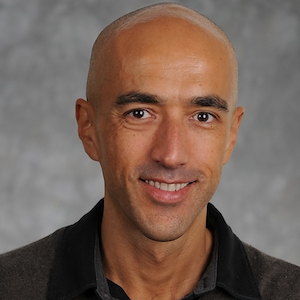}}]{Alessandro Orso} is a Professor in the School of Computer Science and an Associate Dean in the College of Computing at the Georgia Institute of Technology. He received his M.S. degree in Electrical Engineering (1995) and his Ph.D. in Computer Science (1999) from Politecnico di Milano, Italy. From March 2000, he has been at Georgia Tech. His area of research is software engineering, with emphasis on software testing and program analysis. His interests include the development of techniques and tools for improving software reliability, security, and trustworthiness, and the validation of such techniques on real-world systems. Dr. Orso has received funding for his research from both government agencies, such as DARPA, ONR, and NSF, and industry, such as Facebook, Fujitsu Labs, Google, IBM, and Microsoft. He served on the editorial boards of ACM TOSEM and on the Advisory Board of Reflective Corp, and he is serving on the editorial boards of IEEE TSE. Ha also served as program chair for ACM SIGSOFT ISSTA 2010, program co-chair for IEEE ICST 2013, ACM-SIGSOFT FSE 2014, and ACM-SIGSOFT/IEEE ICSE 2017, and as a technical consultant to DARPA. Dr. Orso has received multiple Best Paper and ACM Distinguished Paper Awards and three Impact and Most Influential Paper Awards. Dr. Orso is a Distinguished Member of the ACM and an IEEE Fellow. More information is available at \url{http://www.cc.gatech.edu/~orso/}.
\end{IEEEbiography}

\vskip -2.5\baselineskip plus -1fil

\begin{IEEEbiography}[{\includegraphics[width=1in,height=1.25in,clip,keepaspectratio]{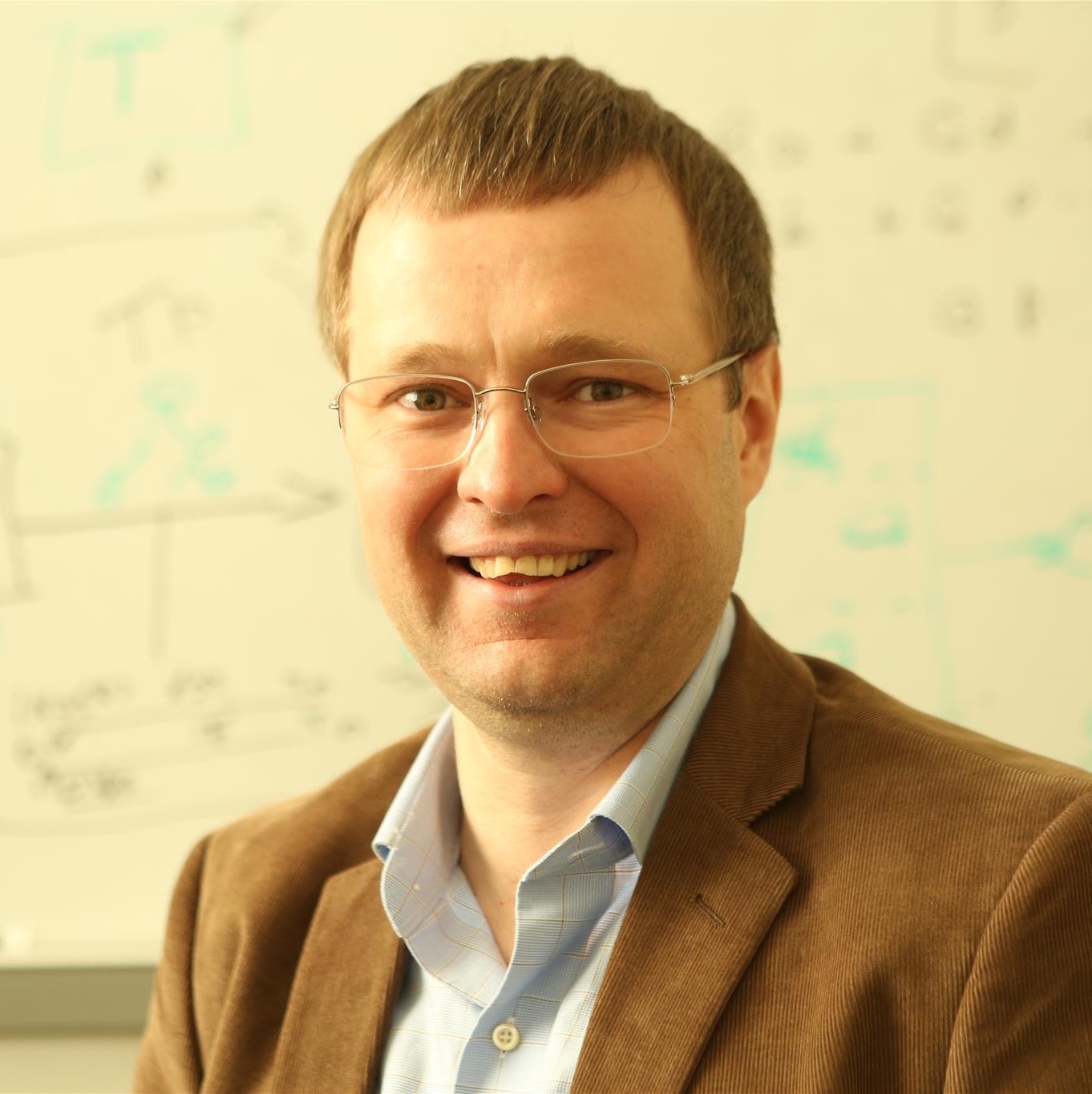}}]{Denys Poshyvanyk} is a Professor of Computer Science at William and Mary. He received the MS and MA degrees in Computer Science from the National University of Kyiv-Mohyla Academy, Ukraine, and Wayne State University in 2003 and 2006, respectively. He received the PhD degree in Computer Science from Wayne State University in 2008. He served as a program co-chair for ASE'21, MobileSoft'19, ICSME'16, ICPC'13, WCRE'12 and WCRE'11. He currently serves on the editorial board of IEEE Transactions on Software Engineering (TSE), Empirical Software Engineering Journal (EMSE, Springer), Journal of Software: Evolution and Process (JSEP, Wiley) and Science of Computer Programming. His research interests include software engineering, software maintenance and evolution, program comprehension, reverse engineering and software repository mining. His research papers received several Best Paper Awards at ICPC'06, ICPC'07, ICSM'10, SCAM'10, ICSM'13, CODAPSY'19 and ACM SIGSOFT Distinguished Paper Awards at ASE'13, ICSE'15, ESEC/FSE'15, ICPC'16, ASE'17, ESEC/FSE'19 and ICSE'20. He also received the Most Influential Paper Awards at ICSME'16, ICPC'17 and ICPC'20. He is a recipient of the NSF CAREER award (2013).  He is a member of the IEEE and ACM. More information is available at: \url{http://www.cs.wm.edu/~denys/}
\end{IEEEbiography}

\balance

\end{document}